\documentstyle[aps,epsf,twocolumn,draft]{revtex}
\def\be{\begin{eqnarray}}
\def\bea{\begin{eqnarray}}
\def\bma{\begin{mathletters}}
\def\ee{\end{eqnarray}}
\def\eea{\end{eqnarray}}
\def\ema{\end{mathletters}}

\begin{document}
\author{V. Vedral$^1$}
\title{The Role of Relative Entropy in Quantum Information Theory}
\address{Centre for Quantum Computation, Clarendon Laboratory, University of Oxford,\\
Parks Road OX1 3PU}
\date{\today}
\maketitle

\begin{abstract}
Quantum mechanics and information theory are among the most
important scientific discoveries of the last century. Although
these two areas initially developed separately it has emerged that
they are in fact intimately related. In this review I will show
how quantum information theory extends traditional information
theory by exploring the limits imposed by quantum, rather than
classical mechanics on information storage and transmission. The
derivation of many key results uniquely differentiates this review
from the "usual" presentation in that they are shown to follow
logically from one crucial property of relative entropy. Within
the review optimal bounds on the speed-up that quantum computers
can achieve over their classical counter-parts are outlined using
information theoretic arguments. In addition important
implications of quantum information theory to thermodynamics and
quantum measurement are intermittently discussed. A number of
simple examples and derivations including quantum super-dense
coding, quantum teleportation, Deutsch's and Grover's algorithms
are also included.
\end{abstract}

\tableofcontents

\vspace*{.5cm}

\stepcounter{footnote}
\footnotetext{Permanent address from October $2000$: Blackett
Laboratory, Imperial College, London SW7 2BZ}

\noindent

\section{Introduction}

Quantum physics not only provides the most complete description of
physical phenomena known to man, it also provides a new
philosophical framework for our understanding of nature. It
enables us to accurately model microscopic systems such as quarks
and atoms to large cosmic objects such as black holes. Information
theory, on the other hand, teaches us about our physical ability
to store and process information.  Without a formalised
information theory many of the recent developments in
telecommunications, computer science and engineering would simply
not have been possible. Although quantum physics and information
theory initially developed separately, their recent integration is
seen as yet another important step towards understanding the
fundamental properties and limitations of nature.

One of the central information theoretic concepts in science is
that of distinguishability. Inevitably an animal's survival
depends on its ability to distinguish a mate from a predator or a
prey. In the same way, physical experiments aim to be sensitive
enough to be able to distinguish one hypothesis from another. It
is however no surprise that the influence of the concept of
distinguishability is felt far beyond science. Life consists of a
series of decisions that have to be made. This we do, consciously
or unconsciously, by evaluating all the alternatives and
distinguishing consequences of various different alternative
actions.

The purpose of this review is to show that the apparently simple
concept of distinguishability is at the root of information
processing. Ultimately, how well we can distinguish different
physical states determines how much information we can encode into
a certain system and how quickly we can manipulate it.
Distinguishability in turn is completely dependent upon the laws
of physics, and quantum physics naturally allows for more
versatile information processing than classical physics.  The
reasoning behind this is that unlike with classical states, two
different quantum states are not necessarily fully
distinguishable.  It is interesting to note that although this at
first seems like a limitation, it in fact presents us with
significantly more possibilities for information encoding and
transmission.

In this review I first plan to argue that the
relative entropy is the most appropriate quantity to measure
distinguishability between different states.
The proper framework to talk about states is, of course,
quantum mechanics, so it is necessary to define the quantum relative entropy. I
prove that the relative entropy, both classical and quantum, does not
increase with time. Thus two states can only become less distinguishable
as they undergo any kind of evolution. This result will be central to my review
as subsequent results will follow from this simple fact.

I then go on to show that the `no increase of relative entropy'
principle tells us about the ability of quantum states to store
and process information. Information has to be encoded and
manipulated in physical systems. Therefore, distinguishability of
different states within a physical system is a prerequisite.
Looking at this from the point of view of communication, what does
it mean to send and receive a message? Sending a message
successfully means encoding the information we wish to send into a
structured format which the receiver must be able to unambiguously
distinguish. The communication capacity can then be thought of to
be the rate at which we can send and receive messages. The rate of
successful transmission is determined by the relative entropy
between various encoding states.

What is less obvious, but nonetheless equally true, is that computation can
also be viewed as a special kind of communication. This will allow the use of
the relative entropy to quantify the efficiency (i.e. speed) of
quantum computation in general.

The role of measurement within
quantum mechanics and therefore information theory is paramount.
Classically the measurement process is implicit because physical quantities have
well defined pre-existing properties. For example, a classical bit is either
in the state $0$ or $1$, whereas a quantum bit can exist in a combination
of the two states. At the end, a measurement is necessary to "collapse" this
combination to a classical result which we can then read.
The very concept of efficiency of a measurement can also be
quantified using the relative entropy. A measurement, like a communication
process, creates correlations between a system and an
apparatus with the purpose of the apparatus receiving an
amount of information from the system. The opposite of this process, namely,
deleting of information, can be seen to be at the root of irreversibility and
this invariably contributes to an increase in the entropy of the
environment. This amount is exactly quantified using
the relative entropy between the environmental state and the apparatus
state and provides an exciting link between information theory, computation,
thermodynamics and quantum mechanics. But, before we reach this exciting stage, our long
journey has to begin with a much simpler question: how do we quantify
uncertainty in a physical state?

\section{Relative entropy}

Fundamental to our understanding of distinguishability is the measure of
{\em uncertainty} in a given probability distribution. This uncertainty can
be quantified by introducing the idea of ``{\em surprise}". Suppose that a
certain event happens with a probability $p$. Then we would like to quantify
how surprised we are when that event {\em does} happen. The first guess would be $%
1/p$: the smaller the probability of an event, the more surprised we are
when the event happens and vice versa. However, an event might be composed of two
independent events which happen with probabilities $q$ and $r$ respectively,
so that the probability of both events occuring is
$p = q \times r$. We would now intuitively expect that the surprise
of $p$ is the same as the surprise of $q$ plus the surprise of $r$. But, $%
1/p \neq 1/q+1/r$, so that $1/p$ is not really a satisfactory
definition from this perspective. Instead, if we define surprise
as $\ln (1/p)$, then the above property called {\em additivity}
is satisfied since $-\ln p_1p_2 = -\ln p_1 -\ln p_2$. With a
probability distribution $\sum_n p_n = 1 $, the total uncertainty
is just the average of all the surprises. Additivity of
uncertainties of statistically independent events is such a
stringent condition that it basically leads to a unique measure
(Shannon and Weaver, 1949) up to a constant and logarithm base.

\noindent {\bf Definition}. The uncertainty in a collection of possible
states $a_i$ with corresponding probability distribution $p(a_i)$ is given
by its {\em entropy}
\begin{eqnarray}
S(p) := -\sum_i p(a_i) \ln p(a_i)
\end{eqnarray}
called the {\em Shannon entropy}. We note that there is no Boltzmann
constant term in this expression, as there is for the physical entropy,
since it is by convention set to unity. This measure is suitable for
the states of systems described by the laws of classical physics, but
it will have to be changed, along with other classical measures when we
present the quantum information theory.

I ultimately wish to be able to talk about storing and processing
information. For this we require a means of comparing two
different probability distributions, which is why I introduce the
notion of {\em relative entropy} (first introduced by Kullback and
Leibler, 1951). Suppose that a collection of events has the
probability distribution $\{p_i\}$, but we {\em mistakingly} think
that this probability distribution is $\{q_i\}$. For example, we
have a coin which we think is fair, i.e. the probability of a head
or a tail is equal. If we toss this coin $n$ times, on average we
expect heads half of the time and tails the other half. In
reality, the coin by virtue of its uneven weight distribution will
not be completely fair so that our expectation will turn out to be
wrong. There will consequently be a discrepancy between our
expected and real probability distribution. This discrepancy is
very frequently the case in real life and it is, in fact, very
rare that we have complete information about any event. Therefore
we can formalise that when a particular outcome $j$ happens, we
associate the surprise $-\ln q_j$ with it. The average surprise,
or information, according to this erroneous belief, is
\begin{eqnarray}
-\sum_i p_i \ln q_i \; . \nonumber
\end{eqnarray}
Since events happen with probabilities $\{p_i\}$ (in spite of
our belief!) these are the correct
ones to feature in the averaging process. However, the real amount of information we
are obtaining is, as defined before, given by the Shannon entropy
$S(p)=-\sum_i p_i \ln p_i$. It is not so difficult to show that
$S(p)\leq -\sum_i p_i \ln q_i$ (equality holds if and only if $p_i=q_i$ for all $i$)
so that there is an "uncertainty deficit" as it were
stemming from our wrong assumption and is equal to the difference between
the two averages. This deficit quantity is called the relative entropy.

\vspace*{.3cm}
%\noindent
%\fbox{\parbox[b]{17.7cm}{
{\bf Definition}.
Suppose that we have two sets of discrete events
$a_{i}$ and $b_{j}$ with the corresponding probability distributions, ${%
p(a_{i})}$ and ${p(b_{j})}$. {\em The relative entropy} between these two
distributions is defined as
\begin{eqnarray}
S(p(a)\,||\,p(b)):=\sum_{i}p(a_{i})\ln \frac{p(a_{i})}{p(b_{i})}\;.
\end{eqnarray}
%}}
%\vspace*{.3cm}

This function is a measure of the `distance' between ${p(a_{i})}$ and ${%
p(b_{j})}$, even though, strictly speaking, it is not a mathematical
metric since it fails to be symmetric $S(p(a)\,||\,p(b))\neq S(p(b)\,||\,p(a))$.
This is interesting since at first it looks as if there should be no difference
between mistaking the probability distribution $p_i$ for $q_i$, or vice versa.
Intuitively this can be explained using our coin example. Suppose that
someone gives us a coin which is either fair or completely unfair, e.g.
always gives heads. Now we have to toss this coin a number of times
and infer which of the two coins we have. If we
toss the fair coin and obtain tails, then our inference will immediately
be that we have the fair coin. If, however, we obtain heads, then it
could be either coin. By tossing more times, the fair coin would
eventually give us a tails. If, however, we were holding the completely
unfair coin from the beginning, then even after $100$ heads we can never really
eliminate the fair coin since this outcome is statistically possible
(although highly unlikely). Therefore how certain we are about which
coin we hold is clearly dependent on whichever coin we hold and
how different it is to the other one. As we will see shortly
our uncertainty is quantified by the relative entropy and
it is thus to be expected that it is asymmeric. I now describe this
statistical approach in more detail.

\subsection{Statistical significance}

A more operational interpretation of both the Shannon entropy and
the relative entropy comes from the statistical point of view. The
generalization of this formalism to the quantum domain will be
presented in the next section and we will offer an operational
interpretation of the measures of quantum correlations to be
introduced therein. I now follow the approaches of Cover and
Thomas (Cover and Thomas, 1991), and Csisz\'ar and K\"orner
(Csiszar and Korner, 1981) and the reader interested in more
detail should consult these two books.

Let $X_1, X_2, ...X_n$ be a sequence of $n$ symbols from an
alphabet $A= \{a_1,a_2,...,a_{|A|}\}$, where $|A|$ is the size of
the alphabet. We denote a sequence $x_1,x_2,...,x_n$ by $x^n$
or, equivalently, by ${\bf x}$. The type $P_{{\bf x}}$ of a sequence $%
x_1,x_2,...,x_n$ will be called the relative proportion of occurences of
each symbol of $A$, i.e. $P_{{\bf x}}(a) = N(a|{\bf x})/n$ for all $a\in A$,
where $N(a|{\bf x)}$ is the number of times the symbol $a$ occurs in the
sequence ${\bf x} \in A^n$. Thus, according to this definition the
sequences $011010$ and $100110$ are of the same type.
${\cal P}_n$ will denote the set of types with
denominator $n$. If $P\in {\cal P}_n$, then the set of sequences of length $n$
and type $P$ is called the {\em type class} of $P$, denoted by $T(P)$,
i.e. mathematically
\begin{eqnarray}
T(P) = \{ {\bf x} \in A^n : P_{{\bf x}} = P \} \; . \nonumber
\end{eqnarray}
We now approach the first theorem about types which is at the heart of
success of this theory and states that the number of types increases only
polynomially with $n$.

\noindent {\bf Theorem 1}.
\begin{eqnarray}
|{\cal P}_n | \le (n+1)^{|A|} \nonumber
\end{eqnarray}
Proof of this is left for the reader, but the rationale is simple.
Suppose that we generate an $n$-string of $0$s and $1$s. The
number of different types is then $n+1$, i.e. polynomial in $n$:
the zeroth type has only one string - all zeros, the first type
has $n$ strings - all strings containing exactly one $1$, the
second type has $n(n-1)/2$ strings - all those containing exactly
two $1$s, and so on, the $n$th type has only one sequence - all
ones. The most important point is that the number of sequences is
exponential in $n$, so that at least one type has exponentially
many sequences in its type class, since there are only
polynomially many different types. A simple example is a coin
tossed $n$ times. If it is a fair coin, then we expect heads half
of the time and tails other half of the time. The number of all
possible sequences for this coin is $2^n$ (i.e. exponential in
$n$) where each sequence is equally likely (with probability
$2^{-n}$). However, the size of the type class where there is an
equal number of heads and tails is $C^n_{n/2}$ (the number of
possible ways of choosing $n/2$ element out of $n$ elements), the
log of which tends to $n$ for large $n$. Hence this type class is
in some sense asymptotically as large as all the type classes
together.

We now arrive at a very important theorem for us, which, in fact,
presents the basis of the statistical interpretation of the
Shannon entropy and relative entropy.

\noindent {\bf Theorem 2}. If $X_1, X_2, ...X_n$ are drawn according to $Q(x)$,
then the probability of ${\bf x}$ depends only on its type and is given by
\begin{eqnarray}
Q^n({\bf x}) = e^{-n(S(P_{{\bf x}}) + S(P_{{\bf x}} || Q))} \nonumber
\end{eqnarray}
\noindent
{\bf Proof}.
\begin{eqnarray}
Q^n({\bf x}) & = & \prod_{i=1}^n Q(x_i) = \prod_{a \in A} Q(a)^{N(a|{\bf x})}\nonumber\\
& = & \prod_{a \in A} Q(a)^{nP_{\bf x}(a)} = \prod_{a \in A} e^{nP_{\bf x}(a)\ln Q(a)} \nonumber\\
& = &
\exp\bigg \{ n\sum_{a\in A} - P_{\bf x}(a)\ln \frac{P_{\bf x}(a)}{Q(a)}
+ P_{\bf x}(a)\ln P_{\bf x}(a)\bigg \} \nonumber\\
& = &  e^{-n(S(P_{\bf x}) + S(P_{\bf x} || Q))} \nonumber\; {}_{\Box}
\end{eqnarray}
Therefore a probability of a sequence becomes exponentially small
as $n$ increases. Indeed, our coin tossing example shows this: a
probability for any particular sequence (such as e.g.
$0000011111$) is $2^{-n}$ (note: the reason that we are using $e$
in our theorems instead of $2$ is because we are also using $\ln$
instead of $\log$). This is explicitly stated in the following
corollary.

\noindent
{\bf Corollary}. If ${\bf x}$ is the type class of $Q$, then
\begin{eqnarray}
Q^n({\bf x}) = e^{-nS(Q)}\nonumber
\end{eqnarray}
The proof is left to the reader.

So, as $n$ gets large, most of the sequences become typical and
they are all equally likely. Therefore the probability of every
typical sequence times the number of typical sequences has to be
equal to unity in order to conserve total probability
($e^{-nS(Q)}N=1$). From this we can see that the number of typical
sequences is $N=e^{nS(Q)}$ (we turn to this point more formally
next). Hence, the above theorem has very important implications in
the theory of statistical inference and distinguishability of
probability distributions. To see how this comes about we state
two theorems that give bounds on the size of and probability of a
particular type class. The proofs follow directly from the above
two theorems and the corollary (Cover and Thomas, 1991; Csiszar
and Korner, 1981).

\noindent {\bf Theorem 3}. For any type $P\in {\cal P}_n$,
\begin{eqnarray}
\frac{1}{(n+1)^{|A|}} e^{nS(P)} \le |T(P)| \le e^{nS(P)} \nonumber
\end{eqnarray}
This theorem provides the exact bounds on the number of "typical"
sequences. Suppose that we have a probability distribution $p_1$ and $p_2$
for heads and tails respectively and we toss the coin $n$ times.
The typical (most likely) sequence will be the one where we
have $p_1n$ heads and $p_2n$ tails. The number of such sequences is
\begin{eqnarray}
C^n_{p_1n} = \frac{n!}{(p_1n)!(p_2n)!}\sim e^{n(-p_1\ln p_1-p_2\ln p_2}),\nonumber
\end{eqnarray}
i.e. an exponential in $n$ (more tosses, more possibilities) and entropy
(higher uncertainty, more possibilities). The next theorem offers a statistical
interpretation to the relative entropy.

\noindent
{\bf Theorem 4}. For any type $P\in {\cal P}_n$, and any
distribution Q, the probability of the type class $T(P)$ under $Q^n$ is $%
e^{-nS(P||Q)}$ to first order in the exponent. More precisely,
\begin{eqnarray}
\frac{1}{(n+1)^{|A|}} e^{-nS(P||Q)} \le Q^n(T(P)) \le e^{-nS(P||Q)} \nonumber
\end{eqnarray}
The meaning of this theorem is that if we draw results according to $Q$
the probability that it will "look" as if was drawn from $P$ is exponentially
decreasing with $n$ and relative entropy between $P$ and $Q$. The closer
$Q$ is to $P$ the higher the probability that their statistics will look the
same. Alternatively, the higher the number of draws, $n$, the smaller the
probability that we will confuse the two. We present an explicit example below.
The above two results can be succinctly written in an exponential fashion
that will be useful to us as
\begin{eqnarray}
|T(P)| & \rightarrow & e^{-nS(P)} \label{21.exp} \\
Q^n(T(P)) &\rightarrow & e^{-nS(P||Q)} \; .  \label{2.exp}
\end{eqnarray}
The first statement also leads to the idea of {\em data
compression}, where a string of length $n$ generated by a source with
entropy $S$ can be encoded into a string of length $nS$. The second
statement says that if we are performing $n$ experiments according to
distribution $Q$, the probability that we will get something that looks as
if it was generated by distribution $P$ decreases exponentially with $n$
depending on the relative entropy between $P$ and $Q$. This idea immediately
leads to Sanov's theorem, whose quantum analogue will provide a statistical
interpretation of one measure of entanglement presented in section IV.
Now we present examples of data compression and introduce Sanov's theorem.

\vspace*{.2cm} \noindent {\bf Classical data compression.} Suppose
that we have a binary source generating 0's with twice as big a
probability as that of 1's, so that the Shannon entropy is $S= \ln
3 - 2/3 \ln 2 = 0.64$. Imagine that we have a string of 15 digits
coming out of this source. Then, according to the above
considerations (eq. (\ref{21.exp})) , the most likely type will be
the one with ten 0's and five 1's. But the size of this class is
only $0.64 \times 15 \approx 10 $. So we can use only 10 digits to
encode all the above sequences of 15 numbers just by assigning the
following conventional mapping: the first sequence of 15 numbers
is to be encoded in $0000000000$, the second sequence is to be
encoded in $0000000001$, ... , the $e^{10}$th sequence is to be
encoded in $1111111111$. This encoding is for obvious reasons
called data compression. This, in fact, offers a statistical
reason for employing the Shannon entropy as a measure of
uncertainty. This result is known as Shannon's lower bound (or
Shannon's First Theorem) on data compression, i.e. a message
cannot be compressed per bit to less than its Shannon entropy
(Shannon and Weaver, 1949). There are a number of different
methods used for compression each with varying degree of success
dependent on the statistical distribution of the message, see e.g.
Cover and Thomas (1991).

Now we look at the distinguishability of two probability distributions.
Suppose we would like to check if a given coin is ``fair", i.e. if it
generates a ``head--tail" distribution of $f=(1/2,1/2)$. When the coin is
biased then it will produce some other distribution, say $uf=(1/3,2/3)$. So,
our question of the coin fairness boils down to how well we can
differentiate between two given probability distributions given a finite, $n$%
, number of experiments to perform on one of the two
distributions. In the case of a coin we would toss it $n$ times
and record the number of 0's and 1's. From simple statistics
(Cover and Thomas, 1991) we know that if the coin is fair than the
number of 0's, $N(0)$, will be roughly $n/2-\sqrt{n} \le N(0) \le
n/2+\sqrt{n}$, for large $n$, and the same for the number of 1's.
So if our experimentally determined values do not fall within the
above limits the coin is not fair. We can look at this from
another point of view which is in the spirit of the method of
types; namely, what is the probability that a fair coin will be
mistaken for an unfair one with the distribution of $(1/3,2/3)$
given n trials of the fair coin? For large $n$ the answer is given
in the previous subsection
\begin{eqnarray}
p(\mbox{fair}\rightarrow \mbox{unfair}) = e^{-nS(uf||f)} \,\, , \nonumber
\end{eqnarray}
where $S_{cl}(uf||f) = 1/3 \ln 1/3 + 2/3 \ln 2/3 - 1/3 \ln 1/2 -2/3 \ln 1/2$
is the Shannon relative entropy for the two distributions. So,
\begin{eqnarray}
p(\mbox{fair}\rightarrow \mbox{unfair}) = 3^n 2^{-\frac{5}{3}n} \,\, , \nonumber
\end{eqnarray}
which tends exponentially to zero with $n \rightarrow \infty$. In
fact we see that already after $\sim 20$ trials the probability of
mistaking the two distributions is vanishingly small, $<
10^{-10}$. This leads to the following important result (Sanov,
1957).

%%SANOV FIG 1

\begin{figure}[ht]
\begin{center}
\hspace{0mm}
\epsfxsize=7.5cm
\epsfbox{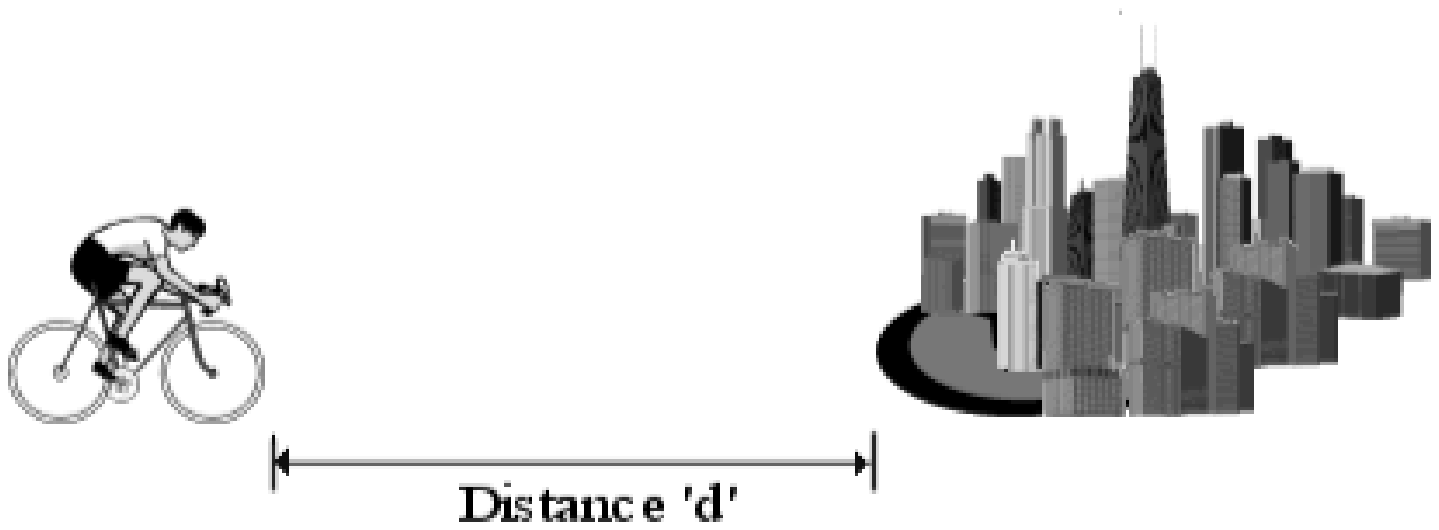}\\[0.2cm]
\begin{caption}
{\narrowtext The concept of distinguishability is illustrated.
What do we mean by the distance from the cyclist to the
city in the figure? It is defined as the distance from the
cyclist to the closest house in the city. Also, which
distance measure should be chosen to appropriately
measure this? In the text I argue that when it
comes to distinguishing between two
or more probability distributions the most appropriate measure
is the relative entropy.}
\end{caption}
\end{center}
\end{figure}

\vspace*{.3cm}
%\noindent
%\fbox{\parbox[b]{17.7cm}{
{\bf Sanov's theorem}.
If we have a probability distribution $Q$ and a set
of distributions $E \subset {\cal P}$ then
\begin{eqnarray}
Q^n(E) \rightarrow e^{-nS(P^* ||Q)}
\label{Sanov}
\end{eqnarray}
where $P^{*}$ is the distribution in $E$ that is closest to $Q$
using the Shannon relative entropy (see Fig. 1).
%}}
%\vspace*{.3cm}

This can also be rephrased in the language of distinguishability:
when we are distinguishing a given distribution from a set of distributions,
then what matters is how well we can distinguish that distribution from the
closest one in the set (see Fig. 1). When we turn to the quantum case later,
the probability distributions will become quantum densities
representing various states of a quantum system, and the question will be how well
we can distinguish between these states. Note that we could also talk about
$Q$ coming form a set of states in which case we would have $S(P||Q^*)$,
$Q^*$ being the state that minimizes the relative entropy (i.e. the
closest state).

\subsection{Other information measures from relative entropy}

Another important concept derived from the relative entropy concerns
gathering information. When one system learns something about another one,
their states become correlated. How correlated they are, or how much information
they have about each other, can be quantified by the mutual information.

\noindent
{\bf Definition}. {\em The Shannon mutual information} between two
random variables $A$ and $B$, having a joint probability distribution ${%
p(a_i,b_j)}$, and therefore marginal probability distributions
${p(a_i)}=\sum_j p(a_i,b_j)$ and ${p(b_j)}=\sum_i p(a_i,b_j)$, is defined as
\begin{eqnarray}
I_S(A:B) := S(p(a))+S(p(b))-S(p(a,b)) \; .
\end{eqnarray}
We now present two very instructive ways of looking at this quantity, which
will form a basis for the review. Mathematically, $I_S$ can be
written in terms of the Shannon relative entropy. In this sense it
represents a distance between the distribution ${p(a,b)}$ and the product of
the marginals ${p(a)}\times{p(b)}$. As such, it is intuitively clear that
this is a good measure of correlations, since it shows how far a joint
distribution is from the product one in which all the correlations have been
destroyed, or alternatively, how distinguishable a correlated state is
form a completely uncorrelated one. So, we have
\begin{eqnarray}
I_S(A:B) = S(p(a,b)\,||\,{p(a)}\times{p(b)}) \; . \nonumber
\end{eqnarray}

%% Venn Fig2

\begin{figure}[ht]
\begin{center}
\hspace{0mm}
\epsfxsize=7.5cm
\epsfbox{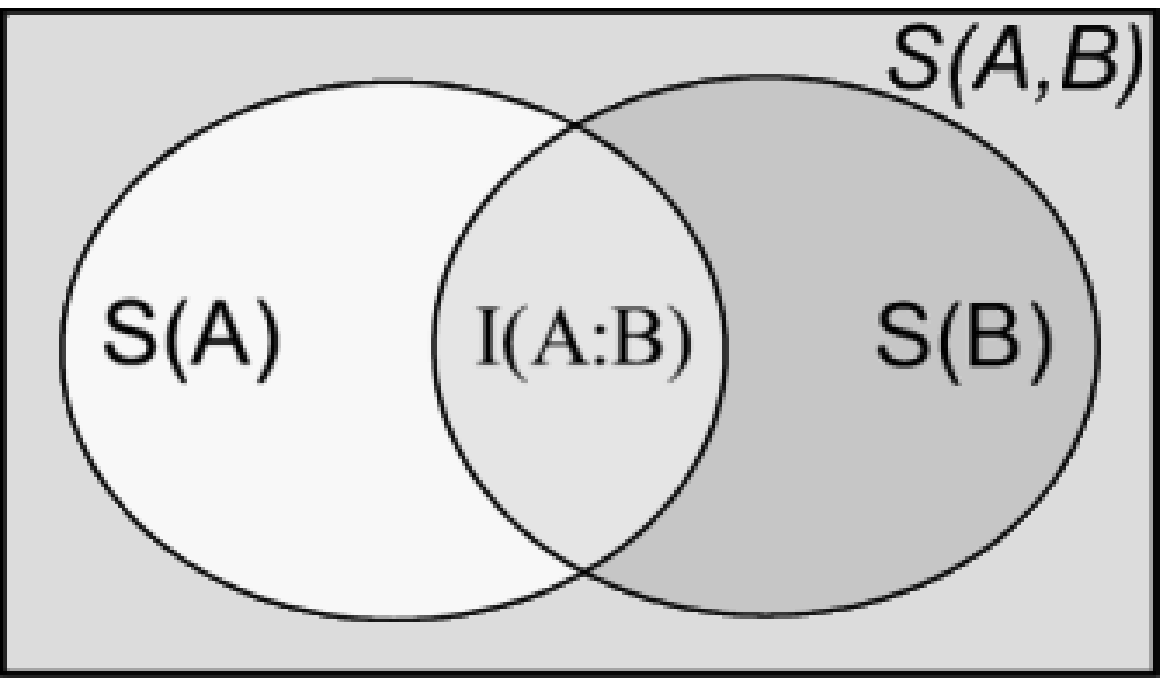}\\[0.2cm]
\begin{caption}
{\narrowtext The Venn diagram representation of the joint Shannon
entropy of two random variables as well as the marginal
Shannon entropies. It is clear that geometrically the
Shannon mutual information is obtained by summing the
marginal entropies and subtracting the total entropy. It is
interesting to note that its generalisation fails for
three or more random variables.}
\end{caption}
\end{center}
\end{figure}

Let us now view this from another angle. Suppose that we wish to know the
probability of observing $b_j$ {\em if $a_i$ has been observed}. This is
called a conditional probability and is given by:
\begin{eqnarray}
p_{a_i}(b_j) := \frac{p(a_i,b_j)}{p(a_i)} \; . \nonumber
\end{eqnarray}
This motivates us to introduce a conditional entropy, $S_A(B)$, as:
\begin{eqnarray}
S_A(B) & = & - \sum_i p(a_i) \sum_j p_{a_i}(b_j) \ln p_{a_i}(b_j) \nonumber\\
& = & -\sum_{ij} p(a_i,b_j) \ln p_{a_i}(b_j) \; .\nonumber
\end{eqnarray}
This quantity tells us how uncertain we are about the value of $B$ once we
have learned about the value of $A$. Now the Shannon mutual information can
be rewritten as
\begin{eqnarray}
I_S(A:B) = S(B) - S_A(B) = S(A) - S_B(A) \; .  \label{clascap}
\end{eqnarray}
So, the Shannon mutual information, as its name indicates,
measures the quantity of information conveyed about the random
variable $A$ ($B$) through measurements of the random variable $B$
($A$). This quantity, being positive, tells us that the initial
uncertainty in $B(A)$ can in no way be increased by making
observations on $A(B)$. Note also that, unlike the Shannon
relative entropy, the Shannon mutual information is symmetric (see
Fig. 2). The following example demonstrates the symmetry of
Shannon's mutual information.

Let us briefly go back to our original idea of a surprise to interpret the
Shannon mutual information as a measure of correlations. Suppose that one of
our friends likes to wear socks of two colours only: red and blue. In
addition we know that her socks are always the same colour and that when she
gets up in the morning, she randomly chooses the colour, but we know that
she prefers blue to red with the ratio $3:1$. So, when we meet our friend,
before we have looked at the colour of her socks, we know that she wears
blue socks with the probability $p(b)=0.75$ and red socks with the
probability $p(r) = 0.25$. However, when we look at one sock and observe,
say, the colour blue, we immediately know that the other sock must be blue,
too. This means that the colours of her two socks are correlated. So, before
we look at one of the socks, we are uncertain about the colour of the other
sock by an amount of $- 0.75\ln 0.75 -0.25 \ln 0.25 $. But then, when we
look at one of them the uncertainty immediately disappears. So, we expect
that the information we gain about one sock by looking at the colour of the
other is given by $- 0.75\ln 0.75 -0.25 \ln 0.25 $. The Shannon mutual
information predicts exactly the same thing. We see that the largest
correlations would be if $p(r)=p(b)=0.5$ and would be $\ln 2$. This, of course,
agrees with our intuitive notion of surprise, since then, before looking at
her one sock, we would be completely uncertain about the colour of the other
sock. Therefore by observing its colour we obtain the largest possible
amount of information (i.e. remove the largest possible uncertainty in this
case).

Although it will be seen that the Shannon mutual information is a good
measure of correlations between two random variables, its {\em natural}
generalization to three and more random variables fails. It is easy to see
that from three random variables the Shannon mutual information should be of
the following form:
\begin{eqnarray}
I_S (A:B:C) & = & S(A,B,C) - S(A,B) - S(A,C) - S(B,C)  \nonumber \\
& + & S(A) + S(B) + S(C)\; .
\end{eqnarray}
However, there exist $A,B,C$ such that $I_S (A:B:C)< 0$ (I leave
this as an exercise for the reader), and since we regard the
amount of correlation as being strictly positive, this is
automatically ruled out as a good measure of correlation. A way to
side-step this difficulty is to define mutual information via the
relative entropy as $S(p(A,B,C)||p(A)p(B)p(C))$. This is a
positive quantity representing the distance between the joint
three random variables probability distribution from the product
of the corresponding marginals. This, of course, immediately
generalizes to any number of random variables. Next I show why the
relative entropy and mutual information are also very useful from
the dynamical perspective.

\subsection{Classical evolution and relative entropy}

The above application of relative entropy to physics via the
concept of distinguishability might seen contrived. This is,
however, not at all the case, and this section shows the great
importance of the relative entropy for the dynamics of classical systems.
A state of a physical system in classical mechanics can be represented
as a vector whose entries are various probabilities for the system
to occupy its different possible states. The evolution of this system
is seen as the change of these probabilities with time. So, the
evolution is a linear transformation of a state into another state,
i.e. of a vector into another vector,
\begin{eqnarray}
q_j = \sum_k P(j|k) p_k  \nonumber
\end{eqnarray}
where $P(j|k)$ is the conditional probability for the system to
change from the state $k$ to the state $j$. Because the probability
has to be conserved ($\sum_j q_j =1$), we have that $\sum_k P(j|k)=1$.
Matrices with this simple property, namely that their entries are positive
and columns sum up to 1, are called {\em stochastic}.
The above can be generalised to continuous systems and
continuous time evolution, but this will not be relevant for the
rest of this review.

A very important property of any measure that aims at quantifying
the amount of correlations between two random variables (i.e two
states of the same or two different systems in classical
mechanics) is the following: if either or both of the variables
undergo a {\em local} stochastic evolution, then the amount of
correlations cannot increase (in fact, it usually decreases). We
now prove this in the case of the Shannon mutual information,
following an approach similar to that given by Everett (1973) (see
also Penrose's excellent book on Statistical mechanics; Penrose,
1973).

First, we establish without proof two inequalities following from
the convex properties of the logarithmic functions (Everett,
1973). Lemma 1 states that entropy is a concave function, whereas
lemma 2 states that the relative entropy is a convex function.

\noindent {\bf Lemma 1.} $\sum_i P_i x_i \ln \sum_i P_i x_i \le \sum_i P_i
x_i \ln x_i$, where $x_i \ge 0$, $P_i \ge 0$ and $\sum_i P_i = 1$.

\noindent
Physically, this inequality means that the average uncertainty (negative of
the right hand side) is less than or equal to the uncertainty of the average
(negative of the left hand side); in other words, mixing probability distributions
increases entropy. This is a very important property of entropy as a
measure of uncertainty since when we mix probability distributions we
expect to increase our uncertainty.

\noindent {\bf Lemma 2.} $\sum_i x_i \ln \frac{\sum_i x_i}{\sum_i a_i} \le
\sum_i x_i \ln \frac{x_i}{a_i}$, where $x_i \ge 0$ and $a_i\ge 0$ for all $i$.

\noindent
This is just a statement of the fact that mixing decreases distinguishability.
Note that this is in accord with the lemma 1, since the more mixed the probability
distributions, the less distinguishable they are.

\noindent
These two simple and self-evident statements lead to a very important
result that the Shannon relative entropy between two probability
distributions decreases when the same two undergo a stochastic process. This
is a very satisfying property from the physical point of view, where two
probability distributions undergoing stochastic changes, in fact, represent two
evolving physical systems. It says that two probability distributions are in
some sense closer to each other (i.e. ``harder to distinguish") after a
stochastic process, or analogously, that two physical systems become more
alike.

So, we consider a sequence of transition-probability matrices $T^n_{ij}:=P_n(i|j)$,
where $\sum_j T^n_{ij} = 1$ for all $n$, $i$, and $0 \le T^n_{ij} \le 1$.
We also introduce a sequence of positive measures (i.e. probability
distributions) $a^n_i$ having the property that
\begin{eqnarray}
a^{n+1}_j = \sum_i a^n_i T^n_{ij} \; . \nonumber
\end{eqnarray}
Transition probabilities $T$ tell us the probability that at the $n$th step
of evolution the system will "jump" from the $j$th to the $i$th state. Thus
constructed transition matrices are stochastic for all $n$.
We further suppose that we have a sequence of probability distributions $p^n_i$
generated by the action of the above stochastic process, such that
\begin{eqnarray}
p^{n+1}_j = \sum_i p^n_i T^n_{ij} \; . \nonumber
\end{eqnarray}
This is the law describing the systems evolution in time, and the state
of the system at time $n$ is given by the probabilities $p^n_i$.
For each of these probability distributions the relative entropy $S^n$
is defined as
\begin{eqnarray}
S^n(p||a) := S(p^n ||a^n) = \sum_i p^n_i \ln \frac{p^n_i}{a^n_i}\; . \nonumber
\end{eqnarray}
We prove the following theorem:

\vspace*{.3cm}
\noindent
%\fbox{\parbox[b]{17.7cm}{
{\bf Distinguishability never increases}.
\begin{eqnarray}
S^{n+1}(p||a) \le S^n(p||a).
\end{eqnarray}
%
%}}
\vspace*{.3cm}

\noindent
{\bf Proof.} Expanding $S^{n+1}(p||a)$ we obtain:
\begin{eqnarray}
S^{n+1}(p||a) & = & \sum_j p^{n+1}_j \ln \frac{p^{n+1}_j}{a^{n+1}_j}
\nonumber\\
& = & \sum_j \{ \sum_i p^n_i T^n_{ij} \} \ln
\frac{\sum_i p^n_i T^n_{ij}}{\sum_i a^n_i T^n_{ij}} \, . \nonumber
\end{eqnarray}
However, using lemma 2 we have the following inequality
\begin{eqnarray}
\sum_i p^n_i T^n_{ij} \ln \frac{\sum_i p^n_i
T^n_{ij}}{\sum_i a^n_i T^n_{ij}} \le \sum_i p^n_i T^n_{ij}
\ln \frac{p^n_i T^n_{ij}}{a^n_i T^n_{ij}}  \;\; . \nonumber
\end{eqnarray}
>From the above two it follows that
\begin{eqnarray}
S^{n+1}(p||a) & \le &  \sum_j \bigg \{ \sum_i p^n_i T^n_{ij} \ln
\frac{P^n_i}{a^n_i} \bigg \} = \sum_i p^n_i T^n_{ij} \ln
\frac{p^n_i}{a^n_i} \nonumber\\
        &  =  & \sum_i p^n_i \ln \frac{p^n_i}{a^n_i} = S^n(p||a) \nonumber
\end{eqnarray}
and the proof is completed ${}_{\Box}$.

\noindent This property means that a distance between two states
cannot increase with time if the states evolve under any
stochastic map. The proof can be immediately specialized to the
cases when $T$ is stationary, i.e. $T$ is independent of $n$, and
when $T$ is {\em doubly stochastic}, i.e. $\sum_i T_{ij} =1 $ for
all $j$. A corollary to this important lemma is the following:

\noindent {\bf Corollary}. If we take $p=p(a,b)$, and $a=p(a)p(b)$, and
suppose that the stochastic process acting separately on $A$ and $B$ are
uncorrelated, we see that the Shannon mutual information does not increase
under these {\em local} stochastic processes (by local we mean that they act
separately on $A$ and $B$).

This is a very important, and physically intuitive, property of
any measure of correlations; its quantum analogue will be of
central importance for quantifying quantum correlations between
entangled subsystems. This corollary, in fact, can be taken as a
guidance for a ``good" measure of correlations. We can state that
any measure of correlations has to be non-increasing under local
stochastic processes. {\em In other words this means that the only
way that the systems can become more correlated, i.e. that they
gain more information about each other, is if they interact.}
Without mutual interaction the correlations can only decrease or
at best stay the same. The nature of quantum local stochastic
processes will form the physical basis for our argument in the
next section. A condition similar to property above, but employing
quantum stochastic processes, will be a key element in our search
for measures of entanglement. When we go to quantum mechanics, the
notion of a probability distribution will be replaced by a quantum
state (i.e. density matrix), and a stochastic process will become
a measurement process in quantum theory. The formulation of
probability theory that is most naturally generalized to quantum
states is provided by Kolmogorov (1950), and the quantum
generalization expressing similarities with von Neumann's Hilbert
Space formulation (von Neumann, 1932) can be found in Mackey
(1963) (c.f. Holevo, 1982). However, knowledge of this approach
will not be necessary for the rest of the review. Finally it is
important to stress that if the local stochastic processes are
correlated they virtually become global, and therefore the
correlations between the systems can increase as well as decrease.

\subsection{Schmidt Decomposition and Quantum Dynamics}

The difference between classical and quantum physics can be seen
in the fact that quantum states are described by a density matrix
$\rho$ (and not just vectors). The density matrix is a positive
semi-definite Hermitian matrix, whose trace is unity (representing
the fact that all the probabilities add up to $1$). An important
class of density matrices is the idempotent one, i.e.
$\rho^2=\rho$. The states these matrices represent are called pure
states. When there is no uncertainty in the knowledge of the state
of the system its state is then pure. Another important notion is
that of a composite system. A composite quantum system is one that
consists of a number of quantum subsystems. When those subsystems
are entangled it is impossible to ascribe a definite state vector
to any one of them. The most often quoted entangled system is a
pair of two photons, being in the ``EPR" state (Einstein et. al,
1935; Bell, 1987). The composite system is then mathematically
described by
\begin{eqnarray}
|\Psi\rangle = \frac{1}{\sqrt{2}} (|\uparrow\rangle|\downarrow\rangle + |\downarrow\rangle|\uparrow\rangle )
\end{eqnarray}
where the first ket in either product belongs to one photon and
the second to the other. The property that is described is the
direction of spin or polarization along the z-axis, which can
either be ``up" ($|\uparrow\rangle$) or ``down"
($|\downarrow\rangle$). A two level system of this type is a
quantum analogue of a bit, which we shall henceforth call a {\em
qubit}. We can immediately see that neither of the photons
possesses a definite state vector. The best that one can say is
that if a measurement is made on one photon, and it is found to be
in the state ``up" for example, then the other photon is certain
to be in the state ``down". This idea cannot be applied to a
general composite system, unless the former is written in a
special form. This motivates us to introduce the so called Schmidt
decomposition (Schmidt, 1907), which not only is mathematically
convenient, but also gives a deeper insight into correlations
between the two subsystems.

According to the rules of quantum mechanics the state vector of a
composite system, consisting of subsystems $U$ and $V$, is represented by a vector
belonging to the tensor product of the two Hilbert Spaces
${\cal H}_U\otimes{\cal H}_V$. The general state of this system
can be written as a linear superposition of products of individual states:
\begin{eqnarray}
|\Psi\rangle = \sum_n\sum_m c_{nm} |u_n\rangle |v_m\rangle
\label{general}
\end{eqnarray}
where $\{|u_n\rangle\}_{n=1}^N$ and $\{|v_m\rangle\}_{m=1}^N$ are
the orthonormal basis of the subsystems $U$ and $V$ respectively,
whose dimensions are {\em dim} $U = N$ and ${\em dim} V = M$.
This state can always be written in the so called Schmidt form:
\begin{eqnarray}
|\Psi\rangle = \sum_n g_{n} |u^{\prime}_n\rangle |v^{\prime}_n\rangle\; ,
\label{Schmidt}
\end{eqnarray}
where $|u^{\prime}_n\rangle $ and $|v^{\prime}_n\rangle $ are orthonormal
basis for $U$ and $V$ respectively. Note that in this form the
correlations between the two subsystems are fully displayed. If
$U$ is found in the state $|u^{\prime}_2\rangle$ for example, then
the state of $V$ is $|v^{\prime}_2\rangle$. This is clearly a multi state
generalization of the EPR state mentioned earlier.

I will now prove this assertion by showing how to derive eq. (\ref{Schmidt}) from
eq. (\ref{general}).
To that end, let us assume that $M>N$, which in no way affects our line
of argument since the procedure is symmetric with respect to the
subsystems. Then we have the following five steps:
\begin{enumerate}
\item First we construct a density matrix describing $|\Psi\rangle=\sum_n\sum_m c_{nm} |u_n\rangle |v_m\rangle$.
Once the density matrix is known all the properties of the system
can be deduced from it. Moreover, ensembles which are prepared differently, but have the same
density matrix are statistically indistinguishable and therefore equivalent.
Generally, if we have a mixed state involving
vectors $|\Psi_1\rangle, |\Psi_2\rangle, \ldots |\Psi_D\rangle $ with
corresponding classical probabilities $w_1, w_2, \ldots , w_3$, then
the density matrix is defined to be:
\begin{eqnarray}
\rho = \sum_{d=1}^D w_d  |\Psi_d\rangle\langle\Psi_d | \; . \nonumber
\end{eqnarray}
Since in our case $|\Psi\rangle$ is a pure state, the density matrix is a
projection operator on to $|\Psi\rangle$, i.e.
\begin{eqnarray}
\rho = |\Psi\rangle\langle\Psi | = \sum_{nm} \sum_{pq} \rho_{nmpq} |u_n\rangle\langle u_p | \otimes  |v_m\rangle\langle v_q| \nonumber
\end{eqnarray}
where $\rho_{nmpq} = c_{nm}c_{pq}^*$. If we, however, wish to deal with
one of the subsystems only, then we employ the concept of the reduced density
matrix.

\item We find the reduced density matrix of the subsystem $U$, obtained by tracing $\rho$ over all states of
the subsystem $V$, so that
\begin{eqnarray}
\rho_U = \sum_q \langle v_q| \rho | v_q\rangle =  \sum_{nm} \sum_{p} \rho_{nmpm}
|u_n\rangle\langle u_p| \; . \nonumber
\end{eqnarray}
Note that the partial trace (or the trace itself) does not depend
on the choice of basis. Partial tracing is analogous to finding
marginal probability distributions from a joint probability
distribution in classical probability theory. The crucial step in
the Schmidt decomposition is diagonalizing the above. I shall call
the eigenvalues of $\rho_U$ $|g_1|^2,|g_2|^2,\ldots,|g_N|^2$, and
the corresponding eigenvectors
$|u_1^{\prime}\rangle,|u_2^{\prime}\rangle,\ldots,|u_N^{\prime}\rangle$.

\item Then I re-express the above in terms of $|u^{\prime}\rangle$'s, i.e
\begin{eqnarray}
|\Psi\rangle = \sum_n\sum_m c^{\prime}_{nm} |u^{\prime}_n\rangle |v_m\rangle\; . \nonumber
\end{eqnarray}

\item Now, we construct a new orthonormal basis of the subsystem $V$ such that
each new vector is a ``clever" linear superposition of the old ones, so that
\begin{eqnarray}
|v^{\prime}_i\rangle = \sum_m \frac{c^{\prime}_{im}}{g_i} |v_m\rangle\; . \nonumber
\end{eqnarray}
The matrix given by the coefficients $c^{\prime}_{im}/g_i$ is
unitary which is why the new basis is orthonormal.

\item The Schmidt decomposition of $|\Psi\rangle$ is now given by
\begin{eqnarray}
|\Psi\rangle = \sum_n g_{n} |u^{\prime}_n\rangle |v^{\prime}_n\rangle\; .\nonumber
\end{eqnarray}
\end{enumerate}

There are two important observations to be made, which are fundamental
to understanding correlations between the two subsystems
in a joint pure state:

\begin{itemize}

\item The reduced density matrices of both subsystems, written in the Schmidt basis, are diagonal and
have the same positive spectrum. in particular, the overall density
matrix is given by
\begin{eqnarray}
\rho = \sum_{nm}  g_{n}g^*_{m} |u^{\prime}_n\rangle\langle u^{\prime}_m | \otimes
|v^{\prime}_n\rangle\langle v^{\prime}_m| \nonumber
\end{eqnarray}
whereas the reduced ones are
\begin{eqnarray}
\rho_U & = & \sum_m \langle v^{\prime}_m| \rho | v^{\prime}_m\rangle =  \sum_{n}
|g_{n}|^2 |u^{\prime}_n\rangle\langle u^{\prime}_n| \nonumber\\
\rho_V & = & \sum_n \langle u^{\prime}_n| \rho | u^{\prime}_n\rangle =  \sum_{m}
|g_{m}|^2 |v^{\prime}_m\rangle\langle v^{\prime}_m| \; .\nonumber
\end{eqnarray}

\item If a subsystem is $N$ dimensional it then can be entangled with no more
than $N$ orthogonal states of another one.

\end{itemize}

I would like to point out that the Schmidt decomposition is,
in general, impossible for more than two entangled subsystems.
To clarify this I consider three entangled subsystem as an example. Here, our
intention would be to write a general state such that by observing
the state of the one of the subsystems we instantaneously and with certainty know the
state of the other two. But, this is impossible in general, for the
presence of the third system makes the prediction uncertain.
Loosely speaking, while we know
the state of one of the subsystems, the other two might still be entangled and
cannot have definite
vectors associated with them (an exception to this general rule is,
for example, a state of the
Greenberger--Horne--Zeilinger (GHZ) type
$(1/\sqrt{2}) (|\uparrow\rangle|\uparrow\rangle|\uparrow\rangle +
|\downarrow\rangle|\downarrow\rangle|\downarrow\rangle )$). Clearly,
involvement of even more subsystems complicates this analysis even
further and produces, so to speak, an even greater mixture and
uncertainty. The same reasoning applies to
mixed states of two or more subsystems (i.e. states whose density operator
is not idempotent $\rho^2 \neq \rho$),
for which we cannot have the Schmidt decomposition in general. This reason
alone is responsible for the fact that
the entanglement of two subsystems in a pure state is simple to understand
and quantify, while for mixed states, or states consisting of more than two
subsystems, the question is much more involved.

We now discuss the way quantum systems evolve. An isolated system,
of course, follows a unitary dynamics generated by Schr\"odinger's
equation (non-relativistic). This evolution is fully reversible
(manifesting itself in the fact that the quantum entropy does not
increase during this process as we will see below). However, we
know that most of the processes in Nature are irreversible (think
of the spontaneous emission and the non-existence of its reverse -
"spontaneous absorption").  These processes are non-unitary and
arise from the interaction of the system with the environment;
thus, the system is no longer closed. Mathematically, the
evolution of a quantum state is then most generally of the form
(Davies, 1976)
\begin{eqnarray}
\rho^{\prime} = \sum_{\alpha} A_{\alpha} \rho A^{\dagger}_{\alpha}
\label{CP}
\end{eqnarray}
where, because of the conservation of probability, or, more
precisely, trace preservation $\sum_{\alpha} A^{\dagger}_{\alpha}
A_{\alpha}=1$. The above map is the most general {\em completely}
positive (trace preserving) linear map (CP-map) (Choi, 1975).
Positivity means that density matrices are mapped into density
matrices (strictly speaking, positive operators are mapped onto
positive operators). To define "complete", we first need to
introduce the idea of an extended state. By extension of a state I
mean any state on a larger Hilbert space that reduces itself to
the original state when the extended part is traced out. In turn,
completeness means that any extension of the density matrix is
also mapped into a density matrix. To clarify this I will present
a few examples of CP-maps:
\begin{itemize}
\item Projectors are Hermitian idempotent operators ($P^{\dagger}=P$ and $P^2=P$)
and the evolution of the form $\rho \rightarrow \sum_i P_i \rho P_i$ is a CP-map;
\item Addition of another system to $\rho$ is also a CP-map, $\rho\rightarrow \rho\otimes \rho_1$;
\item Let $E_i \geq 0$ and $\sum_i E_i = I$. Then, $\rho \rightarrow p_k:=Tr(\rho E_k)$
is a CP map which generates a probability distribution from a density matrix.
\item Unitary evolution is a special case of CP-map, where only
one operator is present in the sum, i.e. $U\rho U^{\dagger}$.
\end{itemize}
I leave it for the reader to show that the above CP maps can indeed be written in the
form in eq. (\ref{CP}). We will meet other examples in the next subsection.

Remarkably not all positive maps are completely positive,
transposition being a well known example. Positivity of
transposition follows from the fact for any state $\rho$, its
transposition $\rho^T\geq 0$. However, a counter example to
completeness comes from, for example, a singlet state of two
sub-systems $A$ and $B$. Namely, if we transpose only $A$ (or
$B$), then the resulting operator is not positive (so that it is
not a physical state), i.e. $\rho_{AB}^{T_A}<0$. Confirmation of
this is left as an exercise.

%% CP maps Fig3

\begin{figure}[ht]
\begin{center}
%%\vspace{-5.2cm} 
\hspace{0mm}
\epsfxsize=6cm
\epsfbox{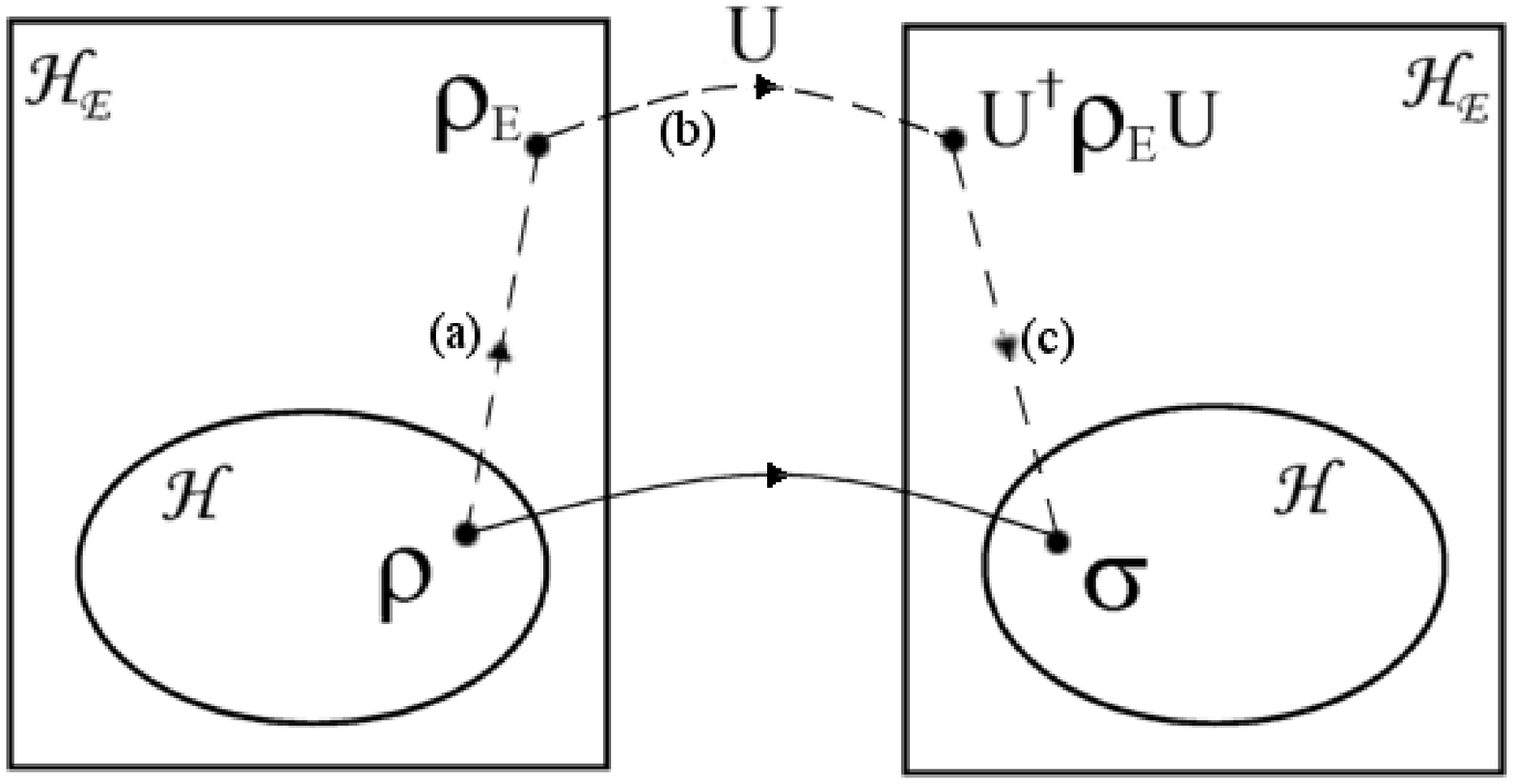}\\[0.2cm] 
%%\vspace{1.5cm}
\caption{\narrowtext The most general evolution in quantum
mechanics is represented by a completely positive trace preserving
map (CP-map). This figure shows two equivalent ways of
representing a CP-map: (a) the canonical form $A (.) A^{\dagger}$,
and (b) via the extension to a larger Hilbert Space ${\cal H}_E$
and an appropriate Unitary transformation there in. The connection
is explained in the text.}
\end{center}
\end{figure}

The reader might wonder as to what the physical implementation of
the canonical form $\sum_{\alpha} A_{\alpha} \rho
A^{\dagger}_{\alpha}$ is? I will now introduce another
representation of the CP-maps that will explain its physical
importance and will be crucial for the rest of the review. Loosely
stated, any CP-map can be represented as a unitary transformation
on a higher Hilbert space (see Fig. 3). Namely, from Schmidt
decomposition we know that a density matrix can be represented as
a "reduction" of a state in an enlarged Hilbert space. Suppose
that $\rho \in {\cal H}$ and that $\rho_E\in {\cal H}\otimes {\cal
H}_a$ is an "extension" of the state $\rho$ such that $Tr_a \rho_E
= \rho$. Then a CP map $\sigma = \Phi (\rho)$ can be represented
as
\begin{eqnarray}
\rho \rightarrow \rho_E \rightarrow U \rho_E U^{\dagger}
\rightarrow Tr_a (U \rho_E U^{\dagger})=\sigma
\end{eqnarray}
Here we have first "lifted" $\rho$ to $\rho_E$, then evolved
$\rho_E$ unitarily into $U \rho_E U^{\dagger}$ which, after
tracing over the Hilbert space extension (i.e. lowering), yields
the final state $\sigma$ as in Fig. 3. The fact that for any
CP-map there exist a unitary operator $U$ which will execute this
map on some higher Hilbert space is guaranteed by a theorem proved
independently by Kraus (1983) and Ozawa (1984), (see Schumacher,
1996 for a modern presentation). I will now only present a
plausibility argument for this correspondence. Let $\rho_E = \rho
\otimes |0\rangle\langle 0|_a$ where $|0\rangle\langle 0|_a \in
{\cal H}_a$. Then
\begin{eqnarray}
\sigma & = & Tr_a (U \rho \otimes |0\rangle\langle 0|_a U^{\dagger})  \nonumber \\
& = & \sum_i \langle i|_a U \rho \otimes |0\rangle\langle 0|_a U^{\dagger} |i\rangle_a \nonumber \\
& = & \sum_i \langle i|U|0\rangle \rho \langle 0|U^{\dagger}|i\rangle \nonumber \; ,
\end{eqnarray}
which has the same form as eq. (\ref{CP}) providing we define $A_i
:=  \langle i|U|0\rangle$. Thus, given a unitary evolution on the
extended Hilbert Space, we can always find the corresponding
positive operators which describe the evolution of the original
system. Note that the choice of the operators is not unique.

Finally, I would like to discuss another frequently used concept
that is in some sense derived from the notion of CP-maps. It can
be loosely stated that the CP-map represents the evolution of a
quantum system when we do not update the knowledge of its state
based on the particular measurement outcome. This is why we have a
summation over all measurements in eq. (\ref{CP}). If, on the
other hand, we know that the outcome corresponding to the operator
$A^{\dagger}_jA_j$ occurs, then the state of the system
immediately afterwards is given by $A_j\rho
A^{\dagger}_j/tr(A^{\dagger}_jA_j\rho)$. This type of measurement
is the most general one and is commonly referred to as the
Positive Operator Valued Measure (POVM). It is positive because
operators of the form $A^{\dagger}A$ are always positive for any
operator $A$ and taking the trace of it together with any density
matrix generates a positive number (i.e. a probability for that
particular measurement outcome). For a more detailed overview of
POVMs see Peres (1993).  The concept of POVM will play a
significant role when defining the quantum relative entropy next.

\subsection{Quantum relative entropy}

When two subsystems become entangled the composite state can be
expressed as a superposition of the product of the corresponding
Schmidt basis vectors. From eq. (\ref{Schmidt}) it follows that
the i-th vector of either subsystem has a probability of $|g_i|^2$
associated with it. We are, therefore, uncertain about the state
of each subsystem, the uncertainty being larger if the
probabilities are evenly distributed. Since the uncertainty in the
probability distribution is naturally described by the Shannon
entropy, this classical measure can also be applied in quantum
theory. In an entangled system this entropy is related to a single
observable. The general state of a quantum system, as I have
already remarked, is described by its density matrix $\rho$. If
$A$ is an observable pertaining to the system described by
${\rho}$, then by the spectral decomposition theorem $A = \sum_i
a_i P_i$, where ${P}_i$ is the projection onto the state with the
eigenvalue $a_i$. The probability of obtaining the
eigenvalue $a_j$ is given by $p_j = \mbox{Tr}({\rho} {P}_j) = \mbox{Tr}({P}%
_j {\rho})$. The uncertainty in a given observable can now be expressed
through the Shannon entropy. Let the observables $A$ and $B$, pertaining to
the subsystems $U$ and $V$ respectively, have a discrete, non-degenerate
spectrum, with corresponding probabilities $p(a_i)$ and $p(b_j)$ of
observables $A$ being $a_i$ and $B$ being $b_j$. Let also the joint
probability be $p(a_i,b_j)$. Then,
\begin{eqnarray}
S(A) & = & - \sum_i p(a_i) \ln p(a_i) \\
& = & -\sum_{ij} p(a_i,b_j) \ln \sum_j p(a_i,b_j) \\
S(B) & = & - \sum_j p(b_j) \ln p(b_j) \\
& = & -\sum_{ij} p(a_i,b_j) \ln \sum_i p(a_i,b_j) \\
S(A,B) & = & -\sum_{ij} p(a_i,b_j) \ln p(a_i,b_j)
\end{eqnarray}
where I have used the fact that $\sum_j p(a_i,b_j) = p(a_i)$ and
$\sum_i p(a_i,b_j) = p(b_j)$. We have seen that a signature of
correlations is that the sum of the uncertainties in the
individual subsystems is greater than the uncertainty in the total
state. So, the Shannon mutual information is a good indicator of
how much the two given observables are correlated. However, this
quantity as it is inherently classical describes the correlations
between single observables only. The quantity that is related to
the correlations in the overall state as a whole is the von
Neumann mutual information. Since it is assigned to the state as a
whole, it is of little surprise that it depends on the density
matrix. First, however, I define the von Neumann entropy (von
Neumann, 1932), which can be considered as the proper quantum
analogue of the Shannon entropy (Ohya and Petz, 1993; Ingarden et.
al, 1997; Wehrl, 1978).

\noindent
{\bf Definition}. {\em The von Neumann entropy} of a quantum
system described by a density matrix ${\rho}$ is defined as
\begin{eqnarray}
S_N(\rho) := -\mbox{Tr}( \rho \ln \rho ) \; .
\end{eqnarray}
(I will drop the subscript $N$ whenever there is no possibility of
confusion). The Shannon entropy is equal to the von Neumann
entropy only when it describes the uncertainties in the values of
the observables that commute with the density matrix, i.e. the
Schmidt observables. Otherwise,
\begin{eqnarray}
S(A) \ge S_N(\rho)  \nonumber
\end{eqnarray}
where $A$ is any observable of a system described by $\rho$. This means that
there is more uncertainty in a single observable than in the whole of the
state, the fact which entirely contradicts our expectations.

I now discuss a relation concerning the entropies of two
subsystems. One part of it is somewhat analogous to its classical
counterpart, but instead of referring to observables it is related
to the two states. This inequality is called the Araki-Lieb
inequality (Araki and Lieb, 1970) and is one of the most important
results in the quantum theory of correlations. Let $\rho_A$ and
$\rho_B$ be the reduced density matrices of subsystems $A$ and $B$
respectively, and $\rho$ be the matrix of a composite system,
then:
\begin{eqnarray}
S_N(\rho_A) + S_N(\rho_B) \ge S_N(\rho) \ge |S_N(\rho_A) - S_N(\rho_B)|\; . \nonumber
\end{eqnarray}
Physically, the left hand side implies that we have more
information (less uncertainty) in an entangled state than if the
two states are treated separately. This arises naturally, since by
treating the subsystems separately we have neglected the
correlations (entanglement). We note that if the composite system
is in a pure state, then $S(\rho)=0$, and from the right hand side
it follows that $S(\rho_A) = S(\rho_B)$ (cf. Schmidt decomposition
eq. (\ref{Schmidt})). To appreciate the extent to which this is a
counter-intuitive result we consider the following example.
Suppose a two level atom is interacting with a single mode of an
EM field as in the Jaynes-Cummings model (Jaynes and Cummings,
1963). If the overall state is initially pure, and the whole
system is isolated then the entropies of the atom and the field
are equally uncertain at all the times. But this is not expected
since the atom has only two degrees of freedom and the field
infinitely many! This, however, is possible, as, by the second
observation, the atom, as a two dimensional subsystem, is only
entangled with two dimensions of the field.

I present without proofs two important properties of entropy which
will be used in the later sections (Wehrl, 1978). These are:
\begin{eqnarray}
&1.&\mbox{\em additivity:} \quad \, \,\,\, S_N({\rho}_A\otimes {\rho}%
_B) = S_N({\rho}_A) + S_N({\rho}_B) ;  \label{ep1} \\
&2.&\mbox{\em concavity:} \quad \, \,\,\, S_N \left ( \sum_i
\lambda_i {\rho}_i \right ) \ge \sum_i \lambda_i S_N({\rho}_i);  \label{ep2}
\end{eqnarray}
The first property is the same as in classical information theory,
namely the entropies of independent systems add up. The concavity simply
reflects the fact that ``mixing increases uncertainty".

Following the definition of the Shannon mutual information I introduce the
von Neumann mutual information, which refers to the correlation between the
whole subsystems rather than relating two observables only.

\noindent
{\bf Definition}. {\em The von Neumann mutual information} between
the two subsystems $\rho_U$ and $\rho_V$ of the joint state ${\rho}_{UV}$ is
defined as
\begin{eqnarray}
I_N(\rho_U:\rho_V ;\rho_{UV}) = S_N(\rho_U) + S_N(\rho_V) -
S_N(\rho_{UV})\;\; .  \label{def7b}
\end{eqnarray}
As in the case of the Shannon mutual information this quantity can
be interpreted as a distance between two quantum states. For this
I first need to define the von Neumann relative entropy, in a
direct analogy with the Shannon relative entropy (in fact, this
quantity was first considered by Umegaki (1962), but for
consistency reasons I name it after von Neumann; I will also refer
to it as the quantum relative entropy).

\vspace*{.3cm}
\noindent
%\fbox{\parbox[b]{17.7cm}{
{\bf Definition}. {\em The von Neumann relative entropy} between
the two states $\sigma$ and $\rho$ is defined as
\begin{eqnarray}
S_N(\sigma ||\rho) = \mbox{Tr} \sigma (\ln \sigma - \ln \rho) \;\; .
\label{def8}
\end{eqnarray}
%}}
\vspace*{.2cm}

\noindent This measure also has the same statistical
interpretation as its classical analogue: it tells us how
difficult it is to distinguish the state $\sigma$ from the state
$\rho$ (Hiai and Petz, 1991). To that end, suppose we have two
states $\sigma$ and $\rho$. How can we distinguish them? We can
chose a POVM $\sum_{i=1}^M A_i ={\bf1}$ which generates two
distributions via

\begin{eqnarray}
p_i & = & tr A_i \sigma \\
q_i & = & tr A_i \rho \;\; ,
\end{eqnarray}
and use classical reasoning to distinguish these two
distributions. However, the choice of POVM's is not unique. It is
therefore best to choose that POVM which distinguishes the
distributions most, i.e. for which the {\em classical} relative
entropy is largest. Thus we arrive at the following quantity

\[
S_1(\sigma ||\rho) := \mbox{sup}_{\mbox{A's}} \{ \sum_{i} tr A_i
\sigma \ln tr A_i \sigma - tr A_i \sigma \ln tr A_i \rho \}  ,
\]
where the supremum is taken over all POVM's. The above is not the
most general measurement that we can make, however. In general we
have $N$ copies of $\sigma$ and $\rho$ in the state
\begin{eqnarray}
\sigma^N & = & \underbrace{\sigma\otimes \sigma ... \otimes \sigma}_{\mbox{total of N terms}}\\
\rho^N & = & \underbrace{\rho\otimes \rho ... \otimes
\rho}_{\mbox{total of N terms}}
\end{eqnarray}
We may now apply a POVM $\sum_{i} A_i ={\bf 1}$ acting on
$\sigma^N$ and $\rho^N$. Consequently, we define a new type of
relative entropy
\begin{eqnarray}
S_{N}(\sigma ||\rho)  & := &  \mbox{sup}_{\mbox{A's}} \{
\frac{1}{N}
\sum_{i} tr A_i \sigma^N \ln tr A_i \sigma^N \nonumber \\
 & - &  tr A_i \sigma^N \ln tr A_i \rho^N \}
\label{lim}
\end{eqnarray}
Now it can be shown that (Donald, 1986)
\begin{equation}
S(\sigma ||\rho) \ge S_{N} \label{ineq}
\end{equation}
where $S(\sigma ||\rho)$ is the quantum relative entropy. (This
really is a consequence of the fact that the relative entropy does
not increase under general CP-maps, a fact that will be proven
later on in this subsection). Equality is achieved in eq.
(\ref{ineq}) iff $\sigma$ and $\rho$ commute (Fuchs, 1996).
However, for any $\sigma$ and $\rho$ it is true that (Hiai and
Petz, 1991)
\[
S(\sigma ||\rho) = \lim_{N\rightarrow \infty} S_{N} \;\; .
\]
In fact, this limit can be achieved by projective measurements
which are independent of $\sigma$ (Hayashi, 1997). From these
considerations it would naturally follow that the probability of
confusing two quantum states $\sigma$ and $\rho$ (after performing
$N$ measurements on $\rho$) is (for large $N$):
\begin{equation}
P_N(\rho \rightarrow \sigma) = e^{-N S(\sigma ||\rho)}  \;\; .
\label{main}
\end{equation}
We would like to stress here that classical statistical reasoning
applied to distinguishing quantum states leads to the above
formula. There are, however, other approaches. Some take eq.
(\ref{main}) for their starting point and then derive the rest of
the formalism thenceforth (Hiai and Petz, 1991). Others, on the
other hand, assume a set of axioms that are necessary to be
satisfied by the quantum analogue of the relative entropy (e.g. it
should reduce to the classical relative entropy if the density
operators commute, i.e. if they are ``classical") and then derive
eq. (\ref{main}) as a consequence (Donald, 1986). In any case, as
we have argued here, there is a strong reason to believe that the
quantum relative entropy $S(\sigma ||\rho)$ plays the same role in
quantum statistics as the classical relative entropy plays in
classical statistics (see also a review by Schumacher and
Westmoreland, 2000).

Now, the von Neumann mutual information can be understood as a distance of
the state $\rho_{UV}$ to the uncorrelated state $\rho_U\otimes\rho_V$,
\begin{eqnarray}
I_N(\rho_U:\rho_V ;\rho_{UV}) = S_N(\rho_{UV} || \rho_U\otimes\rho_V)\; . \nonumber
\end{eqnarray}
The quantum relative entropy will be the most important quantity in
classifying and quantifying quantum correlations. It will be
seen that this quantity does not increase under CP maps,
which are quantum analogues of the stochastic processes. I list three
properties of the relative entropy whose proof is left to the reader:
\begin{description}
\item F1. Unitary operations leave $S(\sigma||\rho)$
invariant, i.e.
$S(\sigma||\rho)=S(U\sigma U^{\dagger}||U \rho U^{\dagger})$. Unitary
transformations represent a change of basis (i.e. a change in our "perspective")
and the distance between two states should not (and does not in this case) change under
this.
\item F2. $S(\mbox{Tr}_p \sigma ||\mbox{Tr}_p \rho) \le S({\sigma}||\rho)$,
where $\mbox{Tr}_p$ is a partial trace. Tracing over a part of the system
leads to a loss of information. The less information we have about two states,
the harder they are to distinguish which is what this inequality says.
\item F3. The relative entropy is additive $S(\sigma_1 \otimes \sigma_2||\rho_1 \otimes
\rho_2)=S(\sigma_1||\rho_1)+S(\sigma_2||\rho_2)$. This inequality is a
consequence of additivity of entropy itself.
\end{description}
These I now show have profound implication for the evolution of quantum systems.

\vspace*{.3cm}
\noindent
%\fbox{\parbox[b]{17.7cm}{
{\bf Quantum distinguishability never increases}.
For any completely positive, trace preserving map
$\Phi$, given by $\Phi \sigma = \sum V_i\sigma V^{\dagger}_i$ and $\sum
V^{\dagger}_i V_i = 1$, we have that $S(\Phi \sigma ||\Phi \rho) \le
S({\sigma}||\rho)$.
%}}
\vspace*{.3cm}

\noindent
I will first present a physical argument as to why we should expect
this theorem to hold. As I have discussed, a CP-map can be represented
as a unitary transformation on an extended Hilbert space. According to
F1, unitary transformations do not change the relative entropy between
two states. However, after this, we have to perform a partial tracing
to go back to the original Hilbert space which, according to F2,
decreases the relative entropy as some information is invariably lost
during this operation. Hence the relative entropy decreases under
any CP-map. I now formalise this proof.

\noindent {\bf Proof}. I have discussed the fact that a CP-map
can {\em always} be represented as a unitary operation+partial
tracing on an extended Hilbert Space ${\cal H} \otimes {\cal
H}_n$, where $\dim {\cal H}_n = n$ (Lindblad, 1974; 1975). Let
$\{|i\rangle \}$ be an orthonormal basis in  ${\cal H}_n$ and
$|\alpha\rangle$ be a unit vector. So I define,
\begin{eqnarray}
W= \sum_i V_i \otimes |i\rangle\langle \alpha | \,\, .
\label{a}
\end{eqnarray}
Then, $W^{\dagger}W = 1 \otimes P_{\alpha}$ where $ P_{\alpha}=
|\alpha \rangle\langle \alpha |$, and there is a unitary operator
$U$ in ${\cal H} \otimes {\cal H}_n$ such that $W=U (1\otimes
P_{\alpha})$ (Reed an Simon, 1980). Consequently,
\begin{eqnarray}
U(A\otimes P_{\alpha})U^{\dagger} = \sum_{ij}
V_iAV^{\dagger}_j \otimes
|i\rangle\langle j|  \,\,\,\,   ,
\label{b}
\end{eqnarray}
so that,
\begin{eqnarray}
\mbox{Tr}_2\{U(A\otimes P_{\alpha})U^{\dagger}\} =
\sum_i V_iAV^{\dagger}_i  \,\,\,\, .\nonumber
\end{eqnarray}
This shows that the unitary and $\sum_i V_i \rho V_i^{\dagger}$ representations are
equivalent. Now using F2, then F1, and finally F3 we find the following
\begin{eqnarray}
S(\mbox{Tr}_2\{U(\sigma\otimes P_{\alpha})U^{\dagger}\} & || &  \mbox{Tr}_2\{U(\rho\otimes
P_{\alpha})U^{\dagger}\}) \nonumber \\
& \le & S(U(\sigma\otimes P_{\alpha})U^{\dagger} ||  U(\rho\otimes P_{\alpha})
U^{\dagger}) \nonumber \\
& = & S(\sigma\otimes P_{\alpha}||\rho \otimes P_{\alpha}) \nonumber \\
& = & S(\sigma||\rho) \,\,\,\,   .
\end{eqnarray}
This proves the result ${}_{\Box}$.

\noindent
{\bf Corollary}. Since for a complete set of orthonormal
projectors $P$, $\sum_i P_i \sigma P_i$ is
a CP map, then
\begin{eqnarray}
\sum_i S(P_i\sigma P_i||P_i \rho P_i) \le S(\sigma||\rho)\,\,\,\ .
\label{1}
\end{eqnarray}
(The sum can be taken outside as it can be easily shown that
$S(\sum_i P_i\sigma P_i || \sum_i P_i \rho P_i) =\sum_i
S(P_i\sigma P_i||P_i\rho P_i)$). Now from F1, F2, F3 and eq.
(\ref{1}) we have the following

\noindent {\bf Theorem 5}. If $\sigma_i = V_i \sigma
V_i^{\dagger}$ then $\sum S(\sigma_i ||\rho_i) \le
S({\sigma}||\rho)$, where $\rho_i = V_i \rho V_i^{\dagger}/tr(V_i
\rho V_i^{\dagger})$.

\noindent {\bf Proof}. Equations (\ref{a}) and (\ref{b}) are
introduced as in the previous proof. From eq. (\ref{b}) we have
that
\begin{eqnarray}
\mbox{Tr}_2\{1 \otimes P_iU(A\otimes P_{\alpha})U^{\dagger} 1
\otimes P_i\} = V_iAV^{\dagger}_i  \,\,\,\, .\nonumber
\end{eqnarray}
where $P_i = |i \rangle\langle i |$. Now, from F2, the
Corollary and F3 it follows that
\begin{eqnarray}
\sum_i & S & (\mbox{Tr}_2\{1  \otimes P_iU(\sigma\otimes P_{\alpha})
U^{\dagger} 1  \otimes P_i\} \nonumber \\
& || & \mbox{Tr}_2\{1  \otimes P_i U(\rho\otimes
P_{\alpha})U^{\dagger} 1 \otimes P_i\})\nonumber \\
& \le & \sum_i S(1  \otimes P_i U(\sigma\otimes P_{\alpha})
U^{\dagger}1  \otimes P_i \nonumber \\
& || & 1 \otimes P_iU(\rho\otimes P_{\alpha})
U^{\dagger} 1 \otimes P_i) \nonumber\\
& \le & S(U(\sigma\otimes P_{\alpha})U^{\dagger}||
U(\rho\otimes P_{\alpha})U^{\dagger}) \nonumber \\
& = & S(\sigma\otimes P_{\alpha}||\rho\otimes P_{\alpha}) \nonumber \\
& = & S(\sigma||\rho) \,\,\,\,   .
\end{eqnarray}
This proves Theorem 5 ${}_{\Box}$. This theorem will be important in the
next section. A simple consequence of the fact that
the quantum relative entropy itself does not increase under
CP-maps is that correlations (as measured by the quantum
mutual information) also cannot increase but now
under {\em local} CP-maps.

\vspace*{.3cm}
\noindent
%\fbox{\parbox[b]{17.7cm}{
{\bf Correlations cannot increase without interaction}.
Correlations, as measured by the von
Neumann mutual information, do not increase during local
complete measurements carried on two entangled quantum systems.
%}}
\vskip 0.3cm

\noindent
The Shannon mutual information, although having this desired property,
does not distinguish between the quantum and classical correlations
(rather, it measures {\em total} correlations). In order to do this
I will have to introduce the possibility of
classical communication between $A$ and $B$. This will allow
classical correlations to increase while leaving quantum correlations
intact, as will be seen in the following section.
Now we put the theory developed so far to practical use: communication.

\noindent
{\bf Digression on the Second Law of Thermodynamics}. The Second
Law of Thermodynamics states that entropy of an isolated
system never decreases. This does not follow directly
from the no increase of the quantum relative entropy under CP-maps.
Strictly speaking, an isolated system in quantum mechanics evolves
unitarily and therefore its entropy never changes. Under CP-maps,
on the other hand, the entropy can both increase as well as
decrease. If, however, the state $\rho$ is maximally mixed $I/n$
for example, then the quantum relative entropy is given by:
\begin{eqnarray}
S(\sigma || \rho) = \ln n - S(\sigma) \; . \noindent
\end{eqnarray}
If in addition the evolution is such that $I/n$ is the equilibrium
state, then the monotone decrease in the quantum relative entropy
implies a monotone increase in $S(\sigma)$, just as in the Second
Law of Thermodynamics. Otherwise, the entropy itself can both
increase as well as decrease. A detailed discussion of the
statistical foundations of the Second Law can be found in
Tollman's classic "The Principles of Statistical Mechanics"
(Tolman, 1938).

\section{Quantum communication: Classical Use}

The central objective of communication theory is to allow a
person, often referred to as Alice, to communicate accurately with another person,
called Bob, even in the presence of noise.
Alice encodes her message into a number of different (distinguishable) states, with
each state representing a different symbol in the message.
For example, Alice encodes the bit value
$1$ into the excited state of a two level atom and sends this atom to Bob.
On its way to Bob the atom may transform into its ground state
due to either stimulated or spontaneous emission
thereby giving Bob the impression that Alice transmitted $0$. This
unwanted state transition is a form of channel noise.

The key question is: what is the largest amount of information
(per symbol) that Alice can send to Bob, i.e. what is the
capacity of the communication channel taking into account any
possible noise? In classical information theory the capacity for
communication is given by the mutual information between Alice's
sent message and Bob's received message (Shannon and Weaver,
1949). This is intuitively clear, since mutual information
quantifies correlations between sent and received messages and it
thus tells us how faithful the transmission is. If we use quantum
states to encode symbols, then the capacity is not given by the
quantum mutual information we introduced before. We derive a new
quantity for this purpose called the Holevo bound (Holevo, 1973).
{\em The benefit of performing the full quantum derivation is that
this is a more fundamental approach to information processing.}
We can then deduce the classical capacity as a special case.

\subsection{Holevo bound}

A quantum communication channel (QCC) consists of a number, $N$, of quantum
systems prepared in states $\rho_1, \rho_2 \ldots \rho_N$ and whatever physical
medium is used to send the states from Alice to Bob.
These states encode $N$ different symbols with certain a priori probabilities, $%
p_1, p_2, \ldots p_N$. Bob then performs a set of measurements to
determine the correct sequence of states comprising Alice's
symbols, which he can then use to reconstruct the entire message
(Ingarden, 1976). If the states suffer no error on the way to the
Bob, then the channel is called noiseless; otherwise it is called
noisy. I only consider the notion of capacity of a noiseless QCC,
since the generalization to a noisy channel is straightforward.

Let $S(\rho) = - \mbox{Tr} \rho \ln \rho$ be the standard von Neumann
entropy of a density matrix $\rho$. Then, the capacity of a QCC is
defined as
\begin{eqnarray}
{\bf C} := \max_{\{p\}} C(\{p\}, \rho) \nonumber
\end{eqnarray}
where
\begin{eqnarray}
C(\{p\}, \rho) = S(\sum_i p_i \rho_i) - \sum_i p_i S(\rho_i) \;\; ,
\end{eqnarray}
is the {\bf Holevo bound}. Note that the above can be expressed succinctly as
\begin{eqnarray}
C(\{p\}, \rho) = \sum_i p_i S(\rho_i || \rho) \;\; ,  \label{caprel}
\end{eqnarray}
where $S(\; ||\; )$ is the von Neumann relative entropy and $\rho = \sum_i
p_i \rho_i$. When there is no possibility of confusion I write $C(\{p\},
\rho) \equiv C(\{p\})$. The reader may ask why we need to maximise
symbol probabilities in order to compute the capacity. This is because
the channel can be used with different input probabilities and the
capacity represents the maximum that can be communicated using this
channel.

To see the physical motivation behind this quantity consider $N$
states $\rho_1,...\rho_N$ sent by Alice to Bob according to
probabilities $p_1,...p_N$ respectively. Bob now performs a set of
complete measurements $\sum_i E_i = I$, where $E_i \geq 0$, in
order to determine which state was sent to him (a complete
measurement is like a CP-map, but where we record each of the
outcomes). The {\em accessible information} to Bob is given by the
mutual information between his measurement and $\rho_1,...\rho_N$
(Holevo, 1973; Davies 1976). This quantity tells us how well Bob's
measurement can distinguish between the message states and is
given by
\begin{eqnarray}
I(E:\rho) & = & \bigg \{ \sum_i - \mbox{Tr}(\rho E_i) \ln (%
\mbox{Tr}(\rho E_i)) \nonumber \\
& + & \sum_j p_j \mbox{Tr}(\rho_j E_i) \ln (\mbox{Tr}(\rho_j E_i))\bigg \} \nonumber
%\\
%& = & \sum_i \sum_j \bigg \{ p_j Tr(\rho_j E_i) \ln \frac{(\mbox{Tr}%
%(\rho_j E_i))}{(\mbox{Tr}(\rho E_i))} \bigg \} \; .
\end{eqnarray}
The rationale behind this expression is that the uncertainty in the message
before any measurement is performed is given by the first term and the
second term represents the uncertainty after the measurement has identified
(partially in general) the message states. The Holevo bound is an upper
bound to the above accessible information, i.e.
\begin{eqnarray}
S(\sum_i p_i \rho_i) - \sum_i p_i S(\rho_i) \ge \max_E I(E:\rho) \; .
\label{H-bound}
\end{eqnarray}
This equality is saturated if and only if $[\rho_i,\rho_j]=0$ for all $i$
and $j$. {\em Therefore, since the Holevo bound is an upper bound
to accessible information that Bob can gain about Alice's message, we identify its
maximum over all possible initial probabilities with the classical capacity
of a quantum channel}.

The Holevo bound has an even more suggestive form:
the uncertainty in the initial message is $S(\rho)$, but after the states
are correctly identified the average uncertainty is $\sum_i p_i S(\rho_i)$.
{\em The difference between these two quantities when maximised over
all $p_i$s is the classical communication
capacity of a quantum channel}. Note that one of the most profound implications of the
Holevo bound is that a quantum bit cannot store more information than a
classical bit. In spite of this limitation, quantum information processing is more
efficient than its classical analogue. This is due to the different nature of
information encoding, which is reflected in the existence of superpositions
of different states as well as entanglement between different
qubits (see also section on dense coding).

\noindent {\bf Proof of the Holevo bound in eq. (\ref{H-bound})}.
The Holevo bound is a direct consequence of the fact that the
quantum relative entropy does not increase under CP maps as in
Theorem 1 (note that Holevo's original proof is much more
complicated and does not involve using the quantum relative
entropy. Here I follow Yuen and Ozawa in spirit, as in the last
reference of Holevo (1973)). One such map is
\begin{eqnarray}
\tau (A) = \frac{1}{n} \mbox{Tr} (A) \nonumber
\end{eqnarray}
where $A$ is any $n \times n$ positive matrix. This leads to the
Pierls - Bogoliubov inequality (PBI) (Bhatia, 1997)
\begin{eqnarray}
\tau (A) (\ln \tau (A)-\ln \tau (B)) \leq \tau (A\ln A-A\ln B)
\end{eqnarray}
To prove the Holevo bound I first use that fact that (Theorem 5)
\begin{eqnarray}
S(\rho_i ||\rho)\geq \sum_j S(A_j\rho_iA^{\dagger}_j || A_j\rho A^{\dagger}_j)\nonumber
\end{eqnarray}
PBI now implies that
\begin{eqnarray}
S(A_j \rho_i A^{\dagger}_j ||A_j\rho A^{\dagger}_j) & \geq & \mbox{Tr} (A_j\rho_i A^{\dagger}_j)
\{ \ln (\mbox{Tr} (A_j\rho_i A^{\dagger}_j))\nonumber\\
& - & \ln (\mbox{Tr} (A_j\rho A^{\dagger}_j))\} \nonumber\\
& = & p(j|i) (\ln p(j|i) - \ln p(j)) \nonumber
\end{eqnarray}
where $p(j|i)=Tr\{A_{j}\rho_i A^{\dagger}_{j}\}$ is the conditional probability
that the message $\rho_i$ will lead to the outcome $E_j=A^{\dagger}_jA_j$ and
$p(j)=\sum_i p(j|i)$. Thus we now have that
\begin{eqnarray}
S(\rho_i||\rho) \geq \sum_j p(j|i) (\ln p(j|i) - \ln p(j)) \nonumber
\end{eqnarray}
Multiplying both sides by the (positive) $p_i$ and summing over all $i$
leads to the Holevo bound ${}_{\Box}$.

Since Holevo's result is one of the key results in quantum
information theory I present another simple way of understanding
it via the quantum mutual information. This, of course, is only an
additional motivation for the Holevo bound and by no means proves
its validity. Namely, if Alice encodes the symbol (sym) $i$ into
the state (st) $\rho_i$, then the total state (sym $+$ st) is
\begin{eqnarray}
\rho_{Sym+St} = \sum_i p_i |i\rangle\langle i| \otimes \rho_i \; ,\nonumber
\end{eqnarray}
where the kets $|i\rangle$ are orthogonal (we can think of these as
representing different states of consciousness of Alice!).
Bob now wants to learn about the symbols by distinguishing the
states $\rho_i$. He cannot learn more about the symbols
than is already stored in the correlations
between the symbols and the message states. This as we know is given by the
quantum mutual information
\begin{eqnarray}
I(\rho_{Sym+St}) & = & S(Sym)+S(St)-S(Sym+St) \nonumber \\
& = & S(\sum_i p_i \rho_i) - \sum_i p_i S(\rho_i)
\label{mutualholevo}
\end{eqnarray}
which is the same as the Holevo bound.

I would like now to derive the capacity of a classical
communication channel from the Holevo bound. I follow Gordon's
reasoning who was, in fact, the first person to conjecture the
Holevo bound (Gordon, 1964). As I mentioned before, the Holevo
bound itself contains the classical capacity of a classical
channel as a special case. This, as we might expect, happens when
all $\rho_i$s are diagonal in the same basis, i.e. they commute
(classically all the states and observables commute because they
can be simultaneously specified and measured which is in contrast
with quantum mechanics). {\em Therefore density matrices are
reduced to classical probability distributions}. Let us call this
basis the $B$ representation, with orthonormal eigenvectors
$|b\rangle$. Then the probability that the measurement of the
symbol represented by $\rho_i$ will yield the value $b$ is just
$\langle b |\rho_i | b \rangle$. This I call the conditional
probability $p_i(b)$, that if $\rho_i$ was sent the result $b$
was obtained. Now the Holevo bound is
\begin{eqnarray}
{\bf C} = S(\rho) - \sum_i p_i S(\rho_i)=S(\rho) - S_B (\rho_i)  \; ,\nonumber
\end{eqnarray}
where $S_B (\rho_i)$ is the conditional
entropy given by
\begin{eqnarray}
S_B (\rho_i) = \sum_i p_i \sum_b \langle b |\rho_i | b \rangle \ln \langle b
|\rho_i | b \rangle = \sum_i p_i S(\rho_i) \; .\nonumber
\end{eqnarray}
Thus, the Holevo bound reduces itself to the Shannon mutual information
between the commuting messages and the measurement in the B representation.

In general, the usual rule of thumb
for obtaining quantum information theoretic quantities from their classical
counter-parts is by the convention
\begin{eqnarray}
& \sum & \longrightarrow  \mbox{Trace} \nonumber\\
& \sum p(a) & \longrightarrow  \rho_A \; , \nonumber
\end{eqnarray}
so that, for example, the Shannon entropy $S(p(a))=-\sum_i p(a_i)\ln p(a_i)$ now
becomes the von Neumann entropy $S(\rho_A)=-\mbox{Tr} \rho_A \ln \rho_A$.

\noindent {\bf Example}. As the first application of the Holevo
bound I will compute the channel capacity of a Bosonic field, e.g.
Electromagnetic field (for an excellent review see Caves and
Drummond, 1994). The message information will now be encoded into
modes of frequency $\omega$ and average photon number
$\bar{m}(\omega)$. The signal power is assumed be $S$. The noise
in the channel is quantified by the average number of excitations
$\bar{n}(\omega)$ and is assumed to be independent of the signal
(i.e. the power of signal and noise is additive). We saw that when
there is no noise in the channel the Holevo bound is equal to the
entropy of the average signal. In order to compute the capacity we
need to maximize this entropy with the constraint that the total
power (or energy) is fixed. It is well known that thermal states
are those that maximize the entropy. We thus assume that both the
noise and signal+noise are in thermal equilibrium and follow the
usual Bose-Einstein statistics. The noise power is
\begin{eqnarray}
N= \frac{\pi(kT)^2}{12\hbar} \nonumber
\end{eqnarray}
The power of the output of the channel (signal+noise) is
\begin{eqnarray}
P=S+N= \frac{\pi(kT_e)^2}{12\hbar} \nonumber
\end{eqnarray}
where $T_e$ is the equilibrium temperature of signal+noise. Therefore it
follows that
\begin{eqnarray}
T_e = (12\hbar S/\pi k^2 + T^2)^{1/2} \nonumber
\end{eqnarray}
The state of the noise in the mode $\omega$ is
\begin{eqnarray}
\rho_N (\omega)=\sum_n \frac{1-e^{-\hbar \omega/kT}}{e^{\bar{n}(\omega)\hbar \omega/kT}} |n\rangle\langle n|\nonumber
\end{eqnarray}
while the state of the output is
\begin{eqnarray}
\rho_{N+S}(\omega)=\sum_n \frac{1-e^{-\hbar \omega/kT_e}}{e^{\bar{n}(\omega)\hbar \omega/kT_e}}
|n\rangle\langle n|\nonumber
\end{eqnarray}
The capacity of the channel is given by the Holevo bound which is
\begin{eqnarray}
C & = & \int_{-\infty}^{\infty} [S(\rho_{S+N}(\omega))-S(\rho_{N}(\omega))]d\omega \\
  & = & \frac{\pi k T}{6\hbar \ln 2} \{(12\hbar S/\pi (kT)^2 + 1)^{1/2}-1\} \nonumber
\end{eqnarray}
The integration is there to take into account all the modes of the field.
Let us look at the two extreme limits of this capacity. In the high temperature limit
we obtain the "classical" capacity
\begin{eqnarray}
C_{C}=\frac{S}{kT\ln 2}\; ,
\end{eqnarray}
a result derived by Shannon and Weaver (1949). This states that in
order to communicate one bit of information with this set-up we
need exactly $kT\ln 2$ amount of energy. In the low temperature
limit, on the other hand, quantum effects become important and the
capacity becomes independent of $T$
\begin{eqnarray}
C_{Q}=\frac{\sqrt{\pi}}{\ln 2} \{\frac{S}{\hbar}\}^{1/2}\; ,
\end{eqnarray}
which was derived by Stern (1960), Gordon (1964), Lebedev and
Levitin (1963) and Yamamoto and Haus (1986) among others. Note
also the appearance of Planck's constant which is a key feature of
quantum mechanics. If we wish to communicate one bit of
information in this limit we need only $\hbar/\pi (\ln2) \sim
10^{-34}$ joules of energy. {\em This is significantly less than
the corresponding energy in the classical limit}. Let us now
compare the classical and quantum capacity limits to the total
energy of $N$ harmonic oscillators (Bosons) in the same two
limits. In the high temperature limit the equipartition theorem
is applicable and the total energy is $3NkT$ (i.e. it depends on
temperature). In the low temperature limit all the Harmonic
oscillators settle down to the ground state so that the total
energy becomes $N\hbar \omega/2$ (i.e. it is independent of
temperature and we see the quantum dependence through Plank's
constant $\hbar$).

\subsection{Schumacher's compression}

The most optimal communication through a noiseless channel using
{\em pure} states is equivalent to data compression. We have seen
in eq. (\ref{21.exp}) that the limit to the classical data
compression is just given by the entropy of the probability
distribution of the data. We would thus guess that the limit to
quantum data compression is given by the von Neumann entropy of
the set of states being compressed. This, in fact, turns out to be
a correct guess as first proven by Schumacher (1995). So, Alice
now encodes letters of her classical message into {\em pure}
quantum states and sends these to Bob. For example if $a
\rightarrow |\psi_a\rangle$ and $b \rightarrow |\psi_b\rangle$,
then Alice's message $aab$ will be sent to Bob as the sequence of
pure quantum states $|\psi_a\rangle|\psi_a\rangle|\psi_b\rangle$.

The exact problem can be phrased in the following equivalent
fashion: suppose a quantum source randomly prepares different
qubit states $|\psi_i\rangle$ with the corresponding probabilities
$p_i$. A random sequence of $n$ such states is produced. By how
much can this be compressed, i.e. how many qubits do we really
need to encode the original sequence (in the limit of large $n$)?
First of all the total density matrix is
\begin{eqnarray}
\rho = \sum_i p_i |\psi_i\rangle\langle \psi_i| \nonumber
\end{eqnarray}
Now, this matrix can be diagonalised
\begin{eqnarray}
\rho = \sum_i r_i |r_i\rangle\langle r_i| \nonumber
\end{eqnarray}
where $r_i$ and $|r_i\rangle$ are the eigenvectors
and eigenvalues. This decomposition is, of course, indistinguishable from
the original one (or any other decomposition for that matter). Thus we can think about
compression in this new basis, which is easier as it behaves completely
classically (since $\langle r_i|r_i\rangle=\delta_{ij}$). We can therefore
invoke results from the previous section on classical typical sequences
and conclude that the limit to compression is $n(-\sum_i r_i \ln r_i)$,
i.e. $n$ qubits can be encoded into $nS(\rho)$ qubits. No matter how the
states are generated, as long as the total state is described by the same
density matrix $\rho$ its compression limit is its von Neumann entropy.
This protocol and result will be very important when we discuss entanglement
measures in the following section.

%% Bloch Sphere Fig4

\begin{figure}[ht]
\begin{center}
\hspace{0mm}
\epsfxsize=7.5cm
\epsfbox{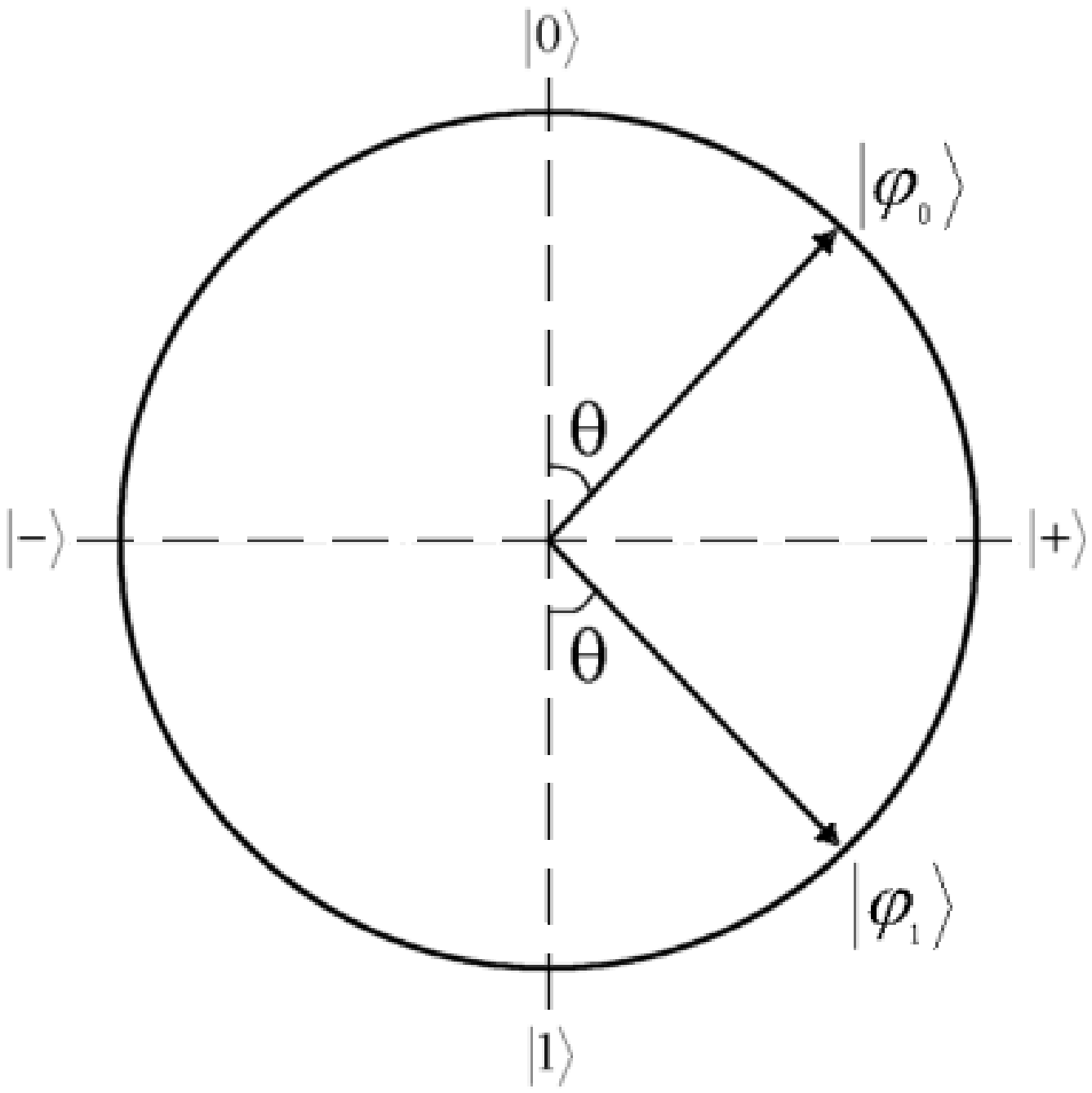}\\[0.2cm]
\begin{caption}
{\narrowtext This figure shows the two non-orthogonal states on the
Bloch sphere which are used to encode a message. The overlap
between them is $\sin \theta$ and the smaller the overlap, the
more the total message can be compressed. In terms of information, the
less distinguishable the states (i.e. the smaller the overlap),
the less information they carry.}
\end{caption}
\end{center}
\end{figure}

\noindent
{\bf Example}. Suppose that Alice encodes her bit into states $|\Psi_0\rangle =
\cos (\theta/2) |0\rangle + \sin (\theta/2) |1\rangle $ and $|\Psi_1\rangle =
\sin (\theta/2) |0\rangle + \cos (\theta/2) |1\rangle$ with $p_0=p_1=1/2$
(see Fig. 4). Classically it is not possible to compress a source that generates $0$ and $1$
with equal probability. Quantum mechanically, however, compression
can be achieved not only by the nature of the probability distribution
but also due to the non-orthogonality of the states encoding symbols of the
message. In our example the overlap between
the two states is $\langle \psi_0|\psi_1 \rangle = sin \theta$ and they
are orthogonal only when $\theta = \pi$ in which case no compression is
possible. Otherwise, the compression ratio is directly proportional to
the overlap between the states. Suppose Alice's messages are only
$3$ qubits long. Then there are $8$ different possibilities, $|\Psi_0\Psi_0\Psi_0\rangle,...
|\Psi_1\Psi_1\Psi_1\rangle$, which are all equally likely with $1/8$ probability.
In general these states will lie with a high probability within a subspace
of the $8$ dimensional Hilbert space. Let us call this likely subspace a "typical"
subspace. Its orthogonal complement will be unlikely and hence called an "atypical"
subspace. In order to find the typical and atypical subspaces we need
to diagonalise the "average" signal
\begin{eqnarray}
\rho = \frac{1}{2} (|\Psi_0\rangle\langle \psi_0| + |\Psi_1\rangle\langle \psi_1|) \nonumber
\end{eqnarray}
Its diagonal form is
\begin{eqnarray}
\rho = \frac{1}{2} (1+\sin \theta )|+\rangle\langle +| + \frac{1}{2} (1-\sin \theta )|-\rangle\langle -| \nonumber
\end{eqnarray}
where $|\pm\rangle = |0\rangle \pm |1\rangle$. Now we look at the probabilities
for each of the $8$ messages to lie along the new orthogonal basis $|+++\rangle,...
|---\rangle$ of the Hilbert space of three qubits:
\begin{eqnarray}
|\langle +++|\psi^{\otimes 3}\rangle|^2 & = & (\cos (\theta/2) + \sin (\theta/2))^6 \nonumber\\
|\langle ++-|\psi^{\otimes 3}\rangle|^2 & = & (\cos (\theta/2) + \sin (\theta/2))^4 \nonumber\\
& + & (\cos (\theta/2) - \sin (\theta/2))^2  \nonumber \\
|\langle +--|\psi^{\otimes 3}\rangle|^2 & = & (\cos (\theta/2) + \sin (\theta/2))^2 \nonumber\\
& + & (\cos (\theta/2) - \sin (\theta/2))^4 \nonumber \\
|\langle ---|\psi^{\otimes 3}\rangle|^2 & = & (\cos (\theta/2) - \sin (\theta/2))^6\; \nonumber
\end{eqnarray}
where $|\psi^{\otimes 3}\rangle$ represents any $3$ qubit sequence of $|\psi_0\rangle$
and $|\psi_1\rangle$. In addition all the probabilities for $|++-\rangle,|+-+\rangle,
|-++\rangle$ are equal and so are the probabilities for $|+--\rangle,|--+\rangle,
|-+-\rangle$. Thus the above equation contains $64$ probabilities in total.
Suppose now that $\cos (\theta/2) \sim \sin (\theta/2)$. Then, we see that
the states containing two or more $+$ become much more likely. This means
that the message states are much more likely to be in this particular
subspace. Therefore the compression would be as follows. First the source
generates three qubits in some state. Then we project this message onto
the typical subspace. If we are successful, then this will lie in that
four dimensional typical subspace for which we need only two qubits rather
than three. Otherwise, our projection will fail and the message will end up
in the atypical subspace in which case Alice does not compress it. The probability
to end up in the atypical space asymptotically goes to zero (the law of large numbers).
Therefore in this example the limit to our compression is given by
$-(1/2(1+\sin \theta ))\ln (1/2(1+\sin \theta )) - (1/2(1-\sin \theta )) \ln (1/2(1-\sin \theta ))$ which is
of course the von Neumann entropy of $\rho$. The number of
dimensions of the typical subspace of the total Hilbert space is likewise
in general equal to $e^{nS(\rho)}$.

Interestingly, if instead of pure states a quantum source
generates mixed states $\rho_i$ with probabilities $p_i$, then the
best compression limit is in general unknown. We can, of course,
use the above protocol to compress the sequence to the von Neumann
entropy of the average signal, $S(\sum_i p_i \rho_i)$. However, in
some cases it is known that a better compression can be achieved.
The lower bound to compression is the Holevo bound, $S(\sum_i p_i
\rho_i)-\sum_i p_i S(\rho_i)$, but it is not known whether this
bound can in general be attained (see Horodecki, 1998b).

Next we look at a protocol for classical communication that involves
entanglement. At first sight this protocol seems to violate the Holevo bound
on classical communication, i.e. that it is possible to
communicate only 1 bit per single qubit. However, a
closer inspection will show that this is not the case.

\subsection{Dense coding}

Now I consider the case of dense coding which was introduced by
Bennett and Wiesner, 1992. In this protocol entanglement plays a
crucial role and this will give us a first indication of the fact
that entanglement can be quantified like any other resource, such
as energy for example. Alice and Bob initially share an entangled
pair of qubits in some state $W_0$, which may be mixed. Alice then
performs local unitary operations on her qubit to put this shared
pair of qubits into either of the states $W_0, W_1, W_2$ or $W_3$.
In general, Alice may use a completely arbitrary set of unitary
operations to generate these states:
\begin{eqnarray}  \label{cg}
W_i={\bf U_i}\otimes {\bf I}~W_0~{\bf U_i^{\dagger}}\otimes {\bf
I},
\end{eqnarray}
and the number of generated states is completely arbitrary. In the
above equation, ${\bf U_i}$ acts on Alice's qubit and ${\bf I}$
acts on Bob's qubit. By sending her encoded qubit to Bob, Alice is
essentially communicating with Bob using the states $W_0, W_1,
W_2$ and $W_3$ as separate letters. The number of bits she can
communicate to Bob using this procedure is thus bounded by the
Holevo bound. Moreover, if some block coding is done on a large
enough collection of qubits in addition to the dense coding, then
the number of bits of information communicated is equal to the
Holevo function. We will thus take
\begin{eqnarray}  \label{see}
\mbox{\bf C}= S(\rho)-\sum_i p_i S(\rho_i),
\end{eqnarray}
assuming that any additional necessary block coding will
automatically be performed to supplement the dense coding. This
coding is essential in order to achieve the capacity given by the
Holevo bound, in the asymptotic limit (The fact that the bound is
achievable follows from a complicated argument and cannot really
be derived using the arguments presented in this review. Hausladen
et. al (1996) have proved this for pure states and Schumacher and
Westmoreland (1997) and independently Holevo (1998) for mixed
states). Exactly the same assumption has been used in Ref.
Hausladen et. al (1996) to calculate the capacity for dense coding
in the case of pure letter states. Eqs.(\ref{cg}) and (\ref{see})
define the most general version of dense coding and I shall refer
to this as completely general dense coding (CGCD).

A simpler example of dense coding is the case when the letter states are
generated from the initial shared state $W_0$ by

\begin{eqnarray}  \label{w0}
W_0={\bf I}\otimes {\bf I}~W_0~{\bf I}\otimes {\bf I}, \\
W_1={\bf \sigma_1}\otimes {\bf I}~W_0~{\bf \sigma_1}\otimes {\bf I}, \\
W_2={\bf \sigma_2}\otimes {\bf I}~W_0~{\bf \sigma_2}\otimes {\bf I}, \\
W_3={\bf \sigma_3}\otimes {\bf I}~W_0~{\bf \sigma_3}\otimes {\bf I}.
\label{w1}
\end{eqnarray}
In the above set of equations, the first operator of the combination ${\bf %
\sigma_i} \otimes {\bf I}$ acts on Alice's qubit and the second operator
acts on Bob's qubit. I shall refer to this case (i.e when the letter states
are generated by Eqs.(\ref{w0})-(\ref{w1})) as simply general dense coding
(GDC). The generality present in GDC is that Alice is allowed to prepare the
different letter states with unequal probabilities.

In the more special case when Alice not only generates the four letter
states according to Eqs.(\ref{w0})-(\ref{w1})) but also with equal
probability, the ensemble is given by
\begin{eqnarray}  \label{avg1}
W = \frac{1}{4} \sum_{i=0}^3 W_i.
\end{eqnarray}
and the capacity becomes
\begin{eqnarray}  \label{c}
\mbox{\bf C} = \frac{1}{4} \sum_{i=0}^3 S(W_i||W).
\end{eqnarray}
I shall call this simplest case special dense coding (SDC). Among
all the possible ways of doing GDC, SDC is the optimal way to
communicate when $W_0$ is a pure state (Bose et. al 2000a) or a
Bell diagonal state.

Now I derive the most general bound on CGDC (Bowen, 2001).
Furthermore, this bound can be attained by the same protocol as
SDC (Bowen, 2001). The proof is achieved by first finding an
upper bound to the capacity for CGDC and then showing that SDC
actually saturates this bound. Suppose that the initial state of
Alice and Bob is $\rho_{AB}$. Then we have:
\begin{eqnarray}
\mbox{\bf C} & = & \max S(\sum_k p_k (U^k\otimes I) \rho_{AB}
((U^k)^{\dagger}\otimes I)) \nonumber \\
& - &  \sum_k p_k S((U^k\otimes I) \rho_{AB}
((U^k)^{\dagger}\otimes I)) \nonumber \\
& = & \max S(\rho'_{AB})-S(\rho_{AB})\nonumber \\
& \leq & S(\rho'_A) + S(\rho'_B) - S(\rho_{AB})\nonumber\\
& \leq & 1 + S(\rho_B) - S(\rho_{AB}) \label{BowenCGDC}
\end{eqnarray}
Since this bound is achievable as shown by Bowen (2001), the
capacity for CGDC is given by eq. (\ref{BowenCGDC}).

I shall now restrict my attention to a calculation of ${\bf C}$
for pure letter states. Consider the initial shared pure state
$W_0$ to be,
\begin{eqnarray}  \label{p0}
|\psi_0\rangle = (a|00\rangle + b|11\rangle ).
\end{eqnarray}
Then, according to Eqs.(\ref{w0})-(\ref{w1}), the other letter states are
given by
\begin{eqnarray}
|\psi_1\rangle = (a|10\rangle + b|01\rangle ), \\
|\psi_2\rangle = -i (a|10\rangle - b|01\rangle ), \\
|\psi_3\rangle = (a|00\rangle - b|11\rangle ),  \label{p3}
\end{eqnarray}
from which we obtain $W_i = |\psi_i\rangle \langle \psi_i|$. As all $W_i$
are pure states we have
\begin{eqnarray}
S(W_i)=0.
\end{eqnarray}
Thus we have
\begin{eqnarray}  \label{sw}
\mbox{\bf C}=S(W).
\end{eqnarray}
I will consider only the case of SDC as it is optimal. Thus the
ensemble used is obtained from Eq.(\ref{avg1}) to be
\begin{eqnarray}
W &=& \frac{|a|^2}{2}|00\rangle\langle 00| + \frac{|b|^2}{2}%
|01\rangle\langle 01|  \nonumber \\
&+& \frac{|a|^2}{2}|10\rangle\langle 10| + \frac{|b|^2}{2}|11\rangle\langle
11|. \nonumber
\end{eqnarray}
Thus from Eq.(\ref{sw}) for the capacity {\bf C}, we get
\begin{eqnarray}
\mbox{\bf C} &=& - (|a|^2 \log \frac{|a|^2}{2}+ |b|^2 \log \frac{|b|^2}{2})
\nonumber\\
&=& 1 - (|a|^2 \log |a|^2+ |b|^2 \log |b|^2).
\end{eqnarray}
(Note that this agrees with eq. (\ref{BowenCGDC}) as for pure
states the total entropy is zero.) Now this implies that a good
measure of entanglement for a pure state of a system composed of
two subsystems A and B can be given by the von Neumann entropy of
the state of either of the subsystems. Let us call this measure
the von Neumann entropy of entanglement and label it by $E_v$
(Popescu and Rohrlich, 1997; Bennett et. al, 1996a). Thus
\begin{eqnarray}
E_v(|\psi\rangle \langle \psi|_{\mbox \scriptsize {A+B}})= S {\large (}%
\mbox{Tr}_{\mbox \scriptsize {A}}(|\psi\rangle \langle \psi|_{\mbox
\scriptsize {A+B}}){\large )}, \nonumber
\end{eqnarray}
where $\mbox{Tr}_{\mbox \scriptsize {A}}$ stands for partial trace over
states of system A. Therefore, for all the states $W_i$,
\begin{eqnarray}
E_v(W_i) = - (|a|^2 \log |a|^2+ |b|^2 \log |b|^2). \nonumber
\end{eqnarray}
Thus,
\begin{eqnarray}  \label{pure1}
\mbox{\bf C}=1+E_v(W_i). \nonumber
\end{eqnarray}

%% dense coding Fig.5

\begin{figure}[ht]
\begin{center}
\hspace{0mm}
\epsfxsize=7.5cm
\epsfbox{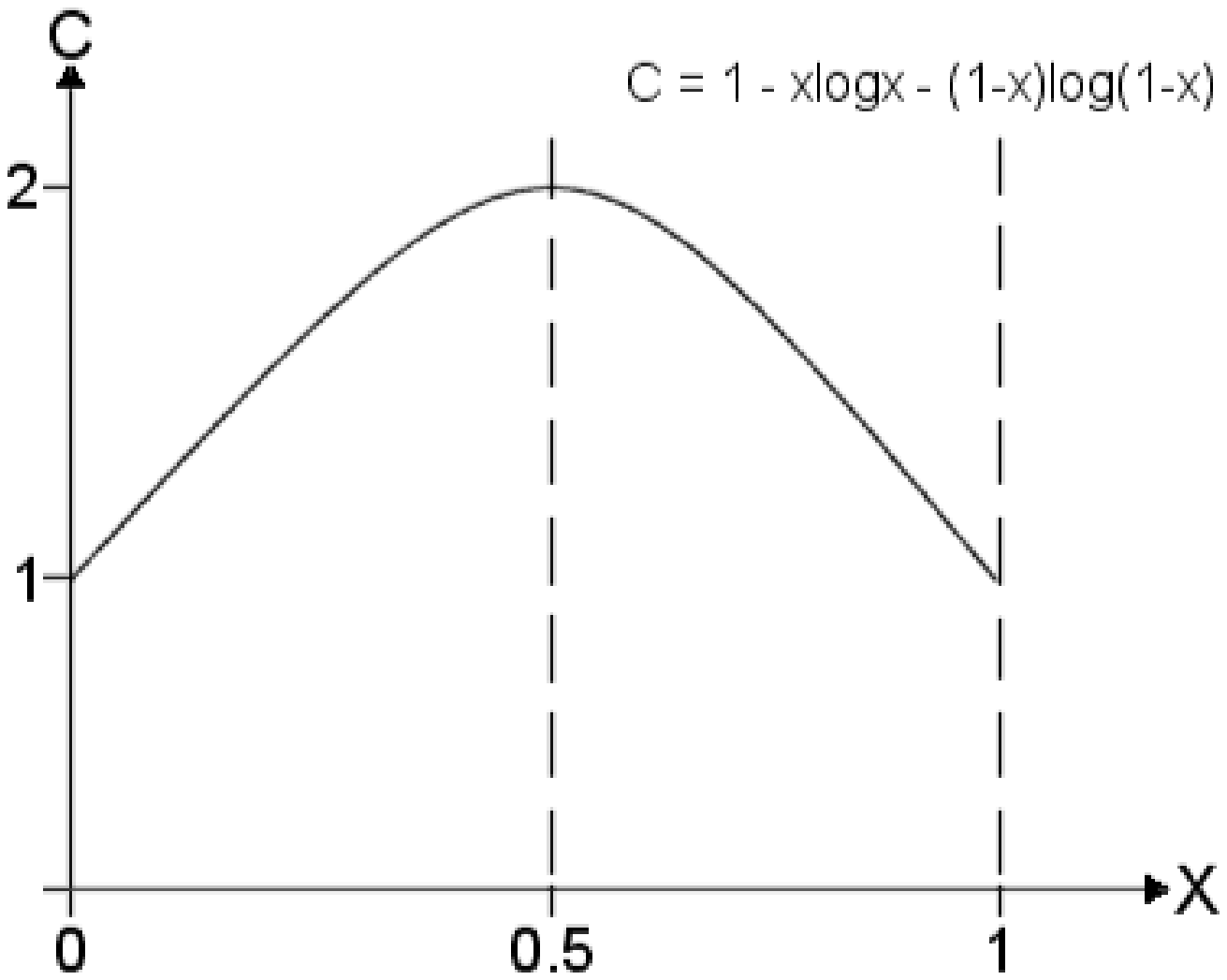}\\[0.2cm]
\begin{caption}
{\narrowtext This figure shows the dependence of capacity for dense
coding for pure states $a|00\rangle+b|11\rangle$ as a function of
its Schmidt coefficient $x=|a|^2$. We see that when the state is
disentangled, i.e. when either $a=0$ or $b=0$, the capacity
becomes 1 bit per qubit, i.e. the same as the "classical capacity".}
\end{caption}
\end{center}
\end{figure}

\noindent
(see Fig. 5). We can prove that for pure states, SDC (using all alphabet states with equal
{\em a priori} probability) is the optimal way to communicate among all
possible ways of doing GDC (i.e when the letter states are generated by Eqs.((%
\ref{w0})-(\ref{w1})) (Bose et. al, 2000a). Important, however, is
the fact that the amount of entanglement determines exactly how
much information Alice can convey to Bob. Note that if there is no
entanglement shared between them, then the amount of information
is exactly one bit per Alice's qubit (which is what can be
achieved classically after all). At the other extreme, when they
share a maximally entangled state, the amount of information is
$2$ bits per Alice's qubit. This is the amount that no purely
classical communication can achieve. However, while the von
Neumann entropy is a good measure of entanglement for pure states
(in fact, there are arguments that it is unique for pure states
(Popescu and Rohrlich, 1997)), it fails when we try to apply it to
mixed states. A possibility is to follow the logic of the pure
state dense coding and call $S(\rho_B) - S(\rho_{AB})$ a measure
of entanglement for mixed states as in eq. (\ref{BowenCGDC}). This
measure has been called the ``coherent information" and is used to
describe information transmission through a noisy quantum channel,
as in e.g. Barnum et.al (1998). But, is this measure consistent
with other natural requirements for quantifying entanglement? This
question will be addressed in the next section. Before this, we
show that in order to delete a certain amount of correlations we
need to increase the entropy of the environment by at least this
amount. This is known as Landauer's erasure (Landauer, 1961) and
is seen to be linked directly to the relative entropy.

\subsection{Relative entropy, thermodynamics and information erasure}

We have seen that communication essentially creates correlations
between the sender and the receiver. Creating correlations is
therefore very important in order to be able to convey any
information. However, I would now like to talk about the opposite
process - deleting correlations. Why would one want to do so? The
reason is that we might want to correlate one system to another
and may need to delete all its previous correlations to be able
to store new ones. I would like to give a more physical statement
of information erasure and link it to the notion of measurement.
I will therefore introduce two correlated parties - a system and
an apparatus. The apparatus will interact with the system thereby
gaining a certain amount of information about it (the full
quantum description of this process will be presented in section
V). Suppose now that the apparatus needs to measure another
system. We first need to delete information about the last system
before we can make another measurement. The most general way of
conducting erasure (resetting) of the apparatus is by employing a
reservoir in thermal equilibrium at certain temperature $T$. To
erase the state of the apparatus we just throw it into the
reservoir and introduce a new pure state. The entropy increase of
the operation now consists of two parts. Firstly, the state of
the apparatus evolves to the state of the reservoir and this
entropy is now added to the reservoir entropy. Secondly, the rest
of the reservoir changes its entropy due to this interaction
which is the difference in the apparatus internal energy before
and after the resetting (no work is done in this process). This
quantum approach to equilibrium was also studied by Partovi
(1989). A good model is obtained by imagining that the reservoir
consists of a great number of systems (of the same "size" as the
apparatus) all in the same quantum equilibrium state $\omega$.
Then the apparatus, which is in some state $\rho$, interacts with
these reservoir systems one at a time. Each time there is an
interaction, the state of the apparatus approaches more closely
the state of the reservoir, while that single reservoir system
also changes its state away from the equilibrium. However, the
systems in the bath are numerous so that after a certain number
of collisions the apparatus state will approach the state of the
reservoir, while the reservoir will not change much since it is
very large (this is equivalent to the so called Born-Markov
approximation that leads to irreversible dynamics of the
apparatus described here).

Bearing all this in mind, we now reset the apparatus by plunging it into a
reservoir in a thermal equilibrium (Gibbs state)
at temperature $T.$ Let the state of the reservoir be
\[
\omega =\frac{e^{-\beta H}}{Z}=\sum_{j}q_{j}\left| \varepsilon
_{j}\right\rangle \left\langle \varepsilon _{j}\right|
\]
\noindent
where $H=$ $\sum_{i}\varepsilon _{i}\left| \varepsilon
_{i}\right\rangle \left\langle \varepsilon _{i}\right| $ is the
Hamiltonian of the reservoir, $Z=Tr(e^{-\beta H})$ is the
partition function and $\beta^{-1}=k T$, where
$k$ is the Boltzmann constant. Now suppose that due to
the measurement the entropy of the apparatus is $S(\rho )$ (and an amount $
S(\rho )$ of information has been gained)$,$ where $\rho
=\sum_{i}r_{i}\left| r_{i}\right\rangle \left\langle r_{i}\right| $ is the
eigen expansion of the apparatus state. Now the total entropy increase in
the erasure is (there are two parts as I argued above: 1. change in the entropy
of the apparatus and 2. change in the entropy of the reservoir)
\[
\Delta S_{er}=\Delta S_{app}+\Delta S_{res}
\]

\noindent We immediately know that $\Delta S_{app}=S(\omega )$, since the
state of apparatus (no matter what state it was before) is now erased to be the same as
that of the reservoir. On the other hand, the entropy change in the
reservoir is the average over all states $\left| r_{i}\right\rangle $ of
heat received by the reservoir divided by the temperature. This is minus the
heat received by the apparatus divided by the temperature; the heat received
by the apparatus is the internal energy after the resetting minus the initial internal
energy $\left\langle r_{i}\right| H$ $\left| r_{i}\right\rangle .$ Thus,
\begin{eqnarray*}
\Delta S_{res} &=&-\sum_{k}r_{k}\frac{Tr(\omega H)-\left\langle r_{k}\right|
H\left| r_{k}\right\rangle }{T} \\
&=&\sum_{k}(r_{k}\sum_{j}|\left\langle r_{k}|\varepsilon _{j}\right\rangle
|^{2}-q_{k})(-\log q_{k}-\log Z) \\
&=&-Tr(\rho -\omega )(\log \omega -\log Z) = Tr(\omega-\rho)\log \omega
\end{eqnarray*}

\noindent
Altogether we have an exact expression on the amount of entropy increase due
to deletion:

\vspace*{.3cm}
\noindent
%\fbox{\parbox[b]{17.7cm}{
{\bf Entropy increase due to Landauer's erasure.}
\[
\Delta S_{er}=-Tr(\rho \log \omega )
\]
%}}
\vspace*{.3cm}
\noindent This result (Vedral, 2000) generalizes Lubkin's result
which applies only when $\left[ \rho ,\omega \right] =0$. In
general, however, the information gain is equal to $S(\rho )$,
the entropy increase in the apparatus. This entropy increase is a
maximum, the information between the system and apparatus is
usually smaller as in eq. ({\ref{mutualholevo}). Thus, we see that
\[
\Delta S_{er}=-Tr(\rho \log \omega )\geq S(\rho )=I
\]

\noindent and Landauer's principle is confirmed (the inequality follows from
the fact that the quantum relative entropy $S(\rho ||\omega )=-Tr(\rho \log
\omega )-S(\rho )$ is non-negative). So the erasure is the least wasteful when $%
\omega =$ $\rho ,$ in which case the entropy of erasure is equal to $S(\rho
),$ the information gain. This is when the reservoir is in the same state as the
state of the apparatus we are trying to erase. In this case we just have a
state swap between the new pure state of the apparatus which is used to replace our old
state $\rho$. Curiously enough, creating correlations is not costly in terms of the
entropy of environment (such as when Alice and Bob communicate).

Landauer's erasure is a statement which is equivalent to the
Second Law of Thermodynamics. If we could delete information
without increasing entropy, then we could construct a machine
that completely converts heat into work with no other effect
which contradicts the Second law. The opposite is also true.
Namely if we could convert heat into work with no other effect
than we could use this energy to delete information with no
entropy increase (Penrose, 1973; Landauer, 1961). Thus, the
relative entropy provides an interesting link between
thermodynamics, information theory and quantum mechanics (also
see Brillouin's excellent book (Brillouin, 1956)).

\noindent {\bf Landauer's erasure and data compression}. I will
now show how Landauer's principle can be used to derive the limit
to quantum data compression. The free energy lost in deleting
information stored in a string of $n$ qubits all is the state
$\rho$ is $n\beta ^{-1}S(\rho )$. However, we could first
compress this string and then delete the resulting information.
The free energy loss after the compression is $m\beta ^{-1}\log
2=m\beta^{-1}$, where the string has been compressed to $m$
qubits. The two free energies before and after compression should
be equal if no information is lost during compression, i.e. if we
wish to have maximal efficiency, and therefore $m/n=$ $S(\rho )$
as shown previously (c.f. (Feynman, 1996)). The equality is, of
course, only achieved asymptotically.

So far we have seen that the entropy plays a pivotal role in communication theory and
data compression as a limit to both communication capacity and compression. It
also quantifies the amount of entanglement in a pure bi-partite state. Finally,
it plays thermodynamical role characterizing the mixedness in a certain quantum
state. This last role was introduced first by von Neumann. Now we go beyond
the classical use of quantum states towards looking at how we can achieve
quantum communication of quantum states.

\section{Quantum Communication: Quantum Use}

In this section the problem of entanglement quantification is
analysed. Previously we have seen that the reduced von Neumann
entropy is a good measure of entanglement for two subsystems in a
joint pure state (see also Bennett et. al (1996a)). This is a
consequence of the Schmidt decomposition procedure introduced
earlier and was exemplified in the dense coding. However, for the
mixed states of two subsystems, or for more than two subsystems
this procedure does not exist in general. Therefore it is not
immediately clear how to understand and quantify correlations for
these states. Initially, we might think that Bell's inequalities
(Bell, 1987; Clauser et. al, 1969; Redhead, 1987) would provide a
good criterion for separating quantum correlations (entanglement)
from classical correlations in a given quantum state. States that
violate Bell's inequalities would be entangled and other states
would be disentangled. However, while it is true that a violation
of Bell's inequalities is a signature of quantum correlations, not
all entangled states violate Bell's inequalities (Gisin; 1996).
So, in order to completely separate quantum from classical
correlations we need a different criterion.

I will present here an approach that has proven to be very fruitful in understanding
entanglement in general. It begins by presenting a set of conditions that
any reasonable measure of entanglement has to satisfy. Then, I discuss
possible candidates based on this criterion.

\subsection{Quantifying entanglement}

In this section we will mainly focus on understanding entanglement
of bi-partite systems, i.e. systems containing two subsystems
only. The term entanglement, or versr\"ankung as it was originally
called, was introduced by Schr\"odinger (1935) to emphasise
bizarre implications of quantum mechanics. The reason for studying
bi-partite entanglement is that it is the simplest and most basic
kind of entanglement and is well understood at present. Also,
starting from bi-partite entanglement we will build up a theory
that can be generalized to any number of systems. So, unless
stated otherwise, the presentation in this subsection is confined
to bi-partite systems only.

To determine the basic properties every "good" entanglement
measure should satisfy (Vedral et. al, 1997a; Vedral and Plenio,
1998), we have to discuss the issues of what we actually mean when
we say that something is "disentangled". By definition a
bi-partite state is disentangled if it can be written in the
separable form $\rho_{AB}=\sum_i p_i \rho_i^A\otimes \rho_i^B$
(Werner, 1989). It is clear why we choose to define disentangled
states in this manner: these are the most general states which can
be can be created by Alice and Bob by local operations and
classical communication (LOCC). Thus these states contain no
entanglement, as entanglement can only be created through global
operations. All other states will be entangled to some degree. In
addition note that the set of all disentangled states is convex: a
convex combination (mixture) of any two disentangled states is
itself disentangled. This fact will be important when we quantify
entanglement later.

So the first question to answer is the following: "{\em when can a
given matrix be written in a separable form?"}. The necessary and
sufficient condition is known in general in terms of positive (but
not  necessarily completely positive) maps (Peres, 1996;
Horodecki et. al, 1996). Suppose that $\Lambda$ is any positive
map; then,
\begin{eqnarray}
I_A\otimes \Lambda_B (\sum_i p_i \rho_i^A\otimes \rho_i^B)=\sum_i p_i \rho_i^A\otimes\Lambda_B (\rho_i^B)
\end{eqnarray}
is always a positive operator. Remarkably, the converse is also
true. If, for all positive maps $\Lambda$, the state $I_A\otimes
\Lambda_B(\rho_{AB})$ is positive, then $\rho_{AB}$ is separable
(disentangled). Therefore, if we want to know whether a given
state $\rho_{AB}$ is entangled, we need to find a positive map
whose action on $B$ will result in a negative operator and hence
not a physical state (Horodecki, 2000a). This condition is still
not operational since there is an infinite number of positive maps
to search. In fact, there is no operational condition in general,
but it only exists in some special cases. For example, for two
qubits and a qubit and a qutrit (three level system), this
condition simplifies to the following (Peres, 1996; Horodecki et.
al, 1996): such a state is entangled iff a transposition of $B$
results in a negative operator, i.e. $\rho_{AB}^{T_B}$. The
relationship between positive maps and entanglement is a very
active field of research and I refer the interested reader to some
papers investigating this issue: Bennett et. al (1999b), Kraus et.
al (2000), DiVicenzo et. al (2000) and Lewenstein et. al (2000).
With this in mind, I turn to quantifying entanglement.

The first property we need from an entanglement measure is that a disentangled
state does not have any quantum correlations.
This gives rise to our first condition:
\begin{description}
\item[E1)] For any separable state $\sigma$ the measure of entanglement should be zero, i.e.
\begin{eqnarray}
    E(\sigma) = 0 \;\; .
    \label{cond1}
\end{eqnarray}
\end{description}
Note that we do not ask the converse, i.e. that if $E(\sigma)=0$,
then $\sigma$ is separable. The reason for this will become clear
below.

The next condition concerns the behavior of the entanglement under
simple local unitary transformations. A local unitary
transformation simply represents a change of the basis in which we
consider the given entangled state. But a change of basis should
not change the amount of entanglement that is accessible to us,
because at any time we could just reverse the basis change (since
unitary transformations are fully reversible).
\begin{description}
\item[E2)] For any state $\sigma$ and any local unitary transformation, i.e. a unitary
transformation of the form $U_A \otimes U_B$, the entanglement remains unchanged. Therefore
\begin{eqnarray}
    E(\sigma) = E(U_A \otimes U_B \sigma U_A^{\dagger} \otimes U_B^{\dagger}) \;\; .
    \label{cond2}
\end{eqnarray}
\end{description}
The third condition is the one that really restricts the class of possible entanglement
measures. Unfortunately it is usually also the property that is the most difficult
to prove for potential measures of entanglement. We have already proved that
no good measure of correlations between two subsystems should increase under local
operations on the subsystems separately. However, quantum entanglement is even
more restrictive in that the total amount of entanglement cannot
increase locally {\em even with the aid of classical
communication}. Classical correlations, on the other hand, can be
increased by LOCC.

\noindent
{\bf Example}. Suppose that Alice and Bob share $n$ uncorrelated
pairs of qubits, for example all in the state $|0\rangle$.
Alice's computer then interacts with each of her qubits such that
it randomly flips each qubit with probability $1/2$. However,
whenever a qubit is flipped, Alice's computer (classically) calls
Bob's computer and informs it to do likewise. After this action
on all the qubits, Alice and Bob end up sharing $n$ (maxiamlly) correlated
qubits in the state $|00\rangle\langle 00| + |11\rangle\langle 11|$,
i.e. whenever Alice's qubit is zero so is Bob's and whenever Alice's
qubit is one so is Bob's. The state of each pair is mixed because
Alice and Bob do not know whether their computers flipped their respective
qubits or not.

We can always calculate the total amount of entanglement by summing up the
entanglement of all systems after we have applied our local operations and classical
communications.
\begin{description}
\item[E3)] Local operations, classical communication and subselection cannot increase the
expected entanglement, i.e. if we start with an ensemble in state $\sigma$ and end up with
probability $p_i$ in subensembles in state $\sigma_i$ then we will have
\begin{eqnarray}
    E(\sigma) \ge  \sum_i p_i E(\sigma_i) \;\; .
\label{cond3}
\end{eqnarray}
\end{description}
where $\sigma_i = A_i\otimes B_i \sigma A_i^{\dagger}\otimes
B_i^{\dagger}/p_i$ and $p_i = \mbox{Tr} (A_i\otimes B_i \sigma
A_i^{\dagger}\otimes B_i^{\dagger})$. The form $A\otimes B$ shows
that Alice and Bob perform their operation locally (i.e. Alice
cannot affect Bob's system and vice versa).  However, Alice's and
Bob's operations can be correlated which is manifested in the fact
that they have the same index. It should be pointed out that
although all the LOCC can be cast in the above product form, the
opposite is not true: not all the operations of the product form
can be executed locally (Bennett et. al, 1999a). This means that
the above condition is more restrictive than necessary, but this
does not have any significant consequences as far as I am aware.
An example of E3 operation is local addition of particles on
Alice's and Bob's side. Note also that E2 operations are a subset
(special case) of E3 operations.

The last condition is there to make sure that our measure is consistent with pure
states.
\begin{description}
\item[E4)] Entanglement of a pure state is equal to the reduced von Neumann
entropy.
\end{description}

The above conditions are natural and easy to understand
physically. However, they can be reduced to simpler and more
elementary conditions, which I now briefly discuss. Suppose that
we ask that the measure of entanglement is:

\begin{enumerate}
\item {\em Weakly additive}, i.e. $E(\rho\otimes\rho) =
2E(\rho)$;
\item {\em Continuous}, i.e. if $\rho$ is close to $\sigma$, then
$E(\rho)$ is close to $E(\sigma)$.
\end{enumerate}
Then, it can be shown (Popescu and Rohrlich, 1997; Vidal, 2000)
that E4 is a consequence of the weak additivity and continuity
(providing we assume that the entanglement of a maximally
entangled state is normalised to $\log 2$). Also, in E3 we use
the most general local POVMs, but we know that these can be
implemented by adding ancillas locally, performing a unitary
transformation on the system and ancilla locally and then tracing
out the ancillas. So, E2 and E3 can be presented in a more
elementary way as was done by Vidal (2000). Thus, E2-E4 can be
written in terms of more fundamental processes and properties.
However, I chose to introduce entanglement measures via E1-E4 as
I think that they are more intuitive and capture the main ideas.
Readers interested in further analysis of the conditions are
advised to read: Vidal (2000) and Horodecki et al. (2000b).

Before I introduce different entanglement measures I would like to
discuss the following question: ``What do we mean by saying that a
state $\sigma$ can be converted into another state $\rho$ by
LOCC?". Strictly speaking, we mean that there exists an LOCC
operation that, given a sufficiently large number of copies, $n$,
of $\sigma$, will convert them arbitrarily close to $m$ copies of
the state $\rho$, i.e.,
\begin{eqnarray}
(\forall \epsilon > 0)(\forall m \in N) (\exists n \in N; \exists
\Phi \in \mbox{LOCC})  \nonumber \\
||\Phi (\sigma^{\otimes n}) - \rho^{\otimes m} || < \epsilon
\label{loccconversion}
\end{eqnarray}
where $||\sigma - \rho||$ is some measure of distance (metric) on
the set of density matrices. Now, if $\sigma$ is more entangled
than $\rho$, we expect that there is an LOCC such that $m>n$;
otherwise, we expect that we can have $n\leq m$. Measuring
entanglement now reduces to finding an appropriate function on the
set of states to order them according to their local
convertibility. This is usually achieved by letting either
$\sigma$ or $\rho$ be a maximally entangled state.

I now introduce three different measures of entanglement, all of
which obey E1--E4. First I discuss the entanglement of formation
(Bennett et al., 1996b). Bennett et al. (1996b) define the
entanglement of formation of a state ${\rho}$ by

\noindent
{\bf Entanglement of formation.}
\begin{eqnarray}
E_F(\rho):= \mbox{min} \sum_i p_i S(\rho_A^i)  \label{ef}
\end{eqnarray}
where $S(\rho_A)= -\mbox{Tr} \rho_A \ln \rho_A$ is the von Neumann entropy
and the minimum is taken over all the possible realizations of the state, $%
\rho_{AB} = \sum_j p_j |{\psi_j}\rangle\langle{\psi_j}|$ with $\rho_A^i = %
\mbox{Tr}_B (|{\psi_i}\rangle\langle{\psi_i}|)$. This measure
satisfies E1-E4. The basis of formation is that Alice and Bob
would like to create an ensemble of $n$ copies of the
non-maximally entangled state, $\rho_{AB}$, using only local
operations, classical communication, and a number, $m$, of
maximally entangled pairs (see Fig. 6). Entanglement of formation
is the asymptotic conversion ratio, $\frac{m}{n}$, in the limit of
infinitely many copies. The form of this measure given in eq.
(\ref{ef}) will be more transparent after the next subsection.
Furthermore, I will analyse the relationship between the
entanglement of formation and other measures proposed in more
detail later. It is worth mentioning that a closed form for this
measure exists for two qubits (Wootters, 1998).

Related to this measure is the entanglement of distillation, also
introduced by Bennett et al. (1996b).

\noindent {\bf Entanglement of distillation.} This measure defines
the amount of entanglement of a state $\sigma$ as the asymptotic
proportion of singlets that can be distilled using a purification
procedure (for a rigorous definition see Rains (1999)). This is
the opposite process to that leading to the entanglement of
formation (Fig. 6), although its value is generally smaller. This
implies that formation of states is in some sense irreversible.
The reason for this irreversibility will be explained in the next
sub-section. This measure fails to satisfy the converse of E1.
Namely, for all disentangled state the entanglement of
distillation is zero, but the converse is not true. There exist
states which are entangled, but no entanglement can be distilled
from them and, for this reason, they are called {\em bound
entangled} (Horodecki et al., 1998a)(see also DiVicenzo (2000)).
This is the reason why the condition E1 is not stated as both the
necessary and sufficient condition.

%% formation distillation Fig6

%\vspace*{-2cm}
\begin{figure}[ht]
\begin{center}
%\vspace*{-6cm}
\hspace{2cm}
\epsfxsize=6.5cm
\epsfbox{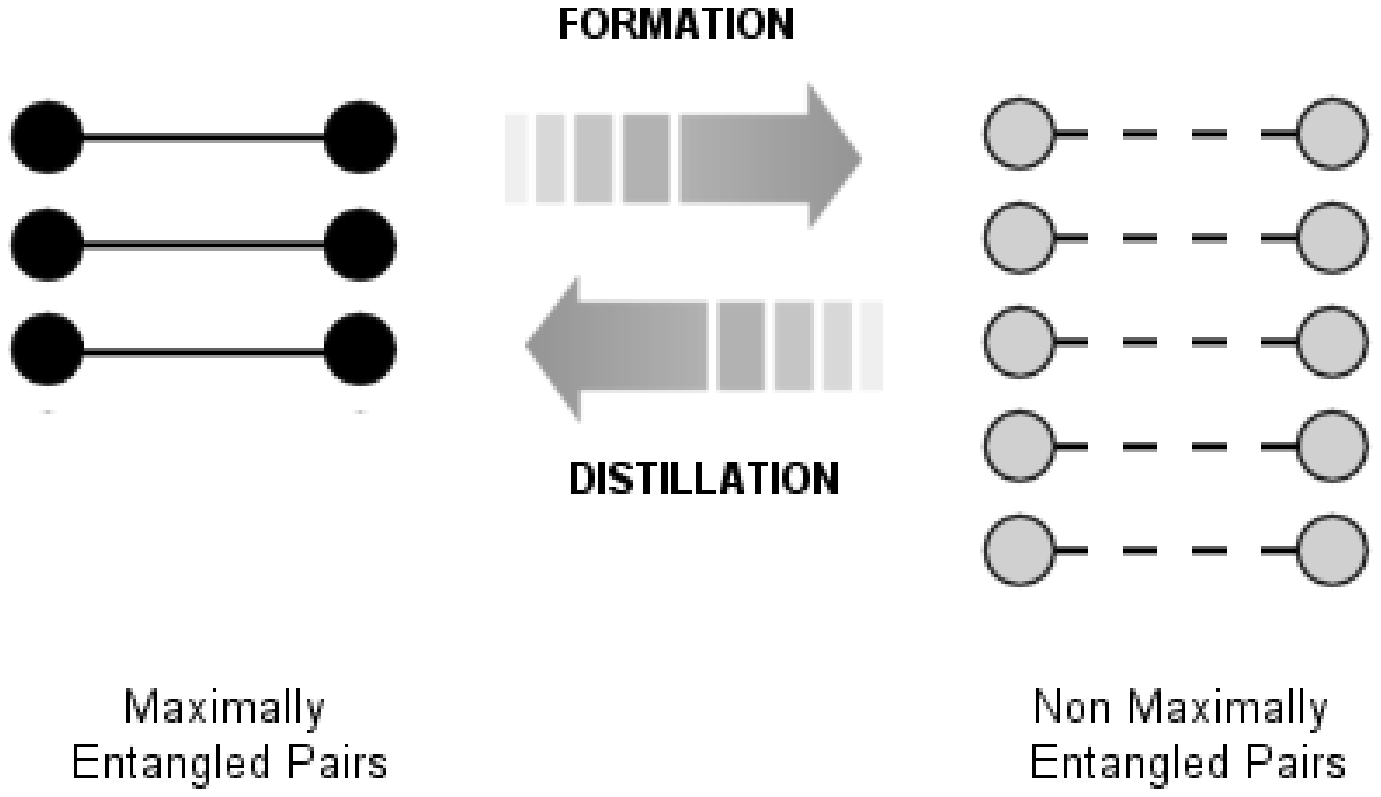}\\[1.5cm]
\begin{caption}
{\narrowtext This figure illustrate formation of entangled states:
a certain number of maximally entangled pairs is manipulated by
LOCC and converted into pairs in some state $\rho$. The asymptotic
conversion ration is known as the entanglement of formation. The
converse of formation is distillation of entanglement. Again, the
asymptotic rate of converting pairs in state $\rho$ into maximally
entangled states is known as the entanglement of distillations.
The two measures of entanglement are in general different,
distillation being greater than or equal to formation. This
surprising irreversibility of entanglement conversion is explained
in the text as a consequence of loss of classical information
about the decomposition of $\rho$.}
\end{caption}
\end{center}
\end{figure}

I now introduce the final measure of entanglement which was first
proposed in Vedral et al. (1997a). This measure is intimately
related to the entanglement of distillation by providing an upper
bound for it. If ${\cal D} $ is the set of all disentangled
states, the measure of entanglement for a state $\sigma$ is then
defined as

\vspace*{.3cm}
\noindent
%\fbox{\parbox[b]{17.7cm}{
{\bf Relative entropy of entanglement.}
\begin{eqnarray}
E({\sigma}):= \min_{\rho \in {\cal D}}\,\,\, S(\sigma || \rho)  \label{6a}
\end{eqnarray}
%}}
\vspace*{.2cm}

\noindent where $S(\sigma || \rho)$ is the quantum relative
entropy. This measure, which I will call the relative entropy of
entanglement, tells us that the amount of entanglement in $\sigma$
is its distance from the disentangled set of states. In
statistical terms introduced in Section II, the more entangled a
state is the more it is distinguishable from a disentangled state
(Vedral et al., 1997b). To understand better all three measures of
entanglement we need to introduce another quantum protocol that
relies fundamentally on entanglement.

Another condition which might be considered intuitive for a
measure of entanglement is {\em convexity}. Namely, we might
require that
\begin{eqnarray}
E(\sum_i p_i \sigma^i)\leq \sum_i p_i E(\sigma^i) \nonumber
\end{eqnarray}
This states that mixing cannot increase entanglement. For example,
an equal mixture of two maximally entangled states $|00\rangle +
|11\rangle$ and $|00\rangle - |11\rangle$ is a separable state and
consequently contains no entanglement. I did not include convexity
as a separate requirement for an entanglement measure as it is not
completely independent from E3. This is because E3 and the strong
additivity ($E(\rho\otimes \sigma) = E(\rho) + E(\sigma)$) imply
convexity,
\begin{eqnarray}
n\sum_i p_i E(\rho_i) & = & E(\rho_1^{\otimes p_1
n}\rho_2^{\otimes
p_2 n}...\rho_N^{\otimes p_N n}) \nonumber \\
& \geq & E((\sum_i p_i \rho_i)^{\otimes n}) = n E((\sum_i p_i
\rho_i), \nonumber
\end{eqnarray}
where the equalities follow from the strong additivity assumption
and the inequality is a consequence of E3. The symbol
$\rho^{\otimes m}$ means that we have $m$ copies of the state
$\rho$. Nevertheless, it is interesting to point out that any
convex measure that satisfies continuity and weak additivity has
to be bounded from below by the entanglement of distillation and
from above by the entanglement of formation (Horodecki et al.,
2000b). We will see that "most" entanglement measures can in fact
be generated using the quantum relative entropy.

It is interesting to note that the relative entropy of
entanglement does in fact satisfy both convexity and continuity
(Donald and Horodecki, 1999) although not additivity (Vollbrecht
and Werner 2000). Furthermore, we can easily show that it is an
upper bound to the entanglement of distillation.  We have that for
any pure state $|\psi\rangle$, $\min_{\omega \in {\cal D}}S(\psi
^{\otimes n}||\omega )=min_{\omega \in {\cal D}}-\langle \psi
^{\otimes n}|\log \omega |\psi ^{\otimes n}\rangle $. But, the
logarithmic function is concave so that
\[
\min_{\omega \in {\cal D}} -\langle \psi ^{\otimes n}|\log \omega
|\psi ^{\otimes n}\rangle \geq \min_{\omega \in {\cal D}} -\log
\langle \psi ^{\otimes n}|\omega |\psi ^{\otimes n}\rangle
\]

\noindent However, according to the recent result of the
Horodeckis (Horodecki et al, 1996), since $\omega $ is a
disentangled state, then its fidelity with the maximally entangled
state cannot be larger than the inverse of the half dimension of
that state, so that $\langle \psi ^{\otimes n}|\omega |\psi
^{\otimes n}\rangle \leq 1/2^{n}.$ Thus,
\begin{eqnarray}
\min_{\omega \in {\cal D}}S(\psi ^{\otimes n}||\omega )\geq -\log
(1/2^{n})=n \label{maxrelent}
\end{eqnarray}

\noindent But we know that this minimum is achievable by the state
$\omega =\rho ^{\otimes n},$ where $\rho $ is obtained from $\psi
$ by removing the off-diagonal elements in the Schmidt basis.
Consequently, if we are starting with $n$ copies of state
$\sigma$, and obtaining $m$ copies of $\psi $ by LOCC, then
\[
D=\frac{m}{n}=\frac{1}{n}\min_{\omega \in {\cal D}}S(\psi
^{\otimes m}||\omega )\leq \frac{1}{n}\min_{\omega \in {\cal
D}}S(\sigma ^{\otimes n}||\omega )
\]

\noindent where the equality follows from eq.(\ref{maxrelent}) and
the inequality from the fact that the relative entropy is
non-increasing under LOCC (strictly speaking,
$D=\lim_{n\rightarrow \infty }\frac{m}{n}$ and, of course, $m$ is
a function of $n$, $m=m(n)$). Thus, the distillable entanglement
is bounded from above by the relative entropy of entanglement.

A similar argument can be given to show that the relative entropy
of entanglement is bounded from the above by the entanglement of
formation (Vedral and Plenio, 1998). Since most of the measures
of entanglement can be derived from the relative entropy they will
possess this similar property. In order to see this, we first
need to introduce quantum teleportation.

\subsection{Teleportation}

Let us begin by describing quantum teleportation in the form
originally proposed by Bennett et al. (1993). Suppose that Alice
and Bob, who are distant from each other, wish to implement a
teleportation procedure. Initially they need to share a maximally
entangled pair of qubits. This means that if Alice and Bob both
have one qubit each, then the joint state may for example be:
\begin{eqnarray}
|\Psi_{AB}\rangle = (|0_A\rangle |0_B\rangle + |1_A\rangle |1_B\rangle)/\sqrt{2}\; ,
\end{eqnarray}
where the first ket (with subscript A) belongs to Alice and second
(with subscript B) to Bob.
Note that this state is maximally entangled and
is different from a statistical mixture
$(|00\rangle\langle 00| + |11\rangle\langle 11|)/2$ which is the
most correlated state allowed by classical physics.

Now suppose that Alice receives a qubit in an unknown state
$|\Phi\rangle = a|0\rangle + b|1\rangle$ and
she wants to teleport it to Bob. The state has to be
unknown to her because otherwise
she can just phone Bob up and tell him all the details of the
state, and he can then recreate it on a particle that he
possesses. Given that Alice does not know the state, she cannot
measure it to obtain all the necessary information to specify it.
If she could, this would lead to a violation of uncertainty
principle. Therefore she has to resort to using the state $|\Psi_{AB}\rangle$ that
she shares with Bob to transfer her state to him without actually learning this
state. This procedure is what we mean by quantum teleportation.

I first write out the total state of all three qubits
\begin{eqnarray}
|\Phi_{AB}\rangle := |\Phi\rangle |\Psi_{AB}\rangle = (a|0\rangle + b|1\rangle)(|00\rangle +
|11\rangle)/\sqrt{2} \;\; . \nonumber
\end{eqnarray}
However, the above state can be conveniently written in a different basis
\begin{eqnarray}
|\Phi_{AB}\rangle & = & (a|000\rangle + a|011\rangle + b|100\rangle + b|111\rangle)/\sqrt{2}
\nonumber \\
& = & \frac{1}{2}[ |\Phi^+\rangle (a|0\rangle + b|1\rangle) + |\Phi^-\rangle
(a|0\rangle - b|1\rangle) \nonumber\\
& + & |\Psi^+\rangle (a|1\rangle + b|0\rangle) +
|\Psi^-\rangle (a|1\rangle - b|0\rangle) ] \; , \nonumber
\end{eqnarray}
where
\begin{eqnarray}
|\Phi^+\rangle &=& (|00\rangle + |11\rangle)/\sqrt{2} \label{Bell1}\\
|\Phi^-\rangle &=& (|00\rangle - |11\rangle)/\sqrt{2} \\
|\Psi^+\rangle &=& (|01\rangle + |10\rangle)/\sqrt{2} \\
|\Psi^-\rangle &=& (|01\rangle - |10\rangle)/\sqrt{2}  \;
\label{Bell}
\end{eqnarray}
form an ortho-normal basis of Alice's two qubits (remember that
the first two qubits belong to Alice and the last qubit belongs
to Bob). The above basis is frequently called the Bell basis.
This is a very useful way of writing the state of Alice's two
qubits and Bob's single qubit because it displays a high degree of
correlations between Alice's and Bob's parts: for every state of
Alice's two qubits (i.e. $|\Phi^+\rangle, |\Phi^-\rangle,
|\Psi^+\rangle, |\Psi^-\rangle$) there is a corresponding state of
Bob's qubit. In addition the state of Bob's qubit in all four
cases "looks very much like" the original qubit that Alice has to
teleport to Bob.  It is now straightforward to see how to proceed
with the teleportation protocol (Bennett et al., 1993):
\begin{enumerate}
\item Upon receiving the unknown qubit in state $|\Phi\rangle$ Alice performs
projective measurements on her two qubits in the Bell
basis. This means that she will obtain one of the
four Bell states randomly, and with equal probability.

\item Suppose Alice obtains the state $|\Psi^+\rangle$. Then
the state of all three qubits (Alice $+$ Bob) collapses
to the following state
\begin{eqnarray}
|\Psi^+\rangle (a|1\rangle + b|0\rangle)\; . \nonumber
\end{eqnarray}
(the last qubit belongs to Bob as usual).
Alice now has to communicate the result of her measurement to Bob
(over the phone, for example). The point of this communication is
to inform Bob how the state of his qubit now differs from the state
of the qubit Alice was holding before the Bell measurement.

\item Now Bob has to apply a unitary
transformation on his qubit which simulates a logical
NOT operation: $|0\rangle \rightarrow |1\rangle$ and
$|1\rangle \rightarrow |0\rangle$. He thereby transforms
the state of his qubit into the state $a|0\rangle + b|1\rangle$,
which is precisely the state that Alice had to teleport to him
initially. This completes the protocol. It is easy to see
that if Alice obtained some other Bell state, then Bob
would have to apply some other simple operation to complete
the teleportation. They can be represented by the Pauli spin matrices.

\end{enumerate}
An important fact to observe in the above protocol is that all
the operations (Alice's measurements and Bob's unitary
transformations) are {\em local} in nature. This means that there
is never any need to perform a (global) transformation or
measurement on all three qubits simultaneously, which is what
allows us to call the above protocol a genuine teleportation. It
is also important that the operations that Bob performs are
independent of the state that Alice tries to teleport to him.
Note also that the classical communication from Alice to Bob in
step 2 above is crucial because otherwise the protocol would be
impossible to execute (there is a deeper reason for this: if we
could perform teleportation without classical communication then
Alice could send messages to Bob faster than the speed of light,
see e.g. Vedral et al. (1997c)).

It is important to observe that the fact that the initial state
to be teleported is destroyed immediately after Alice's
measurement, i.e it becomes maximally mixed of the form
$(|0\rangle\langle 0| + |1\rangle\langle 1|)/2$. This has to
happen since otherwise Alice and Bob would end up with two qubits
in the same state. So, effectively, they would clone an unknown
quantum state, which is impossible by the laws of quantum
mechanics. This is the no-cloning theorem of Wootters and Zurek
(1982), which is a simple consequence of {\em linearity} of
quantum dynamical laws. We also see that at the end of the
protocol the quantum entanglement of $|\Psi_{AB}\rangle$ is
completely destroyed. Does this have to be the case in general or
might we save that state at the end (by perhaps performing a
different teleportation protocol)? The answer is yes (Plenio and
Vedral, 1998), and the reason is that if this was not the case,
then entanglement could increase under LOCC, which as we have
seen is prohibited by definition.

Teleportation has been experimentally performed in three
different set-ups (Bouwmeester et al., 1997). It will now be used
to link the three measures of entanglement. I will show that all
different measures of entanglement can be understood as special
cases of the relative entropy of entanglement (Henderson and
Vedral, 2000). This unification relies on adding an ancilla,
which I will call a memory system and which will help us keep
track of the various decompositions of a given bi-partite density
matrix. How much access is available to this memory determines
which measure of entanglement is used.

\subsection{Measures of entanglement from relative entropy}

Suppose that Alice and Bob share a state described by the density matrix
$\rho_{AB}$. The state $\rho_{AB}$ has an infinite number of different
decompositions $\varepsilon=\{\left|  \psi_{AB}^{i}\right\rangle \langle
\psi_{AB}^{i}|,p_{i}\}$, into pure states $\left|  \psi_{AB}^{i}\right\rangle
$, with probabilities $p_{i}$. We denote the mixed state
$\rho_{AB}$ written in decomposition $\varepsilon$ by
\begin{eqnarray}
\rho_{AB}^{\varepsilon}=\sum_{i}p_{i}\left|  \psi_{AB}^{i}\right\rangle
\langle\psi_{AB}^{i}| \label{eq:decomp}
\end{eqnarray}
As we have seen measures of entanglement are associated with
formation and distillation of pure and mixed entangled states.
The known relationships between the different measures of
entanglement for mixed states are $E_{D}(\rho _{AB})\leq
E_{RE}(\rho_{AB})\leq E_{F}(\rho_{AB})$, (Vedral and Plenio,
1998). Equality holds for pure states, where all the measures
reduce to the Von Neumann entropy, $S(\rho_{A})=S(\rho_{B})$.

%%%%%% Fig. 7 Formation %%%%%%%%%%%%%%%%%

\begin{figure}[ht]
\begin{center}
\hspace{0mm} \epsfxsize=7.5cm
\epsfbox{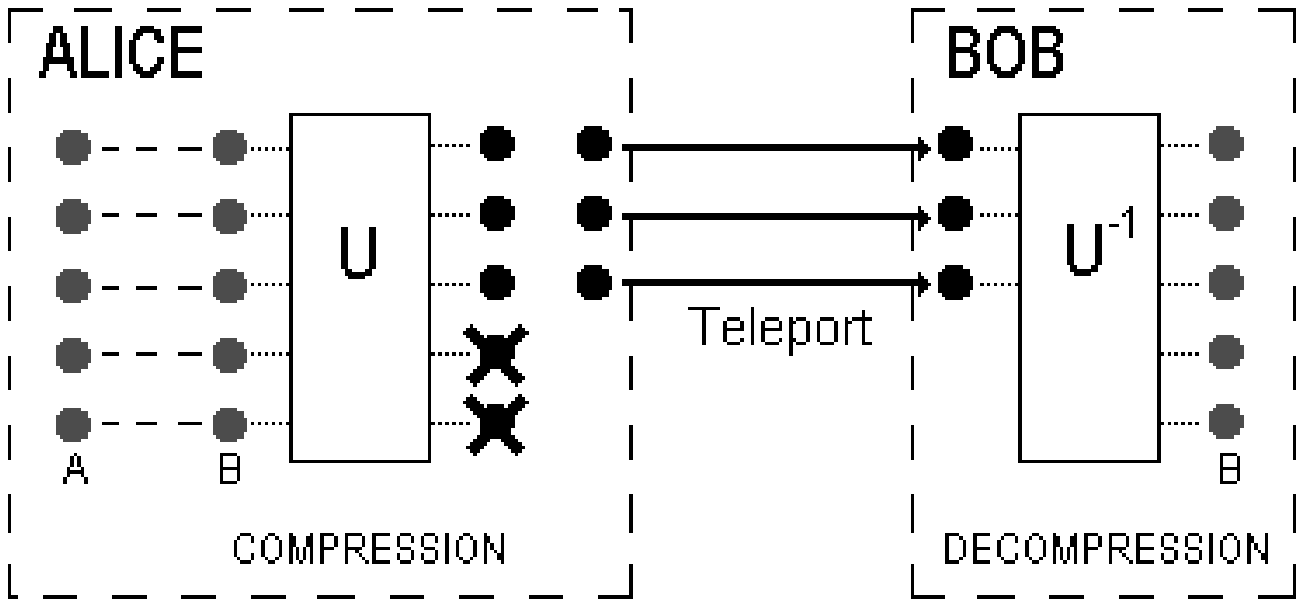}\\[0.2cm]
\begin{caption}
{\narrowtext This figure illustrates formation of a state by LOCC
and with the help of teleportation. First, Alice creates the joint
state of subsystems $A$ and $B$ locally. Then, she performs the
quantum data compression on the subsystem $B$ and teleports the
compressed state to Bob. Finally, Bob decompresses the received
state and hence Alice and Bob end up sharing the joint state of
$A$ and $B$ initially prepared by Alice.}
\end{caption}
\end{center}
\end{figure}

Formation of an ensemble of $n$ non-maximally entangled
\textit{pure} states, $\rho_{AB}=\left|  \psi_{AB}\right\rangle
\langle\psi_{AB}|$ is achieved by the following protocol. Alice
first prepares the states she would like to share with Bob
locally. She then uses Schumacher compression, (Schumacher, 1995),
to compress subsystem $B$ into $nS(\rho_{B})$ states. Subsystem
$B$ is then teleported to Bob using $nS(\rho_{B})$ maximally
entangled pairs. Bob decompresses the states he receives and so
ends up sharing $n$ copies of $\rho_{AB}$ with Alice. The
entanglement of formation is therefore
$E_{F}(\rho_{AB})=S(\rho_{B})$. For pure states, this process
requires no classical communication in the asymptotic limit (Lo
and Popescu, 1999). The reverse process of distillation is
accomplished using the Schmidt projection method (Bennett et al,
1996a), which allows $nS(\rho_{B})$ maximally entangled pairs to
be distilled in the limit as $n$ becomes very large. No classical
communication between the separated parties is required.
Therefore pure states are fully inter-convertible in the
asymptotic limit.

The situation for mixed states is more complex. When any mixed state, denoted
by Eq.(\ref{eq:decomp}), is created, it may be imagined to be part of an
extended system whose state is pure. The pure states $|\psi_{AB}^{i}\rangle$
in the mixture may be regarded as correlated to orthogonal states
$|m_{i}\rangle$ of a memory $M$. The extended system is in the pure state
$\left|  \psi_{MAB}\right\rangle =\sum_{i}\sqrt{p_{i}}|m_{i}\rangle|\psi
_{AB}^{i}\rangle$. If we have no access to the memory system, we trace over it
to obtain the mixed state in Eq.(\ref{eq:decomp}). In fact, the lack of access
to the memory is of a completely general nature. It may be due to interaction
with another inaccessible system, or it may be due to an intrinsic loss of
information. The results I will present are universally valid and do not depend on the nature
of the information loss. We will see that the amount of entanglement involved
in the different entanglement manipulations of mixed states depends on the
accessibility of the information in the memory at different stages. Note that
a unitary operation on $\left|  \psi_{MAB}\right\rangle $ will convert it into
another pure state $\left|  \phi_{MAB}\right\rangle $ with the same
entanglement, and tracing over the memory yields a different decomposition of
the mixed state. Reduction of the pure state to the mixed state may be
regarded as due to a projection-valued measurement on the memory with
operators $\{E_{i}=\left|  m_{i}\right\rangle \langle m_{i}|\}$.

Consider first the protocol of formation by means of which Alice
and Bob come to share an ensemble of $n$ mixed states $\rho_{AB}$
as in Fig. 7. Alice first creates the mixed states locally by
preparing a collection of $n$ states in a particular
decomposition, $\varepsilon=\{\left|  \psi_{AB}^{i}\right\rangle
\langle \psi_{AB}^{i}|,p_{i}\}$ by making $np_{i}$ copies of each
pure state $|\psi_{AB}^{i}\rangle$. At the same time we may
imagine a memory system entangled to the pure states to be
generated, which keeps track of the identity of each member of the
ensemble. I consider first the case where the state of subsystems
$A$ and $B$ together with the memory is pure. Later, I will
consider the situation in which Alice's memory is decohered. There
are then three ways for her to share these states with Bob. First
of all, she may simply compress subsystem $B$ to $nS(\rho_{B})$
states, and teleport these to Bob using $nS(\rho_{B})$ maximally
entangled pairs. The choice of which subsystem to teleport is made
so as to minimise the amount of entanglement required, so that
$S(\rho_{B})\leq S(\rho_{A})$. The teleportation in this case
would require no classical communication in the asymptotic limit,
just as for pure states (Lo and Popescu, 1999). The state of the
whole system which is created by this process is an ensemble of
pure states $\left| \psi_{MAB}\right\rangle $, where subsystems
$M$ and $A$ are on Alice's side and subsystem $B$ is on Bob's
side. In terms of entanglement resources, however, this process is
not the most efficient way for Alice to send the states to Bob.
She may do it more efficiently by using the memory system of
$\left| \psi_{MAB}\right\rangle $ to identify blocks of
$np_{i\text{ }}$members in each pure state $\left|
\psi_{AB}^{i}\right\rangle $, and applying compression to each
block to give $np_{i}S(\rho_{B}^{i})$ states. Then the total
number of maximally entangled pairs required to teleport these
states to Bob is $n\sum_{i}p_{i}S(\rho _{B}^{i})$, which is
clearly less than $nS(\rho_{B})$, by concavity of the entropy. The
amount of entanglement required clearly depends on the
decomposition of the mixed state $\rho_{AB}$. In order to
decompress these states, Bob must also be able to identify which
members of the ensemble are in which state. Therefore Alice must
also send him the memory system. She now has two options. She may
either teleport the memory to Bob, which would use more
entanglement resources. Or she may communicate the information in
the memory classically, with no further use of entanglement. When
Alice uses the minimum entanglement decomposition,
$\varepsilon=\{\left| \psi_{AB}^{i}\right\rangle
\langle\psi_{AB}^{i}|,p_{i}\}$, this process, originally
introduced by Bennett \textit{et al.}, (1996b), makes the most
efficient use of entanglement, consuming only the entanglement of
formation of the mixed state,
$E_{F}(\rho_{AB})=\sum_{i}p_{i}S(\rho_{B}^{i})$. We may think of
the classical communication between Alice and Bob in one of two
equivalent ways. Alice may either measure the memory locally to
decohere it, and then send the result to Bob classically, or she
may send the memory through a completely decohering quantum
channel. Since Alice and Bob have no access to the channel, the
state of the whole system which is created by this process is the
mixed state
\begin{eqnarray}
\rho_{ABM}^{\varepsilon}=\sum_{i}p_{i}|\psi_{AB}^{i}\rangle\langle\psi
_{AB}^{i}|\otimes|m_{i}\rangle\langle m_{i}|\label{eq:mixed}% \nonumber
\end{eqnarray}
where Bob is classically correlated to the $AB$ subsystem. Bob is then able to
decompress his states using the memory to identify members of the ensemble.

Once the collection of $n$ pairs is shared between Alice and Bob,
it is converted into an ensemble of $n$ mixed states $\rho_{AB}$
by destroying access to the memory which contains the information
about the state of any particular member of the ensemble. {\em It
is the loss of this information which is responsible for the fact
that entanglement of distillation is lower than entanglement of
formation, since it is not available to parties carrying out the
distillation.} If Alice and Bob, who do have access to the
memory, were to carry out the distillation, they could obtain as
much entanglement from the ensemble as was required to form it.
In the case where Alice and Bob share an ensemble of\ the pure
state $\left|  \psi _{MAB}\right\rangle $, they would simply
apply the Schmidt projection method, (Bennett et al., 1996a). The
relative entropy of entanglement gives the upper bound to
distillable entanglement, $E_{RE}(\left|
\psi_{(MA):B}\right\rangle \langle\psi_{(MA):B}|)=S(\rho_{B})$,
which is the same as the amount of entanglement required to
create the ensemble of pure states, as described above. Here $MA$
and $B$ are spatially separated subsystems on which joint
operations may not be performed. In my notation, I use a colon to
separate the local subsystems.

On the other hand, if Alice used the least entanglement for producing an
ensemble of the mixed state $\rho_{AB}$, together with classical
communication, the state of the whole system is an ensemble of the mixed state
$\rho_{ABM}^{\varepsilon}$, and the process is still reversible. Because of
the classical correlation to the states $\left|  \psi_{AB}^{i}\right\rangle $,
Alice and Bob may identify blocks of members in each pure state $\left|
\psi_{AB}^{i}\right\rangle $, and apply the Schmidt projection method to them,
giving $np_{i}S(\rho_{B}^{i})$ maximally entangled pairs, and hence a total
entanglement of distillation of $\sum_{i}p_{i}S(\rho_{B}^{i})$. The relative
entropy of entanglement again quantifies the amount of distillable
entanglement from the state $\rho_{ABM}^{\varepsilon}$ and is given by
$E_{RE}(\rho_{A:(BM)}^{\varepsilon})=\min_{\sigma_{ABM}\in D}S(\rho
_{ABM}^{\varepsilon}||\sigma_{ABM})$. The disentangled state which minimises
the relative entropy is \linebreak $\sigma_{ABM}=\sum_{i}p_{i}\sigma_{AB}%
^{i}\otimes|m_{i}\rangle\langle m_{i}|$, \noindent where $\sigma_{AB}^{i}$ is
obtained from $|\psi_{AB}^{i}\rangle\langle\psi_{AB}^{i}|$ by deleting the
off-diagonal elements in the Schmidt basis. This is the minimum because the
state $\rho_{MAB}$ is a mixture of the orthogonal states $\left|
m_{i}\right\rangle |\psi_{AB}^{i}\rangle$, and for a pure state $|\psi
_{AB}^{i}\rangle$, the disentangled state that minimises the relative entropy
is $\sigma_{AB}^{i}$. The minimum relative entropy of the extended system is
then
\begin{eqnarray}
S(\rho_{ABM}^{\varepsilon}||\sigma_{ABM})=\sum_{i}p_{i}S(\rho_{B}%
^{i})\nonumber
\end{eqnarray}
This relative entropy, $E_{RE}(\rho_{A:(BM)}^{\varepsilon})$, has
previously been called the `entanglement of projection' (Garisto
and Hardy, 1999), because the measurement on the memory projects
the pure state of the full system into a particular
decomposition. The minimum of
$E_{RE}(\rho_{A:(BM)}^{\varepsilon})$ over all decompositions is
equal to the entanglement of formation of $\rho_{AB}$. However,
Alice and Bob may choose to create the state $\rho_{AB}$ by using
a decomposition with higher entanglement than the entanglement of
formation. The maximum of $E_{RE}(\rho_{A:(BM)}^{\varepsilon})$
over all possible decompositions is called the `entanglement of
assistance' of $\rho_{AB}$ (DiVicenzo et al., 1998). Because
$E_{RE}(\rho_{A:(BM)}^{\varepsilon })$ is a relative entropy, it
is invariant under local operations and non-increasing under
general operations, properties which are conditions for a
good measure of entanglement (Vedral and Plenio, 1998). However, unlike $E_{RE}%
(\rho_{AB})$ and $E_{F}(\rho_{AB})$, it is not zero for completely
disentangled states. In this sense, the relative entropy of entanglement,
$E_{RE}(\rho_{A:(BM)}^{\varepsilon})$, defines a class of entanglement
measures interpolating between the entanglement of formation and entanglement
of assistance. Note that an upper bound for the entanglement of assistance,
$E_{A}$, can be shown using concavity (DiVicenzo et al., 1998), to be $E_{A}%
(\rho_{AB})\leq\min[S(\rho_{A}),S(\rho_{B})]$. This bound can also be shown
from the fact that the distillable entanglement from any decomposition,
$E_{RE}(\rho_{A:(BM)}^{\varepsilon})\leq E_{A}(\rho_{AB})$ cannot be greater
than the entanglement of the original pure state.

Note that here we are really creating a state $\rho^{\otimes
n}=\rho\otimes\rho \ldots \rho$. The entanglement of formation of
such a state is, strictly speaking, given by $E_F(\rho^{\otimes
n})$; so, the entanglement of formation per one single pair is
$E_F(\rho^{\otimes n})/n$. It is at present not clear if this is
the same as $E_F(\rho)$ in general, i.e. whether the entanglement
of formation is additive. Bearing this in mind we continue our
discussion whose conclusions will not depend on the validity of
the additivity assumption of the entanglement of formation (for
more on this issues see for example Hayden et al, 2000).

We may also derive relative entropy measures that interpolate
between the relative entropy of entanglement and the entanglement
of formation (Horodecki et al., 2000b) by considering
non-orthogonal measurements on the memory. First of all, the fact
that the entanglement of formation is in general greater than the
upper bound for entanglement of distillation, emerges as a
property of the relative entropy, namely that it cannot increase
under the local operation of tracing one subsystem (this is
property F2 of the quantum relative entropy given in Section II)
(Lindblad 1974),
\begin{eqnarray}
E_{F}(\rho_{AB})& = & \min_{\sigma_{ABM}\in {\cal D}}S(\rho_{ABM}||\sigma_{ABM})\nonumber\\
& \geq & \min_{\sigma_{AB}\in {\cal D}} S(\rho_{AB}||\sigma_{AB})
\end{eqnarray}
In general, the loss of the information in the memory may be regarded as a
result of an imperfect classical channel. This is equivalent to Alice making a
non-orthogonal measurement on the memory, and sending the result to Bob. In
the most general case, $\{E_{i}=A_{i}A_{i}^{+}\}$ is a POVM (positive operator
valued measure; loosely speaking, this is a CP map as in eq. (\ref{CP})
where all the individual outcomes are recorded) performed on the
memory. The decomposition corresponding to this measurement is composed of
mixed states, $\xi=\{q_{i},Tr_{M}(A_{i}\rho_{MAB}A_{i}^{+})\}$, where
$q_{i}=Tr(A_{i}\rho_{MAB}A_{i}^{+})$. The relative entropy of entanglement of
the state $\rho_{MAB}^{\xi}$, when $\xi$ is a decomposition of $\rho_{AB}$
resulting from a non-orthogonal measurement on $M$, defines a class of
entanglement measures interpolating between the relative entropy of
entanglement and the entanglement of formation of the state $\rho_{AB}$. In
the extreme case where the measurement gives no information about the state
$\rho_{AB}$, $E_{RE}(\rho_{A:(BM)}^{\varepsilon})$ becomes the relative
entropy of entanglement of the state $\rho_{AB}$ itself. In between, the
measurement gives partial information. So far, I have shown that the measures
interpolating between entanglement of assistance and entanglement of formation
result from making orthogonal measurements on preparations of the pure state
$\left|  \psi_{MAB}\right\rangle $ in different bases. I note that they may
equally be achieved by using the preparation associated with entanglement of
assistance, and making increasingly non-orthogonal measurements.

\subsection{Classical Information and Quantum correlations}

The loss of entanglement may be related to the loss of information in the
memory. There are two stages at which distillable
entanglement is lost. The first is in the conversion of the pure state
$\left|  \psi_{MAB}\right\rangle $ into a mixed state $\rho_{ABM}$. This
happens because Alice uses a \textit{classical} channel to communicate the
memory to Bob. The second is due to the loss of the memory, $M$, taking the
state $\rho_{ABM}$ to $\rho_{AB}$. The amount of information lost may be
quantified by the difference in mutual information between the respective
states. Mutual information is a measure of correlations between the memory $M$
and the system $AB$, giving the amount of information about $AB$ which may be
obtained from a measurement on $M$. The quantum mutual information between $M$
and $AB$ is defined as $I_{Q}(\rho_{M:(AB)})=S(\rho_{M})+S(\rho_{AB}%
)-S(\rho_{MAB})$. The mutual information loss in going from the pure state
$\left|  \psi_{MAB}\right\rangle $ to the mixed state in Eq. (\ref{eq:mixed})
is $\Delta I_{Q}=S(\rho_{AB})$. There is a corresponding reduction in the
relative entropy of entanglement, from the entanglement of the original pure
state, $E_{RE}(\left|  \psi_{(MA):B}\right\rangle \langle\psi_{(MA):B}|)$, to
the entanglement of the mixed state $E_{RE}(\rho_{A:(BM)}^{\varepsilon})$ for
all decompositions $\varepsilon$ arising as the result of an orthogonal
measurement on the memory. It is possible to prove, using the non-increase of
relative entropy under local operations, that when the
mutual information loss is added to the relative entropy of entanglement of
the mixed state $E_{RE}(\rho_{A:(BM)}^{\varepsilon})$, the result is greater
than the relative entropy of entanglement of the original pure state,
$E_{RE}(\left|  \psi_{(MA):B}\right\rangle \langle\psi_{(MA):B}|)$,
(Henderson and Vedral, 2000). \ The strongest case, which occurs when $E_{RE}%
(\rho_{A:(BM)}^{\varepsilon})=E_{F}(\rho_{AB})$, is:
\begin{eqnarray}
E_{RE}(\left|  \psi_{(MA):B}\right\rangle \langle\psi_{(MA):B}|)\leq
E_{F}(\rho_{AB})+S(\rho_{AB}) \label{eq:ptom}%
\end{eqnarray}

A similar result may be proved for the second loss, due to loss of the memory,
(Henderson and Vedral, 2000). Again the mutual information loss is $\Delta I_{Q}%
=S(\rho_{AB})$. The relative entropy of entanglement is reduced from
$E_{RE}(\rho_{A:(BM)}^{\varepsilon})$, for any decomposition $\varepsilon$
resulting from an orthogonal measurement on the memory, to $E_{RE}(\rho_{AB}%
)$, the relative entropy of entanglement of the state $\rho_{AB}$ with no
memory. When the mutual information loss is added to $E_{RE}(\rho_{AB})$, the
result is greater than $E_{RE}(\rho_{A:(BM)}^{\varepsilon})$. In this case,
the result is strongest for $E_{RE}(\rho_{A:(BM)}^{\varepsilon})=E_{A}%
(\rho_{AB})$:
\begin{eqnarray}
E_{A}(\rho_{AB})\leq E_{RE}(\rho_{AB})+S(\rho_{AB}) \label{eq:upper}
\end{eqnarray}
Notice that if $\rho_{AB}$ is a pure state, then
$S(\rho_{AB})=0$, and equality holds. Inequalities
(\ref{eq:ptom}) and (\ref{eq:upper}) provide lower bounds for
$E_{F}(\rho_{AB})$ and $\ E_{RE}(\rho_{AB})$ respectively. They
are of a form typical of irreversible processes in that restoring
the information in $M$ is not sufficient to restore the original
correlations between $M$ and $AB$. In particular, they express
that the loss of entanglement between Alice and Bob at each stage
must be accompanied by an even greater reduction in mutual
information between the memory and subsystems $AB$. The general
result can be derived from Donald's equality (Donald, 1986). We
have in general that for any $\sigma$ and $\rho = \sum_i p_i
\rho_i$ the following is true
\begin{eqnarray}
S(\rho ||\sigma) + \sum_i p_i S(\rho_i||\rho) = \sum_i p_i S(\rho_i||\sigma) \nonumber
\end{eqnarray}
Suppose that $E(\rho)=S(\rho ||\sigma)$. Then, since $E(\rho_i)\leq S(\rho_i||\sigma)$,
we have the following inequality
\begin{eqnarray}
E(\rho)+\sum_i p_i S(\rho_i||\rho)\geq \sum_i p_i E(\rho_i) \nonumber
\end{eqnarray}
Thus, the loss of entanglement in $\{ p_i,\rho_i\}\rightarrow \rho$ is bounded
from above by the Holevo information
\begin{eqnarray}
\sum_i p_i E(\rho_i)-E(\rho) \leq \sum_i p_i S(\rho_i||\rho)
\end{eqnarray}
This is a physically pleasing property of entanglement. {\em It
says that the amount of information lost always exceeds the lost
entanglement, which indicates that entanglement stores only a part
of information - the rest, of course, is stored in classical
correlations.} (see also Eisert et al, 2000, who consider a
similar problem, although not in the full generality of the above
analysis).

In summary, the relative entropy of entanglement of the state $\rho_{AB}$
depends only on the density matrix $\rho_{AB}$, and gives an upper bound to
the entanglement of distillation. The other measures of entanglement, which
are given by relative entropies of an extended system, all depend on how the
information in the memory is used, or how the density matrix is decomposed.
There are numerous decompositions of any bipartite mixed state into a set of
states $\rho_{i}$ with probability $p_{i}$. The average entanglement of states
in each decomposition is given by the relative entropy of entanglement of the
system extended by a memory whose orthogonal states are classically correlated
to the states of the decomposition. This correlation records which state
$\rho_{i}$ any member of an ensemble of mixed states $\rho_{AB}^{\otimes n}$
is in. It is available to parties involved in formation of the mixed state,
but is not accessible to parties carrying out distillation. When the classical
information is fully available, different decompositions give rise to
different amounts of distillable entanglement, the highest being entanglement
of assistance and the lowest, entanglement of formation. If access to the
classical record is reduced, the amount of distillable entanglement is
reduced. In the limit where no information is available, the upper bound to
the distillable entanglement is given by the relative entropy of entanglement
of the state $\rho_{AB}$ itself, without the extension of the classical
memory.

I close this section by discussing generalisations to more than
two subsystems. First of all it is not at all clear how to
perform this in the case of entanglement of formation and
distillation. The former one just does not have a natural
generalisation and, for the later one, it is not clear what
states should we be distilling when we have three or more
parties. The relative entropy of entanglement on the other hand
does not suffer from this problem (Vedral and Plenio, 1998;
Vedral et al., 1997b). Its definition for N parties would be
$E_{RE}(\sigma):= \min_{\rho\in D} S(\sigma||\rho)$ where
$\rho=\sum_i p_i \rho^i_1\otimes \rho^i_2...\otimes \rho^i_N$.

I will now use the knowledge we have gained of classical and quantum correlations to
describe quantum computation. It will be seen, perhaps somewhat surprisingly,
that classical correlations will play a more prominent role than quantum
correlations in the speed-up of certain quantum algorithms.

\section{Quantum computation}

A quantum computer is a physical system that can accept input
states which represent a coherent superposition of many different
possible basis states and subsequently evolve them into a
corresponding superposition of outputs. Computation, {\em i.e.} a
sequence of unitary transformations, affects simultaneously each
element of the superposition, generating a massive parallel data
processing albeit within one piece of quantum hardware.  In this
way quantum computers can efficiently solve some problems that
are believed to be intractable on classical computers (Deutsch
and Josza, 1992) (the best example is Shor's factorisation
algorithm (Shor, 1996)). Therefore the advantage of a quantum
computer lies in the exploitation of the phenomenon of
superposition. The great importance of the quantum theory of
computation is in the fact that it reveals the fundamental
connections between the laws of physics and the nature of
computation (Deutsch, 1998).

In order to understand the efficiency of computer algorithms, we
have to discuss the theory of computational complexity. I will
only mention the basics, but a more detailed account can be found
in e.g. (Papadimitriou, 1995). Computational complexity concerns
the difficulty of solving certain problems, such as for example
multiplication of two numbers, finding the minimum of a given
function and so on. Complexity theory divides problems into two
basic categories:
\begin{enumerate}
\item {\em easy problems}: the time of computation $T$ is a polynomial
function of the size of the input $l$, i.e. $T=c_n l^n + ...+ c_1 l+c_0$,
where the coefficients $c$ are determined by the problem.
\item {\em hard problems}: the time of computation is an exponential
function of the size of the input (e.g. $T = 2^{cl}$, where c is problem
dependent).
\end{enumerate}
The size of the input is always measured in bits (qubits). For
example, if we are to store the number $15$, then we need $4$ bits.
In general, to store a number N we
need about $l=\log N$, where the base of the logarithm is $2$.

The division of problems into `easy' and `hard' is, of course,
very rough. First of all, in computation, apart from time, there
are other resources which might matter, such as space, energy and
so on. If time grows polynomially, but we require an
exponentially increasing energy, then the problem is clearly
difficult. Also, suppose that the time complexity of one problem
is $10^{10} n$ and of another one is $10^{-10} 2^n$. Then for
small $n$ (say $n= 10$), the second algorithm, in spite of being
exponential, is clearly more efficient. These two issues
exemplify that the division into hard and easy problems is not
without its own problems. However, this classification system is
very simple to put into practice and does illuminate many
different aspects of computational problems which is why it is so
widely used. I refer the reader to the book by Garey and Johnson
(1979) which presents an introduction to hard problems and their
detailed classification.

%%%%%%%% Mach Zender Fig8

\begin{figure}[ht]
\begin{center}
\hspace{0mm}
\epsfxsize=7.5cm
\epsfbox{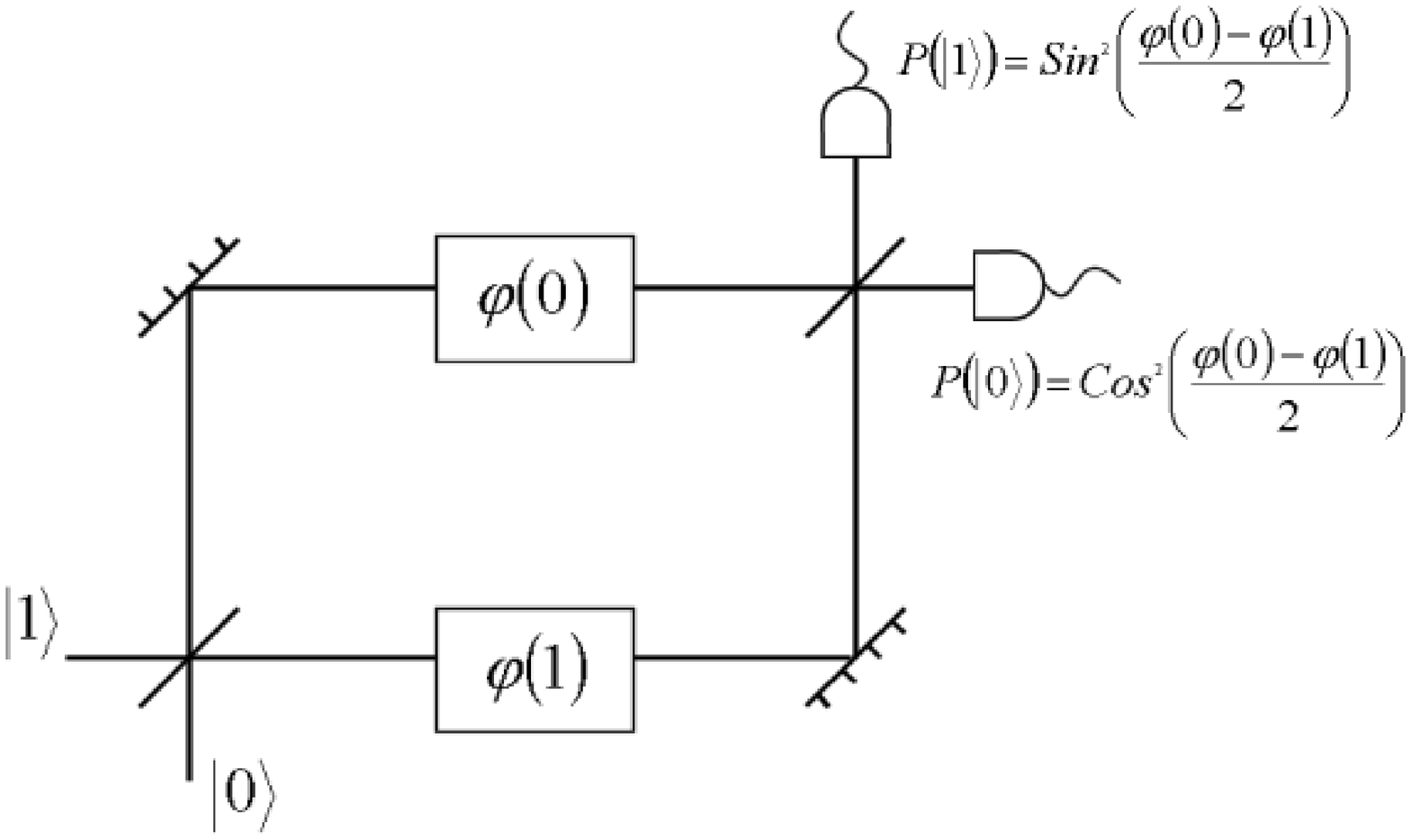}\\[0.2cm]
\begin{caption}
{\narrowtext The Mach-Zender interferometer. A photon is split at
a beam-splitter and can take two different paths. In each of the
paths we have a different phase introduced to the photon
state, so that, after it encounters the second beam-splitter,
the probabilities of detection in two branches have the
sinusoidal dependance on the phase difference. In terms
of quantum computation, the beam-splitter implements the
Hadamard transform and the whole interferometer can be
seen as implementing Deutsch's algorithm (see text for
explanation).}
\end{caption}
\end{center}
\end{figure}

There is a great simplification in understanding quantum
computation: a quantum computer is formally equivalent to a
multiparticle "Mach-Zender like" interferometer (Cleve et al.,
1997). I first present the simplest kind of interferometer in
terms of its function as a simple computer. We see from the Fig. 8
that the path of the photon is in fact a quantum bit in the sense
that the photon can be in a superposition of the two paths. The
first beam splitter acts as the unitary evolution $|0\rangle
\rightarrow |0\rangle+|1\rangle$ which is known as the Hadamard
gate. Next is the phase shift which has the following effect
\begin{eqnarray}
& |0\rangle & \rightarrow e^{i\phi (0)} |0\rangle \nonumber\\
& |1\rangle & \rightarrow e^{i\phi (1)} |1\rangle \nonumber
\end{eqnarray}
At the end we have another beam splitter and two detector measuring contributions
to the state $|0\rangle$ and $|1\rangle$. The corresponding probabilities of
detection are
\begin{eqnarray}
P_0 & = & \cos^2 \frac{\phi (0)-\phi(1)}{2} \nonumber\\
P_1 & = & \sin^2 \frac{\phi (0)-\phi(1)}{2} \nonumber
\end{eqnarray}
If, for example, $\phi (0)=\phi(1)$, then only the detector $0$
will be registering counts. If, on the other hand $\phi
(0)=\phi(1)\pm \pi$, then only detector $1$ will be registering
counts. These two situations are basically identical to what is
known as Deutsch's algorithm (Deutsch, 1985), the first algorithm
to give an indication that quantum computers are more powerful
than their classical counterparts. This algorithm has also
implemented experimentally in Nuclear Magnetic Resonance (NMR)
(Jones and Mosca, 1998a).

\subsection{Deutsch's algorithm}

Deutsch's problem (Deutsch, 1985) is the simplest possible example
which illustrates the advantages of quantum computation. The
problem is the following. Suppose that we are given a binary
function of a binary variable $f: \{ 0,1\} \longrightarrow \{
0,1\}$. Thus, $f(0)$ can either be $0$ or $1$, and $f(1)$
likewise can either be $0$ or $1$, giving altogether four
possibilities. However, suppose that we are not interested in the
particular values of the function at $0$ and $1$, but we need to
know
whether the function is: 1) constant, i.e. $f(0)=f(1)$, or 2) varying, i.e. $%
f(0)\ne f(1)$. Now Deutsch poses the following task: by computing
$f$ only {\em once} determine whether it is constant or varying.
This kind of problem is generally referred to as a {\em promise
algorithm}, because one property out of a certain number of
properties is initially promised to hold, and our task is to
determine computationally which one holds (see also (Deutsch and
Josza, 1992) for other similar types of promise algorithms).

First of all, classically finding out in one step whether a function is constant or
varying is clearly impossible. We need to compute $f(0)$
and then compute $f(1)$ in order to compare them. There is no way out
of this double evaluation. Quantum mechanically, however, there is a simple
method to achieve this task by computing $f$ only once! Two qubits are
needed for the computation. In reality only one qubit is really needed, but
the second qubit is there to implement the necessary transformation.
We can imagine that the first qubit is the input
to the quantum computer whose internal (hardware) part is represented by
the second qubit. The computer itself will implement the following
transformation on the two qubits
(we perform this fully quantum mechanically, i.e. we are now not
using "classical" devices such as beam splitters):
\begin{eqnarray}
|x\rangle |y\rangle \longrightarrow |x\rangle |y \oplus f(x)\rangle \; ,
\label{dd}
\end{eqnarray}
where $x$ is the input qubit and $y$ the hardware, as depicted in Fig. 9.
Note that this transformation is reversible and thus there is a unitary
transformation to implement it (but we will not pay any attention to that at
the moment, as we are only interested here in the basic principle). Note
also that $f$ has been used only once. The trick is to prepare the input in
such a state that we make use of quantum superpositions. Let us have at the
input
\begin{eqnarray}
|x\rangle |y\rangle = (|0\rangle + |1\rangle)(|0\rangle - |1\rangle)\; ,
\label{in}
\end{eqnarray}
where $|x\rangle$ is the actual input and $|y\rangle$ is part of the
computer hardware. Thus before the transformation is implemented, the state
of the computer is in an equal superposition of all four basis states, which
we obtain by simply expanding the state in eq. (\ref{in}),
\begin{eqnarray}
|\Psi_{\mbox{in}}\rangle = |00\rangle - |01\rangle + |10\rangle -
|11\rangle\; . \nonumber
\end{eqnarray}
Note that there are negative phase factors before the second and fourth
term. When this state now undergoes the transformation in eq. (\ref{dd}), we
have the following output state
\begin{eqnarray}
|\Psi_{\mbox{out}}\rangle & = & |0f(0)\rangle - |0\overline{f(0)}\rangle +
|1f(1)\rangle - |1\overline{f(1)}\rangle \nonumber\\
& = & |0\rangle (|f(0)\rangle - |\overline{f(0)}\rangle) + |1\rangle
(|f(1)\rangle - |\overline{f(1)}\rangle ) \; ,\nonumber
\end{eqnarray}
where the bar indicates the {\em opposite} of that value, so
that, for example, $\overline{0}=1$. Now we see where the power
of quantum computers is fully realised: {\em each of the
components in the superposition of $|\Psi_{\mbox{in}}\rangle$
underwent the same evolution of eq. (\ref{dd})
``simultaneously"}, leading to the powerful ``quantum
parallelism" (Deutsch, 1985). This feature is true for quantum
computation in general. Let us look at the two possibilities now:

\begin{enumerate}
\item  if $f$ is constant then
\begin{eqnarray}
|\Psi _{\mbox{out}}\rangle =(|0\rangle +|1\rangle )(|f(0)\rangle -|\overline{%
f(0)}\rangle )\;.\nonumber
\end{eqnarray}

\item  if $f$ is varying then
\begin{eqnarray}
|\Psi _{\mbox{out}}\rangle =(|0\rangle -|1\rangle )(|f(0)\rangle -|\overline{%
f(0)}\rangle )\;.\nonumber
\end{eqnarray}
\end{enumerate}
Note that the output qubit (the first qubit) emerges in two different
orthogonal states, depending on the type of $f$. These two states can be
distinguished with $100$ percent efficiency. This is easy to see if we first
perform a Hadamard transformation on this qubit, leading to the state $%
|0\rangle$ if the function is constant, and to the state $|1\rangle$ if the
function is varying. Now a single projective measurement in $0,1$ basis
determines the type of the function. Therefore unlike their classical
counterparts quantum computers can solve Deutsch's problem.

Let us now rephrase this in terms of phase shifts to emphasise its underlying
identity with the above Mach-Zender interferometer. The transformation of the two
registers is the following
\begin{eqnarray}
|x\rangle |-\rangle \Rightarrow e^{i\pi f(x)} |x\rangle |-\rangle \nonumber
\end{eqnarray}
where $x=0,1$ and $|-\rangle = |0\rangle - |1 \rangle$. Thus, the
first qubit is like a photon in the interferometer, receiving a
conditional phase shift depending on its state ($0$ or $1$). It is
left to the reader to show that this transformation is formally
identical to the above analysis. The second qubit is there just
to implement the phase shift quantum mechanically. It should be
emphasised that this quantum computation, although extremely
simple, contains all the main features of successful quantum
algorithms: it can be shown that all quantum computations are
just more complicated variations of Deutsch's problem (Cleve et
al, 1997). We will use the introduction of a phase shift as a
basic element of a quantum computer and relate this to the notion
of distinguishability and relative entropy.

Note one important aspect: the input could also be of the form $|-\rangle |-\rangle$.
A constant function would then lead to the state $|-\rangle |-\rangle$ and a
varying function would lead to $|+\rangle |-\rangle$. So, the $|+\rangle$ and
$|-\rangle$ are equally good as input states of the first qubit and both lead
to quantum speed-up. Their equal mixture, on the other hand, is not. This means
that the output would be an equal mixture $|+\rangle\langle +| +|-\rangle\langle -|$
no matter whether $f(0)=f(1)$ or $f(0)\neq f(1)$, i.e. the two possibilities would
be indistinguishable. Thus for quantum algorithm to work well, we need the
first register to be highly correlated to the two different types of functions.
So, if the output state of the first qubit $\rho_1$ indicates that we have a
constant function and $\rho_2$ that we
have a varying function, then the efficiency of Deutsch's algorithm depends
on how well we can distinguish the two states $\rho_1$ and $\rho_2$. This
is given by the Holevo bound
\begin{eqnarray}
H = S(\rho)-\frac{1}{2}(S(\rho_1) + S(\rho_2)) \nonumber
\end{eqnarray}
where $\rho=1/2(\rho_1+\rho_2)$. Thus if $\rho_1 = \rho_2$, then $H=0$ and the
quantum algorithm has no speed up over the classical one. One the other extreme,
if $\rho_1$ and $\rho_2$ are pure and orthogonal, then $H=1$ and the computation
gives the right result in one step. In between these two extremes lie all other
computations with varying degree of efficiency as quantified by the Holevo bound.
Note that these are purely classical correlations and that there is no
entanglement between the first and the second qubit. In fact the Holevo bound is
the same as the formula I suggested for classical correlations in the previous
section. The key to understanding the efficiency of Deutsch's algorithm is
therefore through the mixedness of the first register. If the initial state
has the entropy of $S_0$, then the final Holevo bound is
\begin{eqnarray}
S(\rho) - S_0 \nonumber
\end{eqnarray}
So the more mixed the first qubit the less efficient the computation. Note that
the quantum mutual information between the first two qubits is zero throughout
the entire computation (so there are neither classical nor
quantum correlations between them).

\subsection{Computation: Communication in time}

Can we extend the above entropic analysis to other algorithms as
well? The answer is yes and this is exactly what I will describe
next (Bose et al, 2000b). To explain why this is so, I first need
to introduce a few definitions and a communication model of
quantum computation.  We have two programmers, the sender and the
receiver and two registers, the memory ($M$) register and the
computational ($C$) register.  The sender prepares the memory
register in a certain quantum state $|i\rangle_{\scriptsize M}$
which encodes the problem to be solved. For example, in the case
of factorization (Shor, 1996), this register will store the number
to be factored. In case of a search (Grover, 1996), this register
will store the state of the list to be searched. The number $N$ of
possible states $|i\rangle_{\scriptsize M}$ will, of course, be
limited by the greatest number that the given computer could
factor or the largest list that it could search. The receiver
then prepares the computational register in some initial state
$\rho^0_{\scriptsize C}$. Both the sender and the receiver feed
the registers (prepared by them) to the quantum computer. The
quantum computer implements the following general transformation
on the registers
\begin{equation}
 (|i\rangle \langle i|)_{\scriptsize M} \otimes \rho^0_{\scriptsize C} \rightarrow
 (|i\rangle \langle i|)_{\scriptsize M} \otimes U_i \rho_{\scriptsize C}^{0} U_i^{\dagger}.
\end{equation}
The resulting state  $\rho_{\scriptsize C}(i)
= U_i \rho^0_{\scriptsize C} U_i^{\dagger}$ of the computational register
contains the answer to the computation and is measured by the receiver.
As the quantum computation should work for any $|i\rangle_M$, it should
also work for any mixture $\sum_i^N p_i (|i\rangle\langle i|)_{\scriptsize M}$,
where $p_i$ are probabilities.
For the sender to use the above
computation as a communication protocol, he has to prepare
any one of the states $|i\rangle_{\scriptsize M}$ with an {\em apriori} probability
$p_i$. The entire input ensemble is thus $\sum_i^N p_i (|i\rangle\langle i|)_{\scriptsize M}
\otimes \rho^0_{\scriptsize C}$. Due to the quantum computation, this becomes
\begin{equation}
 \sum_i^N p_i (|i\rangle\langle i|)_{\scriptsize M}
\otimes \rho^0_{\scriptsize C} \rightarrow \sum_i^N p_i (|i\rangle\langle i|)_{\scriptsize M}
\otimes \rho_{\scriptsize C}(i).
\end{equation}
Whereas before the quantum computation,
the two registers where completely uncorrelated (mutual information
is zero), at the end, the
mutual information becomes
\begin{eqnarray}
\label{mut}
I_{MC}:&=& S(\rho_{\scriptsize M}) + S(\rho_{\scriptsize C}) -S (\rho_{\scriptsize MC}) \nonumber \\
   &=& S(\rho_{\scriptsize C}) - \sum_i^N p_i S(\rho_{\scriptsize C}(i)),
\end{eqnarray}
where $\rho_{\scriptsize M}$ and $\rho_{\scriptsize C}$ are the
reduced density operators for the two registers,
$\rho_{\scriptsize MC}$ is the density operator of entire $M+C$
system and $S(\rho)=-\mbox{Tr}\rho \log \rho$ is the von Neumann
entropy (for conventional reasons we will use $\log_2$ in all
calculations). Notice that the value of the mutual information
(i.e correlations) is equal to the Holevo bound
$H=S(\rho_{\scriptsize C}) - \sum_i^N p_i S(\rho_{\scriptsize
C}(i))$ for the classical capacity of a quantum communication
channel (Holevo, 1973) (Note that $\rho_{\scriptsize C}=\sum_i^N
p_i \rho_{\scriptsize C}(i)$). This tells us how much information
the receiver can obtain about the choice $|i\rangle_M$ made by the
sender by measuring the computational register. The maximum value
of $H$ is obtained when the states $\rho_{\scriptsize C}(i)$ are
pure and orthogonal. Moreover, the sender conveys the maximum
information when all the message states have equal {\em apriori}
probability (which also maximizes the channel capacity). In that
case the mutual information (channel capacity) at the end of the
computation is $\log{N}$. Thus the communication capacity
$I_{MC}$ (given by Eq.(\ref{mut})) gives an index of the
efficiency of a quantum computation. {\em A necessary target of a
quantum computation is to achieve the maximum possible
communication capacity consistent with given initial states of
the quantum computer}. We cannot give a sufficiency criterion
from our general approach as this depends on the specifics of an
algorithm. If one breaks down the general unitary transformation
$U_i$ of a quantum algorithm into a number of successive unitary
blocks, then the maximum capacity may be achieved only after
number of applications of the block. In each of the smaller
unitary blocks, the mutual information between the $M$ and the
$C$ registers (i.e the communication capacity) increases by a
certain amount. When its total value reaches the maximum possible
value consistent with a given initial state of the quantum
computer, the computation is regarded as being complete.

\subsection{Black box complexity}

Any general quantum algorithm has to have a certain number of
queries into the memory register (Bennett et al., 1997; Beals et
al, 1998; Ambainis, 2000) (this is necessitated by the fact that
the transformation on the computational register has to depend on
the problem at hand, encoded in $|i\rangle_{\scriptsize M}$).
These queries can be considered to be implemented by a black box
into which the states of both the memory and the computational
registers are fed. The number of such queries needed in a certain
quantum algorithm gives the black box complexity of that algorithm
(Bennett et al., 1997; Beals et al., 1998; Ambainis, 2000) and is
a lower bound on the complexity of the whole algorithm. The black
box approach is a simplification for looking at the complexity of
an algorithm. A black box allows us to perform a certain
computation without having its exact details. It is possible that
physical implementations of a particular black box may prove to be
difficult. So when we estimate the complexity of an algorithm by
counting the number of applications of a black box, we have to
bear in mind that there might an additional complexity component
arising in physical implementation.

%%%%%%%%%%%%%%%%%%  Fig. 9 Phase Kick Back %%%%%%%%%%%%%%%%%%%%%%%%%%

\begin{figure}[ht]
\begin{center}
\hspace{0mm} \epsfxsize=7.5cm
\epsfbox{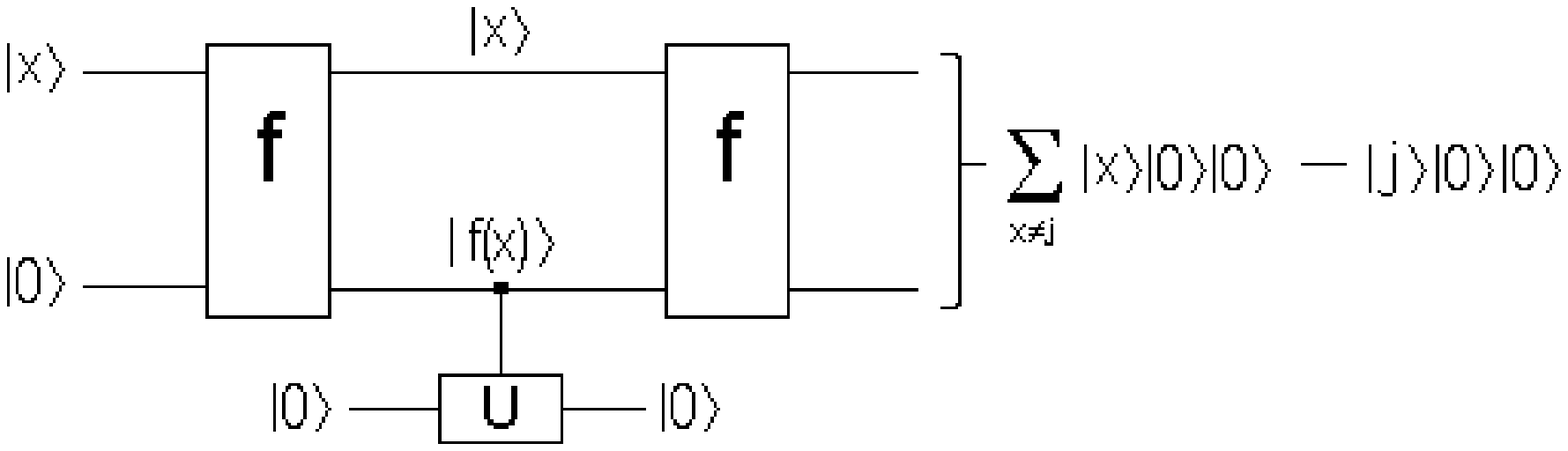}\\[0.2cm]
\begin{caption}
{\narrowtext This figure represents a network that implements the
phase flip operation given the black box computing the function
$f(x)$. The unitary transformation $U$ implements $|0\rangle
\rightarrow - |0\rangle$ conditionally on the value of $f(x)$.}
\end{caption}
\end{center}
\end{figure}

In general we have a function $f: \{0,1\}^n \rightarrow \{0,1\}$
(so the function maps $n$-bit values to either $0$ or $1$).
Quantum algorithms, such as database search, can be expressed in
this form (in the case of database search, all the values of $f$
are $0$ apart from one which is equal to $1$; the task is to find
this value). The black box is assumed to be able to perform the
transformation $|x\rangle|y\rangle \rightarrow
|x\rangle|f(x)\oplus y\rangle$, just like in Deutsch's algorithm.
We have the freedom to represent this black box transformation as
a phase flip which is equivalent in power (up to a constant factor
as seen in Fig. 9),
\begin{eqnarray}
|x\rangle|y\rangle \rightarrow (-1)^{f(x)\oplus y}|x\rangle|y\rangle \nonumber
\end{eqnarray}
Recently, Ambainis (2000) showed in a very elegant paper that if
the memory register was prepared initially in the superposition
$\sum_i^N |i\rangle_{\scriptsize M}$, then, in a search
algorithm, $O(\sqrt{N})$ queries would be needed to completely
entangle it with the computational register. This gives a lower
bound on the number of queries in a search algorithm. In a manner
analogous to his, we will calculate the change in mutual
information between the memory and the computational registers
(from Eq.(\ref{mut})) in one query step. The number of queries
needed to increase the mutual information to $\log{N}$ (for
perfect communication between the sender and the receiver), is
then a lower bound on the complexity of the algorithm.

\subsection{Database search}

Any search algorithm (whether quantum or classical, irrespective
of its explicit form), will have to find a match for the state
$|i\rangle_{\scriptsize M}$ of the $M$ register among the states
$|j\rangle_C$ of the $C$ register and associate a marker to the
state that matches (Here, $|j\rangle_C$ is a complete orthonormal
basis for the $C$ register). The most general way of doing such a
query in the quantum case is the black box unitary transformation
(Ambainis, 2000)
\begin{equation}
 U_{\scriptsize B}|i\rangle_M |j\rangle_C = (-1)^{\delta_{ij}} |i\rangle_M |j\rangle_C.
\end{equation}
Any other unitary transformation performing a query matching the
states of the $M$ and the $C$ registers, could be constructed from
the above type of query. {\em Note that the black box is able to
recognize if a value in the $C$ register is the same as the
solution, but is unable to explicitly provide that solution for
us.} For example, imagine that Socrates goes to visit the
all-knowing Ancient Greek oracle (black box) who is only able to
answer with "yes" or "no". Suppose further that Socrates wants to
know who the wisest person in the world is. He would then have to
ask something like "Is Plato the wisest person in the world?" and
would not be able to ask directly "Who is the wisest person in the
world?". This "yes-no" approach is typical for any black box
analysis. The advantage of using this black box quantum
mechanically is that we can query all the individual elements of
the superposition simultaneously. Although we can identify the
solution in one step quantum mechanically, further computations
are required to amplify the right solution so that the subsequent
measurement is more likely to reveal it.

I would like to put a bound on the change of the mutual
information in one such black box step. Let the memory states
$|i\rangle_M$ be available to the sender with equal apriori
probability so that the communication capacity is a maximum. His
initial ensemble is then $\frac{1}{N}\sum_{i}^N (|i\rangle \langle
i|)_{\scriptsize M}$. Let the receiver prepare the $C$ register in
an initial pure state $\psi^0$ (in fact, the power of quantum
computation stems from the ability of the receiver to prepare pure
state superpositions of form $\frac{1}{N}\sum_{j}^N |j\rangle_C$).
This is an equal weight superposition of all $|j\rangle_C$ as
there is no {\em apriori} information about the right
$|j\rangle_C$. This can be done by performing a Hadamard
transformation to each qubit of the $C$ register. In general,
there will be many black box steps on the initial ensemble before
a perfect correlation is set up between the $M$ and the $C$
registers. Let, after the $k$th black box step, the state of the
system be
\begin{equation}
\rho^k = \frac{1}{N}\sum_{i}^N (|i\rangle \langle i|)_{\scriptsize M} \otimes (|\psi^k(i)\rangle
\langle \psi^k(i)|)_{\scriptsize C}
\end{equation}
where
\begin{equation}
|\psi^k(i)\rangle_{\scriptsize C} = \sum_j \alpha_{ij}^k |j\rangle_{\scriptsize C}.
\end{equation}
The $(k+1)$th black box step changes this state to $\rho^{k+1}=\frac{1}{N}\sum_{i}^N
(|i\rangle \langle i|)_{\scriptsize M} \otimes (|\psi^{k+1}(i)\rangle \langle \psi^{k+1}(i)|)_{\scriptsize C}$ with
  \begin{equation}
|\psi^{(k+1)}(i)\rangle = \sum_{i,j}^{N} \alpha_{ij}^k (-1)^{\delta_{ij}} |j\rangle_{\scriptsize C}.
\end{equation}
Thus we only have to evaluate the difference of mutual
information between the $M$ and the $C$ register for the states.
This difference of mutual information (when computed from
Eq.(\ref{mut})) can be shown to be the difference
$|S(\rho^{k+1}_{\scriptsize C})-S(\rho^k_{\scriptsize C})|$
(Henderson and Vedral, 2000). This quantity is bounded from the
above by (Fannes, 1999)
\begin{eqnarray}
 |S(\rho^{k+1}_{\scriptsize C}) &-& S(\rho^k_{\scriptsize C})| \leq
d_{\scriptsize B}(\rho^k_{\scriptsize C},\rho^{k+1}_{\scriptsize C}) \log{N} \nonumber \\
&-& d_{\scriptsize B}(\rho^k_{\scriptsize C},\rho^{k+1}_{\scriptsize C})
\log{d_{\scriptsize B}(\rho^k_{\scriptsize C},\rho^{k+1}_{\scriptsize C})}
\end{eqnarray}
where, $d_{\scriptsize B}(\sigma,\rho)=\sqrt{1-F^2(\sigma,\rho)}$
is the Bures metric (Bures, 1969) and
$F(\sigma,\rho)=\mbox{Tr}\sqrt{\sqrt{\rho}\sigma \sqrt{\rho}}$ is
the fidelity. Using methods similar to Ambainis (2000), it can be
shown that $F(\rho^0_{\scriptsize C},\rho^{1}_{\scriptsize C})
\geq \frac{N-2}{N}$ from which it follows that the change in the
first step
\begin{equation}
\label{step}
 |S(\rho^{0}_{\scriptsize C})-S(\rho^1_{\scriptsize C})| \leq \frac{3}{\sqrt{N}} \log{N}.
\end{equation}
The change $|S(\rho^{k}_{\scriptsize C})-S(\rho^{k+1}_{\scriptsize C})|$ in
the subsequent steps has to be less than or equal to the change in the first step.
This is because the Bures metric does not increase under general completely
positive maps (which is what the query represents when we trace out the $M$
register). Any other operations performed only on the $C$ register in between
two queries can only reduce the mutual information between the $C$ and the
$M$ register.
 This means that at least $O(\sqrt{N})$ steps are needed to produce
full correlations (maximum mutual information of value $\log{N}$)
between the two registers. This gives the black box lower bound
on the complexity of any quantum search algorithm. Of course, we
know that there also exists an algorithm achieving this bound due
to Grover (1996) and this has been proven to be optimal (Bennett
et al., 1996a; Ambainis, 2000; Zalka, 1999). However, the proof
presented here is the most general as it holds even when any type
of completely positive map is allowed between the queries (only in
Zalka (1999) a heuristic argument was made for the optimality of
Grover's algorithm under general operations). Grover's algorithm
has also been implemented experimentally (Jones and Mosca, 1998b).

%% Grover network Fig 10

\begin{figure}[ht]
\begin{center}
\hspace{0mm}
\epsfxsize=7.5cm
\epsfbox{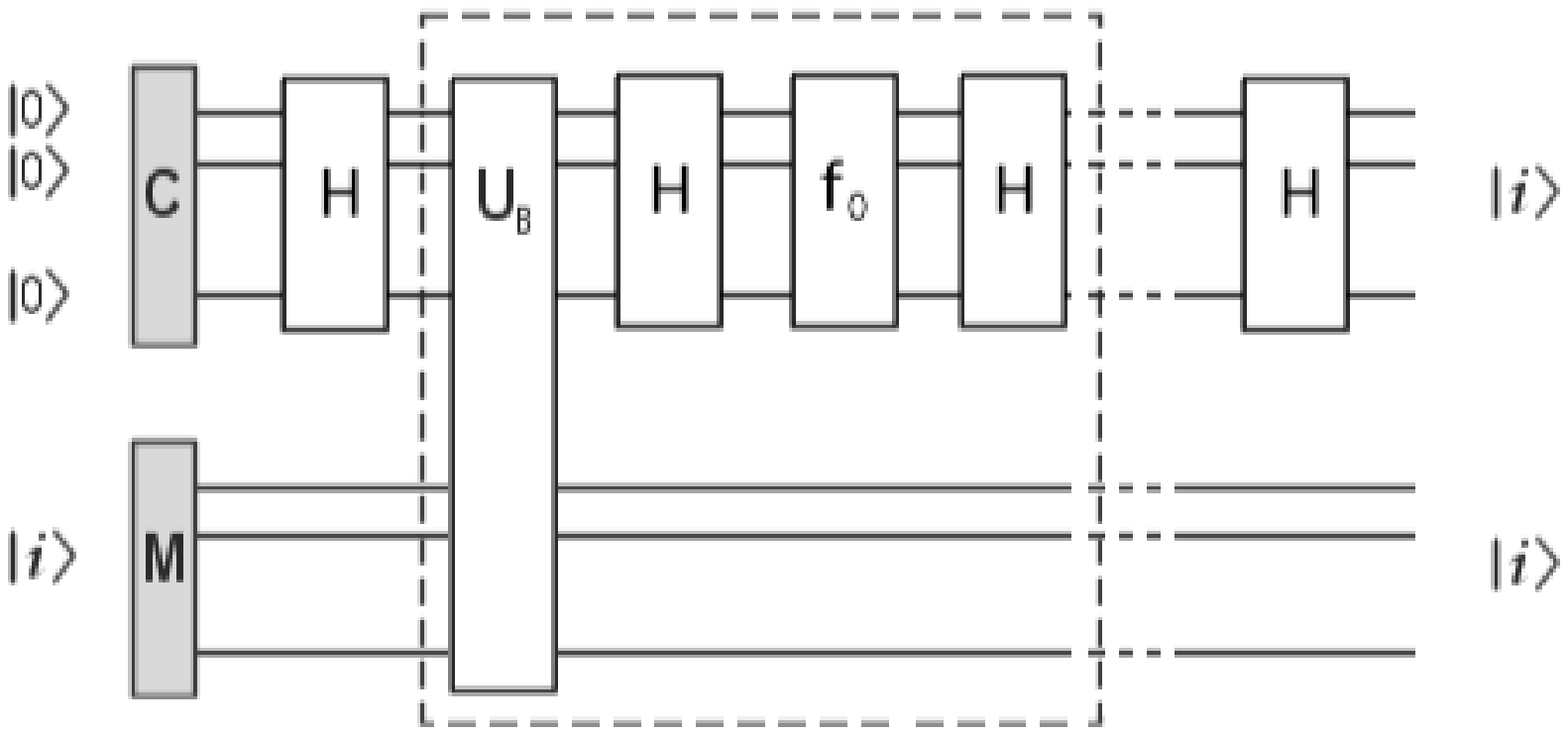}\\[0.2cm]
\begin{caption}
{\narrowtext The figure shows the circuit for Grover's algorithm.
$C$ is the computational register and $M$ is the memory register.
$U_{\scriptsize B}$ is the black box query transformation, $H$ is
a Hadamard transformation on every qubit of the $C$ register and
$f_0$ is a phase flip in front of the $|00...0\rangle_{\scriptsize C}$.
The block consisting of $H,U_{\scriptsize B}, H$ and $f_0$ is
repeated a number of times.}
\end{caption}
\end{center}
\end{figure}

I now use Grover's algorithm to show how the mutual information
varies with time in a quantum search. The general sequence
described by Cleve et. al (1997) for Grover's algorithm will be
used in this letter.  The algorithm consists of repeated blocks,
each consisting of a Hadamard transform on each qubit of the $C$
register, followed by a $U_{\scriptsize B}$ (our black box
transformation), followed by another Hadamard transform on each
qubit of the $C$ register and finally a phase flip $f_0$ of the
$|00...0\rangle_{\scriptsize C}$ state of the $C$ register (See
Fig. 10). This block can then be repeated as many times as is
necessary to bring the mutual information to its maximum value of
$\log{N}$, which, as I have shown in Eq.(\ref{step}) to be
$O(\sqrt{N})$. Note that the only transformation correlating the
$M$ and $C$ registers is the black box transformation
$U_{\scriptsize B}$ and all the other transformations are done
{\em only} on the $C$ register and therefore do not change the
mutual information between the two registers. In Fig. 8 I have
plotted the variation of mutual information between the $M$ and
the $C$ registers (i.e the communication capacity of the quantum
computation) with the number of iterations of the block in
Grover's algorithm. It is seen that the mutual information
oscillates with the number of iterations. Fig. 11 is plotted for a
four qubit computational register which can search a database of
$16$ entries. It is seen that the period is roughly $6$, which
means that the number of steps needed to achieve maximum mutual
information is roughly $3$. This is well above our bound for the
minimum number of steps, which is $4/3$ in this case.

%% Mutual ifo graph Fig11

\begin{figure}[ht]
\begin{center}
\hspace{0mm}
\epsfxsize=7.5cm
\epsfbox{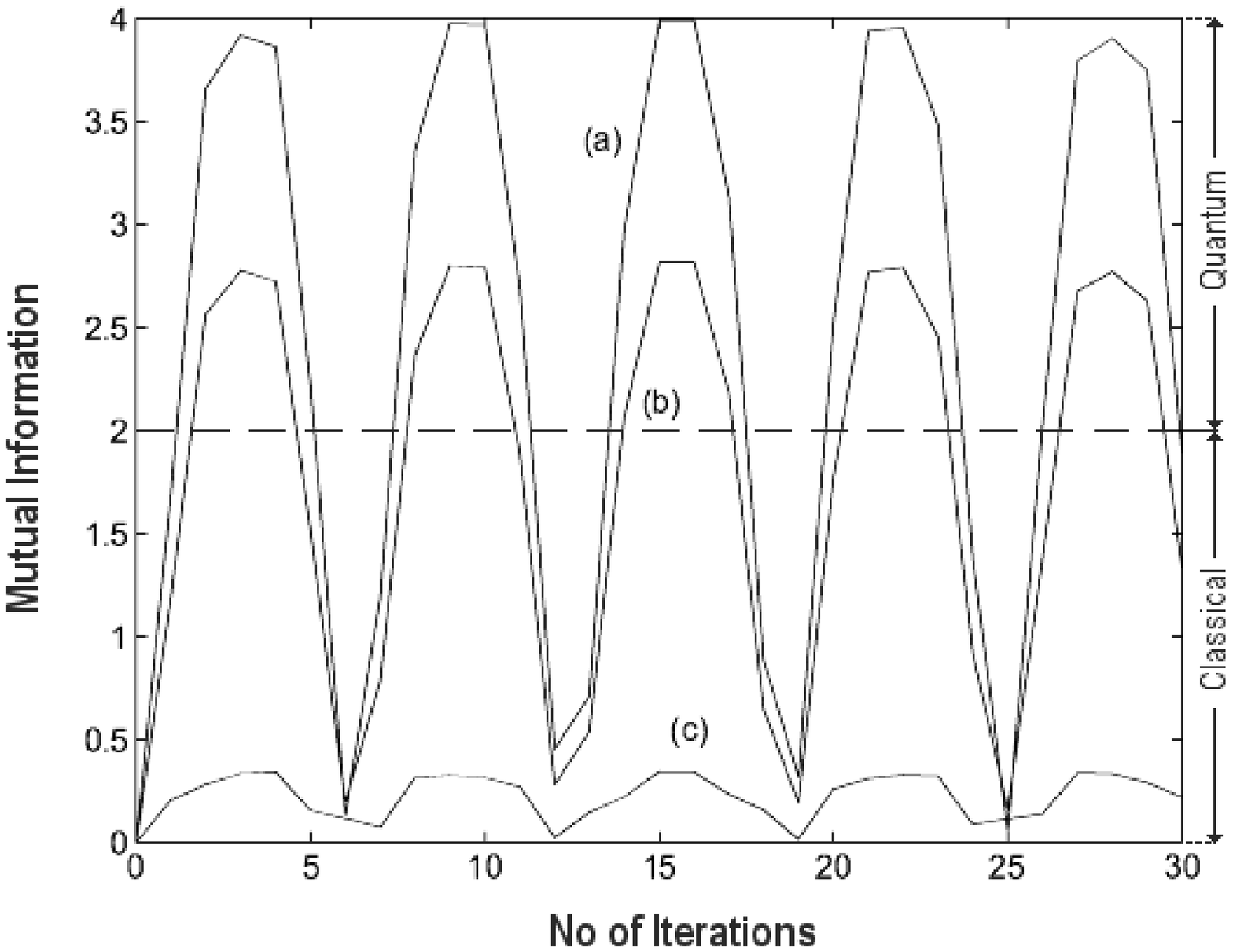}\\[0.2cm]
\begin{caption}
{\narrowtext The figure shows the dependence of the mutual information
between the $M$ and the $C$ registers as a function of the number of times
the block in Grover's algorithm is iterated for various values of initial
mixedness of the $C$ register. Each qubit of the $C$ register is initially
in the state $p|0\rangle\langle 0|+(1-p)|1\rangle \langle 1|$,
(a) $p=1$, (b) $p=0.95$
and (c) $p=0.7$. The (a) and (b) computations achieve higher
mutual information  than classically
allowed in the order of root $N$ steps, while (c) does not.}
\end{caption}
\end{center}
\end{figure}

The fact that the mutual information oscillates periodically (or
more precisely, "quasi"-periodically) follows from the {\em
quantum Poincare recurrence theorem} (Hogg and Huberman, 1983).
Namely, if the system has a discrete spectrum and is "driven" by a
periodic potential (as is in Grover's case, where we repeat the
same operation time and again), then its wavefunction $\psi (t)$
will undergo a quasi periodic motion, i.e. for any $\epsilon >0$,
there exists a relatively dense set $\{T_{\epsilon}\}$ such that
\begin{eqnarray}
|| \psi (t+T_{\epsilon})-\psi (t)|| < \epsilon \nonumber
\end{eqnarray}
for all time $t$ and for each $T_{\epsilon}$ in the set. This is exactly
the behavior seem in the Fig. 9. The distance between the two
states $|\psi\rangle = \sum_i a_i |i\rangle$ and $|\phi\rangle = \sum_j b_j |j\rangle$
is defined in the usual way
\begin{eqnarray}
||\psi-\phi || :=\sum_i |a_i - b_i| \nonumber
\end{eqnarray}
The three graphs (a), (b) and (c) in Fig. 9 are for different values
of initial mixedness of the $C$ register.
We find that the mutual information fails to rise to the maximum
value of $\log{N}$ when the state of the computational register is
mixed. Our formalism thus allows us
to calculate the performance of a quantum computation as a function of the
mixedness (quantified by the von Neumann entropy) of the computational
register. We can put a bound on the entropy of the second register
after which the quantum search becomes as inefficient as the classical search.
If the initial entropy $S(\rho^{0}_{\scriptsize C})$ of the $C$ register exceeds
$\frac{1}{2}\log{N}$, then the change in mutual information between
the $M$ and the $C$ registers in the course of the entire
quantum computation would be at most $\log{\sqrt{N}}$. This can be
achieved by a classical database search in $\sqrt{N}$ steps. So there is
no advantage in using quantum evolution when the initial state is too
mixed. Note that our condition

\vspace*{.3cm}
\noindent
%\fbox{\parbox[b]{17.7cm}{
{\bf Sufficient condition for no quantum speed-up.}
\begin{eqnarray}
S(\rho^{0}_{\scriptsize C}) \geq \frac{1}{2}\log{N} \nonumber
\end{eqnarray}
%}}
\vspace*{.2cm}

\noindent
Note that this is only a sufficient and {\em not} a necessary condition.

I also point out that the states of the $M$ register need not be a
mixture, but could be an arbitrary superposition of states
$|i\rangle_{\scriptsize M}$ (such a state was used by Ambainis in
his argument (Ambainis, 2000)). All the above arguments still
hold in that case, and the $M$ and the $C$ registers become
quantum mechanically entangled and not just classically
correlated. Thus our analysis implies that any quantum
computation is mathematically identical to a measurement process
(Everett, 1973). The system being measured is the $M$ register
and the apparatus is the $C$ register of the quantum computer. As
the time progresses the apparatus (register $C$) becomes more and
more correlated (or entangled) to the system (register $M$). This
means that the states of register $C$ become more and more
distinguishable which allows us to extract more information about
the $M$ register by measuring the $C$ register. The analysis in
the last paragraph, where I showed the limitations on the
efficiency of quantum computation imposed by the mixedness of the
$C$ register, applies also to the efficiency of a quantum
measurement when the apparatus is in a mixed state. Mixedness of
an apparatus, to the best of our knowledge, has never been
considered in the analysis of quantum measurement. In general
practice, any apparatus, however macroscopic, is considered to be
in a pure quantum state before the measurement. Our approach
highlighting the formal analogy between measurement and
computation offers a way to analyse measurement in a much more
general context.

Finally, I would like to discuss what would happen if we decided
to change the nature of the Black box. Suppose that instead of
being able to recognize the right solution, the black box is much
more powerful and it can compare if the individual bit values
coincide with the bit values of the solution. So, for all $k$,
\begin{eqnarray}
&   & |i_0i_1...i_k...i_n\rangle |j_0i_j...j_k...j_n\rangle
\rightarrow \nonumber\\
&   & (-1)^{\delta_{i_kj_k}} |i_0i_1...i_k...i_n\rangle
|j_0i_j...j_k...j_n\rangle ,
\end{eqnarray}
where $i=i_0i_1...i_k...i_n$ and $j=j_0i_j...j_k...j_n$ are the
binary representations of $i$ and $j$ respectively. Then, it can
easily be checked that this gate has the power to correlate the
$C$ and the $M$ register by the amount of $\log 2$. Therefore the
search algorithm would take $\log N$ steps (instead of
$\sqrt{N}$), i.e. it would be polynomial instead of exponential!
There is, of course, a hidden complexity here which is in
constructing the new Black box from the original Black Box. It can
be shown that this requires exponential increase in time (or space
which can always be traded for time) and this then compensates the
exponential decrease in the number of applications of the new
Black Box. In fact, this new black box is equivalent to the
Ancient Greek oracle being able to answer to Socrates "Who is the
wisest person in the world?".

Can we use entropic measures of the above form to quantify
complexity of other quantum algorithms? The answer is unclear at
present. The only algorithm that presently achieves an
exponential speed-up over its classical counterpart, Shor's
factorisation algorithm (Shor, 1996), cannot be usefully
re-phrased in terms of black box operations (more precisely, it
is rather trivial, as it requires only one black box operation!).
However, this does not prevent us from deriving fundamental
bounds on information storage and the speed of its processing
based on the Uncertainty principle. In the next subsection, I
show the ultimate limits of processing power no matter what model
of computation is used so long as it uses quantum systems
(particles or fields alike).

\vspace*{.2cm}

\noindent {\bf Quantum computation and quantum measurement}. I
now show that quantum computation is formally identical to a
quantum measurement as described by von Neumann (1955). The
analysis will be performed in the most general continuous case.
Suppose that we have a system $S$ (described by a continuous
variable $x$) and an apparatus $A$ (described by a continuous
variable $y$ interacting via a Hamiltonian $H=xp$, where $p$ is
the momentum of $A$ (we will assume that $\hbar =1$). Suppose in
addition that the initial state of the total system is
\begin{eqnarray}
|\Psi (0)\rangle = \int_x \phi (x) |x\rangle dx \otimes \eta (y) |y\rangle dy
\nonumber
\end{eqnarray}
in an uncorrelated state. The action of the above Hamiltonian
then transforms the state into an entangled state. In order to
calculate this transformation it will be beneficial to introduce
the (continuous) Fourier transform
\begin{eqnarray}
\mbox{F}_y: |y\rangle \rightarrow \int e^{-iyp} |p\rangle dp \nonumber
\end{eqnarray}
which takes us from the position space of $A$ into the momentum
space of $A$. This is important because we know the effect of the
Hamiltonian in the momentum basis. Now, the action of the unitary
transformation generated by $H$ is
\begin{eqnarray}
|\Psi (t)\rangle & = & e^{-ixpt}|\Psi (0)\rangle \nonumber\\
                 & = & \mbox{F}_y e^{-ixpt} \mbox{F}_y |\Psi (0)\rangle\nonumber\\
                 & = & \int_x \int_y \phi (x) \eta (y-xy) |x\rangle |y\rangle dx dy
\nonumber
\end{eqnarray}
and we see that $S$ and $A$ are now correlated in $x$ and $y$.
This means that by measuring $A$ we can obtain some information
about the state of $S$. The mutual information
$I_{AS}=H(x)+H(y)-H(x,y)$ can be shown to satisfy (Everett, 1973)
\begin{eqnarray}
I_{AS} \geq \ln t \nonumber
\end{eqnarray}
i.e. it is growing at a rate faster than logarithm of time
passage during the measurement. This gives us a lower bound to
exactly how quickly correlations can be established between the
system and the apparatus. This is analogous to the way I derived
the upper bound on the efficiency of quantum search algorithms in
Section V.

I now show the detailed calculation of the effect of measurement
Hamiltonian. Let us define
\begin{eqnarray}
\xi (p) := \mbox{F}_y \{\eta(y)\} \nonumber
\end{eqnarray}
The evolution then proceeds as follows,
\begin{eqnarray}
|\Psi (t)\rangle & = & e^{-xpt} \int_x \phi (x) |x\rangle dx \otimes \eta (y) |y\rangle dy\nonumber\\
                 & = & e^{-xpt} \int_x \phi (x) \int_p \{ \int_y \eta (y) e^{-iyp} dy \} |x\rangle |p\rangle dxdp \nonumber\\
                 & = & e^{-xpt} \int_x \int_p \ \phi (x) \xi (p) |x\rangle |p\rangle dxdp\nonumber\\
                 & = & \int_x \ \phi (x) \int_y \{ \int_p \xi (p) e^{-ixpt} e^{iyp} dp\} |x\rangle |y\rangle dxdy\nonumber\\
                 & = & \int_x \int_y \phi (x) \eta (y-xy) |x\rangle |y\rangle dx dy\nonumber
\end{eqnarray}
This result has the same formal structure of quantum algorithms
presented before: a Fourier transform, followed by a conditional
phase shift and then followed by another Fourier transform (c.f.
Deutsch's and Grover's algorithms). {\em Therefore we can see that
how efficiently we can measure something is the same as how
efficiently we can compute, both of which depend on how quickly we
can establish correlations}.

\subsection{Ultimate limits of computation: The Bekenstein bound}

Given a computer enclosed in a sphere of radius $R$ and having
available the total amount of energy $E$ what is the amount of
information that it can store and how quickly can this
information be processed? The Holevo bound gives us the ultimate
answer. The amount of information that can be written into this
volume is bounded from the above by the entropy, i.e. the number
of distinguishable states that this volume can support. I will
now use a simple, informal argument to obtain this ultimate bound
(Tipler, 1994), but the rigorous derivation can be found in
(Bekenstein, 1981). The bound on energy implies a bound on
momentum and the total number of states in the phase space is
\begin{eqnarray}
N=\frac{PR}{\Delta P \Delta R}\leq\frac{PR}{\hbar}   \nonumber
\end{eqnarray}
where the inequality follows from the Heisenberg uncertainty relations
$\Delta P \Delta R\geq \hbar$ which limits the size of the smallest
volume in the phase space to $\hbar$ in each
of the three spatial directions. From relativity we have that for
any particle $p\leq E/c$, so that
\begin{eqnarray}
I\leq \ln N \leq N \leq \frac{E}{c}\frac{R}{\hbar}\leq \frac{ER}{\hbar c} \nonumber
\end{eqnarray}
which is known as the Bekenstein bound. In reality this inequality will most likely be a
huge over-estimate, but it is important to know that no matter how we encode
information we cannot perform better than is given by our most accurate
present theory - quantum mechanics. As an example consider a nucleus of
the Hydrogen - according to the above result it can encode about $100$ bits
of information (I assumed that $E=mc^2$ and that $R=10^{-15}$m).
At present, NMR quantum computation achieves "only" one
bit per nucleus (and not per nucleon!)- spin "up" and spin "down" are
the two states.

From the Bekenstein bound we can derive a bound on the efficiency
of information processing. Again my derivation will be loose, and
a much more careful calculation confirms what I will present
(Bekenstein, 1984). All the bits in the volume $V$ cannot be
processed faster than it takes light to travel across $V=4/3\pi
R^3$, which is $2R/c$. This gives
\begin{eqnarray}
\frac{dI}{dt}\leq \frac{E}{2\hbar}\nonumber
\end{eqnarray}
Again a Hydrogen nucleus can process $10^{24}$ bits per second, which is also
in sharp contrast with NMR quantum computation where a
NOT gate takes roughly a few milliseconds leading to a maximum of
$10^3$ bits per second.

The Bekenstein bound shows that there is a potentially great
number of under-used degrees of freedom in any physical system.
This provides hope that quantum computation will be an
experimentally realisable goal. At present, there is a number of
different practical implementations of quantum computation, but
none of them can store and manipulate more than $10$ qubits at a
time ($5$ was the largest number (Vandersypen, 2000) manipulated
in a genuine quantum computation process when this review was
finished in the summer of $2000$). The above calculation,
however, does not take into account the environmental influence
on computation nor the experimental precision. I have not at all
touched on the practical possibility of building a quantum
computer. This is partly for reasons of space, partly because it
would spoil the flow of exposition and partly because there is
already a number of excellent reviews of this subject (Steane,
1997). It is generally acknowledged that the difficulties in
building a quantum computer are only of practical nature and
there are no fundamental limits that prohibit such a device. I
hope that this section offers convincing arguments that building
a quantum computer is a very much worthwhile adventure, both from
the technological as well as fundamental perspective. In any case
we see that there is a great deal of currently unused potential in
physical systems in which to store and encode information. As our
level of technology improves we will find more and more ways of
getting closed to the Bekenstein bound.

\section{Conclusions}

We have seen how distinguishability of different physical states
is at the heart of information processing, which we quantified
using the relative entropy. The relative entropy told us about the
possibility of confusing two probability distributions, or, in
the quantum case, two density matrices. We have seen that
relative entropy never increases under any general quantum
evolution, meaning that states can become only less
distinguishable as time progresses. The most important
consequence of this was shown to be the Holevo bound, which is
the bound on the capacity for classical communication using
quantum states. This basically told us that $n$ qubits cannot
store more than $n$ classical bits of information. While this
appears to be a severe limitation on quantum information
processing, I showed with the aid of dense coding that quantum
communication is in some sense more efficient than its classical
counterpart. Dense coding involved the use of entangled states
and I therefore showed how the quantum relative entropy can be
used to quantify entanglement. Moreover, I used the Holevo bound
to put limits on the efficiency of quantum computation by
treating it as a communication protocol. Quantum algorithms were
shown to be considerably more efficient for some problems than
classical algorithms. In particular, I have shown in a new way
that the quantum database search has a square root speed up over
the classical database search. Efficiency of quantum computation
stems from the trade-off between two opposite effects: on the one
hand, superpositions allow us to compute in parallel, while, on
the other hand, the Holevo bound limits the amount of information
we can extract from a quantum state. I also emphasised links
between the black box quantum computation and quantum measurement
and I showed that there is a fundamental limit to deleting
information, leading to Landauer's principle that $1$ bit erased
increases the environment information by $k_B \ln 2$.

With every new physical theory comes a new understanding of the
world we live in. Through Newtonian physics we understood the
Universe as a clockwork mechanism. With the subsequent
development of thermodynamics the Universe became a big Carnot
engine, slowly evolving towards its final equilibrium state after
which no useful work could be obtained - the heat death.
Presently, we see Universe as an information processing machine -
a computer. Limits to the amount of information it can contain and
process are given by the most accurate theory we have, quantum
mechanics, giving rise to quantum information theory.

If there is a single moral to be drawn from the relationship between information
and physics it is that, as we dig deeper into the fundamental laws on physics, we also push
back the boundaries of information processing. It will not be surprising if
all the results presented in this review are superseded by higher level
generalisation of which they become an approximation in the same way that today
classical information theory approximates quantum information theory.

\vspace*{1cm}

\noindent {\bf Acknowledgments}. I would like to thank S. Bose, D.
Deutsch, D. P. Divincenzo, M. J. Donald, A. Ekert, L. Henderson,
P. Hayden, J. A. Jones, E. Kashefi, P. L. Knight, G. Lindblad, M.
Murao, M. Ozawa, M. B. Plenio, L. Rallan, B. Schumacher and G.
Vidal for many stimulating discussions on the subject of quantum
information. In particular, I thank L. Rallan for very thorough
reading of this manuscript and helping me draw some of the
figures. Work for this review has been supported by the Knight
Trust and ORS Award, Elsag-Bailey spa, Hewlett-Packard company,
EPSRC and the European Union project EQUIP (contract
IST-1999-11053).


\begin{thebibliography}{99}
%
\bibitem{1} Ambainis, A., 2000, {\em Quantum lower bounds by quantum
arguments}, quant-ph/0002066.
%
\bibitem{2} Araki, H., and E. H. Lieb, 1970, Comm. Math. Phys., {\bf 18}, 160.
%
\bibitem{3} Barnum, H., M. A. Neilsen and B. Schumacher, (1998), Phys. Rev. A
{\bf 57}, 4153.
%
\bibitem{4} Beals, R., H. Buhrman, R. Cleve, M. Mosca and R. de Wolf, 1998,
{\em Quantum lower bounds by polynomials},  Proceedings of
FOCS'98, pages 352-361. Also quant-ph/9802049.
%
\bibitem{5} Bekenstein, J. D. , 1981, Physical Review D {\bf 23},  287.
%
\bibitem{6} Bekenstein, J. D. , 1984, Physical Review D {\bf 30}, 1669.
%
\bibitem{7} Bell, J., 1987, "Speakable and Unspeakable in Quantum Mechanics",
(Cambridge Univ. Press, Cambridge).
%
\bibitem{8} Bennett, C. H., and S. Wiesner, 1992, Phys. Rev. Lett. {\bf 69}
2881.
%
\bibitem{9} Bennett, C. H., G. Brassard, C. Crepeau, R. Jozsa, A. Peres, and
W.K. Wootters, 1993, Phys. Rev. Lett. {\bf 70}, 1895.
%
\bibitem{10} Bennett, C. H., H. J. Bernstein, S. Popescu, and B. Schumacher,
1996a, Phys. Rev. A {\bf 53}, 2046.
%
\bibitem{11} Bennett, C. H., D. P. DiVincenzo, J. A. Smolin, and W. K.
Wootters, 1996b, Phys. Rev. A {\bf 54}, 3824.
%
\bibitem{12} Bennett, C. H., E. Bernstein, G. Brassard and U. Vazirani, 1997,
SIAM Journal on Computing, {\bf 26}, 1510. Also quant-ph/9701001.
%
\bibitem{13} Bennett, C. H., D. P. Divincenzo, C. A. Fuchs, T. Mor, E. Rains,
P. W. Shor, J. A. Smolin and W. K. Wootters, 1999a, Phys. Rev. A.
{\bf 59} 1070.
%
\bibitem{14} Bennett, C. H., D. P. DiVincenzo, T. Mor, P. W. Shor, J. A.
Smolin and B. M. Terhal, 1999b, Phys. Rev. Lett. {\bf 82} 5385.
%
\bibitem{15} Bhatia, R., 1997, ``Matrix Analysis", (Springer-Verlag, Berlin).
%
\bibitem{16} Bose, S., M. B. Plenio and V. Vedral, 2000a, J. Mod. Opt {\bf 47},
291.
%
\bibitem{17} Bose, S., L. Rallan and V. Vedral, 2000b, Phys. Rev. Lett.
{\bf 85} (25) 5448.
%
\bibitem{18} Bouwmeester, D., J.W. Pan, K. Mattle, M. Eibl, H. Weinfurter, and
A. Zeilinger, 1997, Nature {\bf 390}, 575; D. Boschi, S. Branca,
F. DeMartini, L. Hardy, and S. Popescu, Phys. Rev. Lett. {\bf 80},
1121 (1998); A. Furusawa {\em et al.}, Science {\bf 282}, 706
(1998).
%
\bibitem{19} Bowen, G., 2001, Phys. Rev. A {\bf 63} 022302.
%
\bibitem{20} Brillouin, L., 1956, ``Science and Information Theory", (Academic
Press, New York).
%
\bibitem{21} Bures, D., 1969, {Trans. Am. Math. Soc.} {\bf 135}, 199; see also
A. Uhlmann, {Rep. Math. Phys.} {\bf 9}, 273 (1976); {\em ibid}
{\bf 24}, 229 (1986).
%
\bibitem{22} Caves; C. M., and P. D. Drummond, 1994, Rev. Mod. Phys. {\bf 66},
481.
%
\bibitem{23} Choi, M. D., 1975, Lin. Algebra and Appl. {\bf 10}, 285.
%
\bibitem{24} Clauser, J. F., M. A. Horne, A. Shimony, and R. A. Holt, 1969,
Phys. Rev. Lett. {\bf A 23}, 880.
%
\bibitem{25} Cleve, R., A. Ekert, C. Macchiavello and M. Mosca, 1997, Phil.
Trans. R. Soc. Lond. A.
%
\bibitem{26} Cover, T. M., and J.A. Thomas, 1991, ``Elements of Information
Theory" (A Wiley-Interscience Publication).
%
\bibitem{27} Csisz\'ar, I., and J. K\"orner, 1981, {``Coding Theorems for
Discrete Memoryless Systems"}, (Academic Press, New York).
%
\bibitem{28} Davies, E. B., 1976, "Quantum Theory of Open Systems", (Academic
Press, London).
%
\bibitem{29} Deutsch, D., 1985, Proc. R. Soc. Lond. A  {\bf 400}, 97; D.
Deutsch,  Proc. R. Soc. Lond. A {\bf 425}, 73 (1989).
%
\bibitem{30} Deutsch, D., and R. Jozsa, 1992, Proc. R. Soc. Lond. A {\bf 439},
553; E.~Bernstein and U.~Vazirani, in {\em Proc. 25th ACM
Symposium on the Theory of Computation\/}, 11 (1993); D.S.~Simon,
{\em Proceedings of the 35th Annual Symposium on the Foundations
of Computer Science\/}, edited by S.~Goldwasser (IEEE Computer
Society Press, Los Alamitos, CA), 16 (1994).
%
\bibitem{31} Deutsch, D., 1998, {``The Fabric of Reality"}, (Viking--Penguin
Publishers, London).
%
\bibitem{32} DiVincenzo, D. P., 1998 {\it et~al.}, quant-ph/9803033.
%
\bibitem{33} DiVincenzo, D.P., P.W. Shor, J. A. Smolin, B. M. Terhal and A. V.
Thapliyal, 2000, Phys. Rev. {\bf A 61} 062312.
%
\bibitem{34} Donald, M. J., 1986, Comm. Math. Phys. {\bf 105}, 13; M.J.
Donald, Math. Proc. Camb. Phil. Soc., {\bf 101}, 363 (1987).
%
\bibitem{35} Donald, M. J., 1992, Found. Phys.,{\bf 22}, 1111.
%
\bibitem{36} Donald, M. J., and M. Horodecki, 1999, Physics Letters A {\bf
264}, 257.
%
\bibitem{37} Einstein, A., B. Podolsky and N. Rosen, 1935, Phys. Rev. {\bf 47},
777.
%
\bibitem{38} Eisert, J., T. Felbinger, P. Papadopoulos, M. B. Plenio, and M.
Wilkens, 2000, Phys. Rev. Lett. {\bf 84} 1611.
%
\bibitem{39} Everett, H.,  III, 1973, ``The Theory of The Universal
Wavefunction ", in ``The Many--Worlds Interpretation of Quantum
Mechanics" edited by B. DeWitt and N. Graham (Princeton
University Press).
%
\bibitem{40} Fannes, M., 1973, Commun. Math. Phys. {\bf 31}, 291; M. Nielsen,
{\em Continuity bounds for entanglement}, quant-ph/9908086 (1999).
%
\bibitem{41} Feynman, R.P., 1996, ``Feynmann Lectures on Computation", edited
by A. J. G. Hey and R. W. Allen, (Addison-Wesley Publishing
Company, Inc.).
%
\bibitem{42} Fuchs, C. A., 1996, ``Distinguishability and Accessible
Information in Quantum Theory'', PhD thesis, The University of
New Mexico, Albuquerque, NM (lanl e-print server:
quant-ph/9601020).
%
\bibitem{43} Garey, M., and D. Johnson, 1979, {``Computers and intractability:
a guide to the theory of NP-completeness}, (Freeman, San
Francisco, 1979).
%
\bibitem{44} Garisto, R., and L. Hardy, 1999, Phys. Rev. A {\bf 60},  827.
%
\bibitem{45} Gisin, N., 1996, Phys. Lett. {\bf A 210}, 151, and references
therein.
%
\bibitem{46} Gordon, J. P., 1964, {\em ``Noise at Optical Frequences;
Information Theory}, Quantum Electronics and Coherent Light,
Proc. Int. School Phys. ``Enrico Fermi, Course XXXI, ed. P. A.
Miles, p.p. 156 (Academic Press, New York).
%
\bibitem{47} Grover, L., 1996, {\em A fast quantum mechanical algorithm for
database search}, Proceedings of the 28th ACM Symposium on Theory
of Computing, p. 212. Also quant-ph/9605043.
%
\bibitem{48} Hausladen, P., R. Jozsa, B. Schumacher, M. Westmoreland and W. K.
Wootters, 1996, Phys. Rev. A {\bf 54}, 1869.
%
\bibitem{49} Hayashi, M., 1997, ``Asymptotic Attainment for Quantum Relative
Entropy", lanl e-print server: quant-ph/9704040.
%
\bibitem{49b} Hayden, P. M. , Horodecki, M. Terhal, B. M.,  2000, ``The
asymptotic entanglement cost of preparing a quantum state", lanl
e-print server no. 0008134.
%
\bibitem{50} Henderson, L., and V. Vedral, 2000, Phys. Rev. Lett. {\bf 84},
2263.
%
\bibitem{51} Hiai, F., and D. Petz, 1991, Comm. Math. Phys. {\bf 143}, 99;
%
\bibitem{52} Hogg, T., and B. A. Huberman, 1983, Phys. Rev. A {\bf 28}, 22.
%
\bibitem{53} Holevo, A. S., 1973, {\em Problemy Peredachi Informatsii}, {\bf
9}, 3 (1973) [A. S. Kholevo, Problems of Information Transmission,
{\bf 9}, 177]; for the continuous case see H. P. Yuen and M.
Ozawa, Phys. Rev. Lett. {\bf 70}, 363 (1993).
%
\bibitem{54} Holevo, A. S., 1982, {``Probabilistic and Statistical Aspects of
Quantum Theory"}, (North-Holland Publishing Company, Amsterdam).
%
\bibitem{55} Holevo, A. S., 1998, IEEE. T. Inform. Theory {\bf 44} 269.
%
\bibitem{56} Horodecki, M., P. Horodecki, and R. Horodecki, 1996, Phys. Lett. A
{\bf 223}, 1.
%
\bibitem{57} Horodecki, M., P. Horodecki and R. Horodecki, 1998a, Phys. Rev.
Lett. {\bf 80} 5239.
%
\bibitem{58} Horodecki, M., 1998b, Phys. Rev. A {\bf 57} 3364.
%
\bibitem{59} Horodecki, M., P. Horodecki and R. Horodecki, 2000a, lanl gov
e-print no quant-ph/0006071.
%
\bibitem{60} Horodecki, M., P. Horodecki, R. Horodecki, 2000b, Phys. Rev. Lett.
{\bf 84}, 2014.
%
\bibitem{61} Ingarden, R. S., A. Kossakowski and M. Ohya, 1997, {``Information
Dynamics and Open Systems - Classical and Quantum Approach"},
(Kluwer Academic Publishers, Dordrecht).
%
\bibitem{62} Ingarden, R. S., 1976, Rep. Math. Phys. {\bf 10}, 43.
%
\bibitem{63} Jaynes, E. T., and F. W. Cummings, 1963, Proc. IEEE {\bf 51}, 89.
%
\bibitem{64} Jones, J. A. and M. Mosca, 1998a, J. Chem. Phys. {\bf 109} 1648.
%
\bibitem{65} Jones, J. A., M. Mosca and R. H. Hansen, 1998b, Nature {\bf 393}
344. I. L. Chuang, N. Gershenfeld, and M. Kubinec, Phys. Rev.
Lett. {\bf 80}, 3408.
%
\bibitem{66} Kolmogorov, A. N., 1950, {``Foundations of The Probability
Theory"}, (Chelsea pub. company, New York).
%
\bibitem{67} Kraus, B., J. I. Cirac, S. Karnas and M. Lewenstein, 2000, Phys.
Rev. A {\bf 61} 062302.
%
\bibitem{68} Kraus, K., 1983, {``States, Effects and Operations: Fundamental
Notions of Quantum Theory"}, Lecture Notes in Physics {\bf 180}
(Springer, Berlin, 1983); M. Ozawa, J. Math. Phys. {\bf 25}, 79
(1984).
%
\bibitem{69} Kullback, S., and R.A. Leibler, 1951, Ann. Math. Stat. {\bf 22},
79.
%
\bibitem{70} Landauer, R., 1961, IBM J. Res. Dev. {\bf 5}, 183; C.H. Bennett,
IBM J. Res. Dev. {\bf 32}, 16 (1988); T.~Toffoli, Math. Systems
Theory {\bf 14}, 13 (1981).
%
\bibitem{71} Lebedev, D. S., and L. B. Levitin, 1963, Sov. Phys. Dokl. {\bf 8}
377.
%
\bibitem{72} Lewenstein, M., D. Bruss, J. I. Cirac, B. Kraus, M. Kus, J.
Samsonowicz, A. Sanpera and R. Tarrach, 2000, Separability and
Distillability in composite quantum systems - a primer, lanl gov
e-print server no. 0006064 and references therein.
%
\bibitem{73} Lindblad, G., 1974, Comm. Math. Phys. {\bf 39}, 111.
%
\bibitem{74} Lindblad, G., 1975, Comm. Math. Phys. {\bf 40}, 147.
%
\bibitem{75} Lo, H., and S. Popescu, 1999, Phys. Rev. Lett. {\bf 83},  1459.
%
\bibitem{76} Mackey, G. W., 1963, {``Mathematical Foundations of Quantum
Mechanics"}, (New York, W. A. Benjamin Inc.).
%
\bibitem{77} von Neumann, J., 1955, {``Mathematische Grundlagen der
Quantenmechanic"} (Springer, Berlin, 1932; English Translation,
Princeton University Press, Princeton).
%
\bibitem{78} Ohya, M., and D. Petz, 1993, {``Quantum Entropy and Its Use"},
Texts and Monographs in Physics, (Berlin: Springer-Verlag).
%
\bibitem{79} Papadimitriou, C. H., 1995, ``Computational complexity",
(Addison-Wesley publishing company, New York).
%
\bibitem{80} Partovi, M. H., 1989, Phys. Lett. A {\bf 137}, 445, and the
references therein.
%
\bibitem{81} Penrose, O., 1973, {``Foundations of Statistical Mechanics"},
(Oxford University Press, Oxford).
%
\bibitem{82} Peres, A., 1996, Phys. Rev. Lett. {\bf 77}, 1413.
%
\bibitem{83} Peres, A., 1993,``Quantum Theory: Concepts and Methods", (Kluwer
Academic Publishers).
%
\bibitem{84} Plenio, M. B., and V. Vedral, 1998, Cont. Phys. {\bf 38}, 431.
%
\bibitem{85} Popescu, S., and D. Rohrlich, 1997, Phys. Rev. A {\bf 56}, R3319.
%
\bibitem{86} Rains, E. M., 1999, Phys. Rev. A {\bf 60} 173 ; E. M.
Rains, 1999, Phys. Rev. A {\bf 60} 179.
%
\bibitem{87} Redhead, M., 1987, ``Incompleteness, Nonlocality and Realism",
(Clarendon Press, Oxford).
%
\bibitem{88} Reed, M., and B. Simon, 1980, ``Methods of Modern Mathematical
Physics--Functional Analysis", (Academic Press, New York).
%
\bibitem{89} Sanov, I. N., 1957, Mat. Sbornik (Moscow) {\bf 42}, 11.
%
\bibitem{90} Schmidt, E., 1907, The original reference  is E. Schmidt, ``Zur
Theorie der linearen und nicht linearen Integralgleichungen",
Math. Annalen {\bf 63}, 433 (1907), in the context of quantum
theory see H.~Everett III, in {\em The Many-World Interpretation
of Quantum Mechanics}, ed. B.S.~DeWitt and N.~Graham (Princeton
University Press, Princeton, 1973) p. 3, and H.~Everett III,
Rev.~Mod.~Phys. {\bf 29} 454 (1957). A graduate level textbook by
A.~Peres, {``Quantum Theory: Concepts and Methods"}, (Kluwer,
Dordrecht, 1993), Chapt.~5 includes a brief description of the
Schmidt decomposition.
%
\bibitem{91} Schr\"{o}dinger, E., 1935, Naturwissenschaften {\bf 23}, 807, 823,
844.
%
\bibitem{92} Schumacher, B., 1995, Phys. Rev. A {\bf 51}, 2738; R. Jozsa and
B. Schumacher, J. Mod. Opt. {\bf 41}, 2343 (1994).
%
\bibitem{93} Schumacher, B., 1996, Phys. Rev. A {\bf 54} 2614.
%
\bibitem{94} Schumacher, B., and M. D. Westmoreland, 1997, Phys. Rev. A, {\bf
56}, 131;
%
\bibitem{95} Schumacher, B. W., and M. D. Westmoreland, 2000, {\em Relative
entropy in quantum information theory}, lanl e-print no.
quant-ph/0004045.
%
\bibitem{96} Shannon, C. E., and W. Weaver, 1949, ``The Mathematical Theory of
Communication", (University of Illinois Press, Urbana, IL).
%
\bibitem{97} Shor, P. W., 1996, {\em In Proc. $35th$ Annual Symposium on
Foundations of Computer Science}, ed. S. Goldwasser. (IEEE
Computer Society Press, Nov. 1994) pp. 124-134; for a review of
this algorithm see A. Ekert and R. Jozsa, 1996, Rev. Mod. Phys.
{\bf 68}, 733.
%
\bibitem{98} Steane, A., 1997, Appl. Phys. B. {\bf 64}, 623; D. G. Cory, R.
Laflamme, E. Knill, L. Viola, T. F. Havel, N. Boulant, G. Boutis,
E. Fortunato, S. Lloyd, R. Martinez, C. Negrevergne, M. Pravia,
Y. Sharf, G. Teklemariam, Y. S. Weinstein, W. H. Zurek, {\em "NMR
Based Quantum Information Processing: Achievements and
Prospects"}, Fortschritte der Physik special issue (2000).
%
\bibitem{99} Stern, T. E., 1960, IEEE Trans. Info. Theory {\bf 6}, 435.
%
\bibitem{100} Tipler, F. J., 1994, ``The Physics of Immortality", (Bantam
Doubleday Dell Publishing Group, Inc., New York).
%
\bibitem{101} Tolman, R. C., 1938, {``The Principles of Statistical Mechanics"},
(Oxford University Press, Oxford).
%
\bibitem{102} Umegaki, H., 1962, Kodai Math. Sem. Rep. {\bf 14}, 59.
%
\bibitem{103} Vandersypen, L. M. K., et al., 2000, {\em Experimental realization
of order-finding with a quantum computer}, lanl e-print no.
quant-ph/0007017.
%
\bibitem{104} Vedral, V., M. B. Plenio, M. A. Rippin, and P. L. Knight, 1997a,
Phys. Rev. Lett. {\bf 78}, 2275.
%
\bibitem{105} Vedral, V., M. B. Plenio, K. Jacobs, and P. L. Knight, 1997b,
Phys. Rev. A {\bf 56}, 4452.
%
\bibitem{106} Vedral, V., M. A. Rippin and M. B. Plenio, 1997c, J. Mod. Opt.
{\bf 44}, 2185.
%
\bibitem{107} Vedral, V., and M. B. Plenio, 1998, Phys. Rev. A, {\bf 57}, 1619.
%
\bibitem{108 }Vedral, V., 2000, Proc. Roy. Soc. Lond. A {\bf 456}, 969.
%
\bibitem{109} Vidal, G., 2000, J. Mod. Opt. {\bf 47} 355.
%
\bibitem{110} Vollbrecht, K. G. H., and R. F. Werner, 2000, "Entanglement
Measures under Symmetry", lanl e-print no. quant-ph/0010095.
%
\bibitem{111} Wehrl, A., 1978, Rev. Mod. Phys. {\bf 50}, 221.
%
\bibitem{112} Werner, R. F., 1989, Phys. Rev. A {\bf 40}, 4277.
%
\bibitem{113} Wootters, W.K., and W.H. Zurek, 1982, Nature {\bf 299}, 802.
%
\bibitem{114} Wootters, W. K., 1998, Phys. Rev. Lett {\bf 80}, 2245.
%
\bibitem{115} Yamamoto, Y., and H. A. Haus, 1986, Rev. Mod. Phys. {\bf 58}.
%
\bibitem{116} Zalka, C., 1999, Phys. Rev. A {\bf 60}, 2746.

\end{thebibliography}
\end{document}